\documentclass{JAIS}
\journal{JAIS-ID}
\articletype{Review}
\vol{2023}
\received{xx November 2023}
\published{xx December 2023}

\usepackage[pdftitle={Review},pdfauthor={PK,KN}]{hyperref}
\usepackage{amsmath,amsthm,amsfonts,amssymb}

\usepackage{atlasphysics} 
\usepackage{paralist}
\usepackage[tight]{subfigure}
\usepackage[version=4]{mhchem} 
\usepackage{wrapfig}
\usepackage{siunitx}

\usepackage{verbatim}
 
\usepackage{graphicx} 
\usepackage[shortcuts]{extdash}

\usepackage{xargs}                   
\usepackage[colorinlistoftodos,prependcaption,textsize=tiny]{todonotes}
\newcounter{todocounter}
\newcommandx{\unsure}[2][1=]{\stepcounter{todocounter}\todo[linecolor=red,backgroundcolor=red!25,bordercolor=red,#1]{\thetodocounter:#2}\PackageWarning{TODO:}{#1!}}
\newcommandx{\change}[2][1=]{\stepcounter{todocounter}\todo[linecolor=blue,backgroundcolor=blue!25,bordercolor=blue,#1]{\thetodocounter:#2}\PackageWarning{TODO:}{#1!}}
\newcommandx{\info}[2][1=]{\stepcounter{todocounter}\todo[linecolor=green,backgroundcolor=green!25,bordercolor=green,#1]{\thetodocounter:#2}\PackageWarning{TODO:}{#1!}}
\newcommandx{\improvement}[2][1=]{\stepcounter{todocounter}\todo[linecolor=purple,backgroundcolor=purple!25,bordercolor=purple,#1]{\thetodocounter:#2}\PackageWarning{TODO:}{#1!}}
\newcommandx{\thiswillnotshow}[2][1=]{\stepcounter{todocounter}\todo[disable,#1]{#2}}

\usepackage{cooltooltips}

\usepackage{caption}
\usepackage{lineno}

\begin{document}

\title{Searches for light Dark Matter with Spherical Proportional Counters}

\author{Patrick Knights\auno{1} and Konstantinos Nikolopoulos\auno{1,2} }
\address{$^1$ School of Physics and Astronomy, University of Birmingham, B15 2TT, Birmingham, United Kingdom}
\address{$^2$ Institute for Experimental Physics, University of Hamburg, 22761, Hamburg, Germany\\
\vspace{0.2cm}
Corresponding author: Konstantinos Nikolopoulos\\
Email address: konstantinos.nikolopoulos@cern.ch
}

\begin{abstract}
Elucidating the nature of dark matter is a key priority that would involve discovering new fundamental physics and is essential for understanding the structure and evolution of the universe.
Despite the decades-long ever-more-sensitive searches, the particle content of dark matter remains elusive. 
Direct searches for dark matter candidates, to-date, focused mainly on candidates in the $10\;\gev$ to $1\;\tev$, however, more recently lighter candidates with sub-GeV mass have been brought to the spotlight. 
This is an experimentally challenging mass region, which remains largely uncharted. 
The spherical proportional counter is a new type of gaseous detector which exhibits several features that make it ideally suited for the exploration of this mass range. In this article the invention and development of the spherical proportional counter are presented, its applications in the search for particle dark matter and beyond are reviewed, and possible future directions are discussed. 
\end{abstract}

\maketitle


\section{Introduction}
\label{sec:introduction}
Diverse astronomical~\cite{Bertone:2010zza,Clowe:2006eq} and cosmological~\cite{Planck2015} observations, on scales ranging from galaxies to the entire universe, provide powerful evidence that approximately 85\% of the matter content of the universe consists of non-baryonic cold dark matter (DM).
As reflected by the number of on-going and planned activities, which span orders of magnitude in scale and complexity~\cite{Battaglieri:2017aum,Billard:2021uyg}, elucidating the nature of DM is a key priority, that would involve discovering new fundamental physics and is essential for understanding the structure and  evolution of the universe.

Although the particle nature of DM is currently unknown, many theories predict that it consists of stable, or extremely long-lived, neutral particles with mass from a few eV to a few TeV and small couplings to the  Standard Model (SM) particles. These couplings may be directly involved in DM production in the early universe, or play a role 
in a cosmological scenario that results in the observed DM abundance. In many theories DM-SM interactions are also motivated by aspirations to address other theoretical or observational challenges, e.g. the electroweak hierarchy problem,  the strong CP problem, or the neutrino masses. Over the past decades, the scrutiny fell on DM candidates with masses between 10~\gev\ and 1~\tev~\cite{Liu:2017drf}, because these are predicted in models addressing the hierarchy problem and because they may simply be a non-relativistic relic of the thermal bath of the early universe~\cite{Goodman:1984dc}.  Recently, DM candidates in a wider mass range have been brought into focus. These may have been produced thermally or non-thermally in the early universe and may have a variety of coupling types and strengths to the SM. 
	
Direct detection experiments aspire to detect DM particles from the Milky Way DM halo~\cite{Evans:2018bqy} via their coherent 
scattering off a target nucleus~\cite{MarrodanUndagoitia:2015veg,Schumann:2019eaa},
which may proceed through spin-independent or spin-dependent couplings.
The momentum transferred to the target gives rise to a nuclear recoil and results into the production  of heat,  the release of scintillation photons, or the direct ionisation of atomic electrons. Typically, detection strategies focus on one or two of these signals.
The expected event rates are orders of magnitude lower than
natural radioactivity, requiring large exposures with detectors made
of radiopure materials, shielded and operated deep underground~\cite{Ianni:2017vqi} to
protect against ambient radioactivity and cosmic rays, respectively. Presently, the state-of-the-art constraints on DM-nucleus interactions come from liquid noble element experiments using xenon~\cite{LZ:2022lsv,PandaX-4T:2021bab,XENON:2018voc,PandaX-II:2017hlx,LUX:2016ggv} and argon~\cite{DEAP:2019yzn,Agnes:2018ves}, with maximum sensitivity to DM mass of approximately $30\;\gev$. 

For light DM candidates of mass $m_{\text{DM}}$ the energy transferred to a nuclear recoil is given by 
\begin{equation}
E_{nr} = \frac{1}{2}m_{\text{DM}}v_{\text{DM}}^{2} \frac{4 m_{\text{DM}} m_N}{\left( m_{\text{DM}}+m_N\right)^2}\frac{1+\text{cos}\theta}{2},
\end{equation}
with the mean energy transfer approximately given by,
\begin{equation}
E_{nr}\simeq1\;\eV\times \left(m_{\text{DM}}/100 \MeV\right)^2\left(10 \GeV/m_N\right),
\end{equation}
where $v_{\text{DM}}$ is the DM velocity, $\theta$ is the scattering angle, and $m_N$ is the mass of the target nucleus.
Thus, liquid noble element detectors are not ideal for light DM searches and efforts are invested in 
recovering some sensitivity via the Migdal effect~\cite{Migdal1941, Bernabei:2007jz, Ibe:2017yqa, Dolan:2017xbu}, where, with a small probability, an atomic electron can be emitted following the sudden perturbation of the nucleus by the scattering with an electrically neutral projectile, and ``bremsstrahlung'' photon emission~\cite{Kouvaris:2016afs}. The former leads to significantly better sensitivity than the latter~\cite{Aprile:2019jmx, LUX:2018akb}, but 
still several orders of magnitude weaker constraints than those obtained in the $30\;\gev$ mass region, cf. Refs~\cite{LZ:2022lsv} and~\cite{LZ:2023poo}. 
Neither effect has been observed in nuclear scattering yet, and experimental efforts to establish the Migdal effect are on-going~\cite{LZ:2023poo,Xu:2023wev,Araujo:2022wjh,Adams:2022zvg,Nakamura:2020kex,Bell:2021ihi}.
Regardless, the Migdal effect is by now regularly employed by experiments to extend their sensitivity to lower DM masses~\cite{LUX:2018akb, EDELWEISS:2019vjv,CDEX:2019hzn,XENON:2019zpr,SENSEI:2020dpa,CDEX:2021cll,EDELWEISS:2022ktt}.
Alternatively, the doping of the the liquid noble elements with hydrogen is being considered~\cite{Bell:2023uvf}, but it is still at the research and development stage.

Other approaches to probe light DM candidates involve  experiments relying on cryogenic solid-state detectors to measure -- with extremely low energy detection thresholds -- the heat produced as a result of the DM interaction, potentially combined with either ionisation or light readout. 
Prominent examples are  CRESST\=/III~\cite{Angloher:2015ewa, Abdelhameed:2019hmk},
featuring scintillating \ce{CaWO4}-crystals -- also doped with $^{6}$Li~\cite{CRESST:2022jig} for enhanced sensitivity to spin-dependent interactions --  with light and phonon
readout, and SuperCDMS~\cite{Agnese:2018gze,Agnese:2016cpb}, 
utilising Ge and Si-crystals with ionisation and phonon readout. 
However, the relatively heavy target atoms, 
the large quenching of the nuclear recoil signals in the
ionisation and scintillation channels~\cite{Lindhard1963}, and the reduced background rejection capabilities for low recoil energies, limits their sensitivity.

For DM candidates with mass in the \keV\ to \MeV\ range, their scattering off atomic electrons may also be utilised~\cite{Essig:2011nj}. The resulting ionised electron gives rise to a small signal, potentially, observable in detectors with low energy detection threshold, down to single-electron sensitivity. 
These searches probe different potential DM couplings with respect to the searches relying on nuclear scattering, and are thus complementary.
Typically, these DM-electron interactions,
are parametrised by a cross section and a momentum transfer-dependent DM form factor,  typically, taken to correspond to effective contact interactions through a heavy mediator or a very light scalar or vector mediator. 

Following this brief introduction to the searches for DM particles with mass in the sub-\GeV\ range, the development of the spherical proportional counter, a novel and versatile gaseous detector concept, will be reviewed in detail.
In Section~\ref{sec:invention} the early development of the detector in the middle of the 2000s is presented and is placed in the broader context of particle physics of that time. 
The application of the spherical proportional counter in field of direct DM searches is discussed in Section~\ref{sec:spcDM}, along with its main features.
In Section~\ref{sec:simulation}, the developments in simulation techniques for gaseous detectors, and the spherical proportional counter in particular, are briefly summarised. The availability of reliable, fast, and flexible simulation tools is of paramount importance for the fast progress of the detector development.
In Section~\ref{sec:sensor} the development of the read-out sensor of the spherical proportional counter, which constitutes the corner-stone of detector performance, is presented comprehensively. Emphasis is placed on the requirements leading to the various design choices, and the compromises that were required. 
Section~\ref{sec:quenching} is dedicated to a brief exposition of the ionisation quenching factor and its measurement in the context of gas mixtures relevant for DM searches.
The detector radiopurity considerations are discussed in Section~\ref{sec:radiopurity}, alongside the relevant techniques. 
In Section~\ref{sec:opportunities} the physics potential of the next generation fully electroformed underground detector is presented. 
The application of the spherical proportional counter as a neutron detector for in-situ spectroscopic measurements of the background in underground laboratories and in industrial applications is discussed in Section~\ref{sec:neutrons}.
Finally, a summary and future research and developments directions are discussed in Section~\ref{sec:summary}.

\section{Invention of the spherical proportional counter}
\label{sec:invention}

Neutrinos, by virtue of being the only particles of the Standard Model to interact solely through weak interactions and their extremely small mass, have been a continuous source of exciting experimental and theoretical developments in physics. In 1964, soon after the 1956 first observation of neutrinos~\cite{Cowan:1956rrn}, Davies and his collaborators tried to observe neutrinos produced in the Sun~\cite{Davis:1964hf,Davis:1968cp} giving rise to the solar neutrino problem~\cite{Trimble:1973ca,Haxton:1995hv}. This long-standing puzzle comprised in a factor of three  discrepancy between the rate of electron neutrinos from the Sun observed by detectors on earth and the rate predicted by the Standard Solar Model. This was followed by a period of intense scientific activity, which yielded several new ideas for dedicated detectors, e.g. the HELLAZ concept for a 2000$\;\si{\meter}^3$ high-pressure helium gaseous time projection chamber (TPC)~\cite{Nygren:1974nfi} operating at liquid nitrogen temperatures~\cite{BONVICINI1994438,Gorodetzky:1999ty, Dolbeau:2005kt} to detect recoil electrons from the elastic scattering of solar neutrinos. These efforts culminated to the observation of neutrino oscillations in atmospheric neutrinos, interpreted as $\nu_\mu\to\nu_\tau$ oscillations, by  Super-Kamiokande~\cite{Super-Kamiokande:1998kpq} and the demonstration by SNO that the, until then, mysteriously missing electron neutrinos were observed as muon and $\tau$-lepton neutrinos~\cite{SNO:2001kpb}. This resolved the solar neutrino problem and initiated a period of ever more precise study of the neutrino properties.

\begin{figure}[h]
  \centering
 \includegraphics[angle=-90,width=0.65\linewidth]{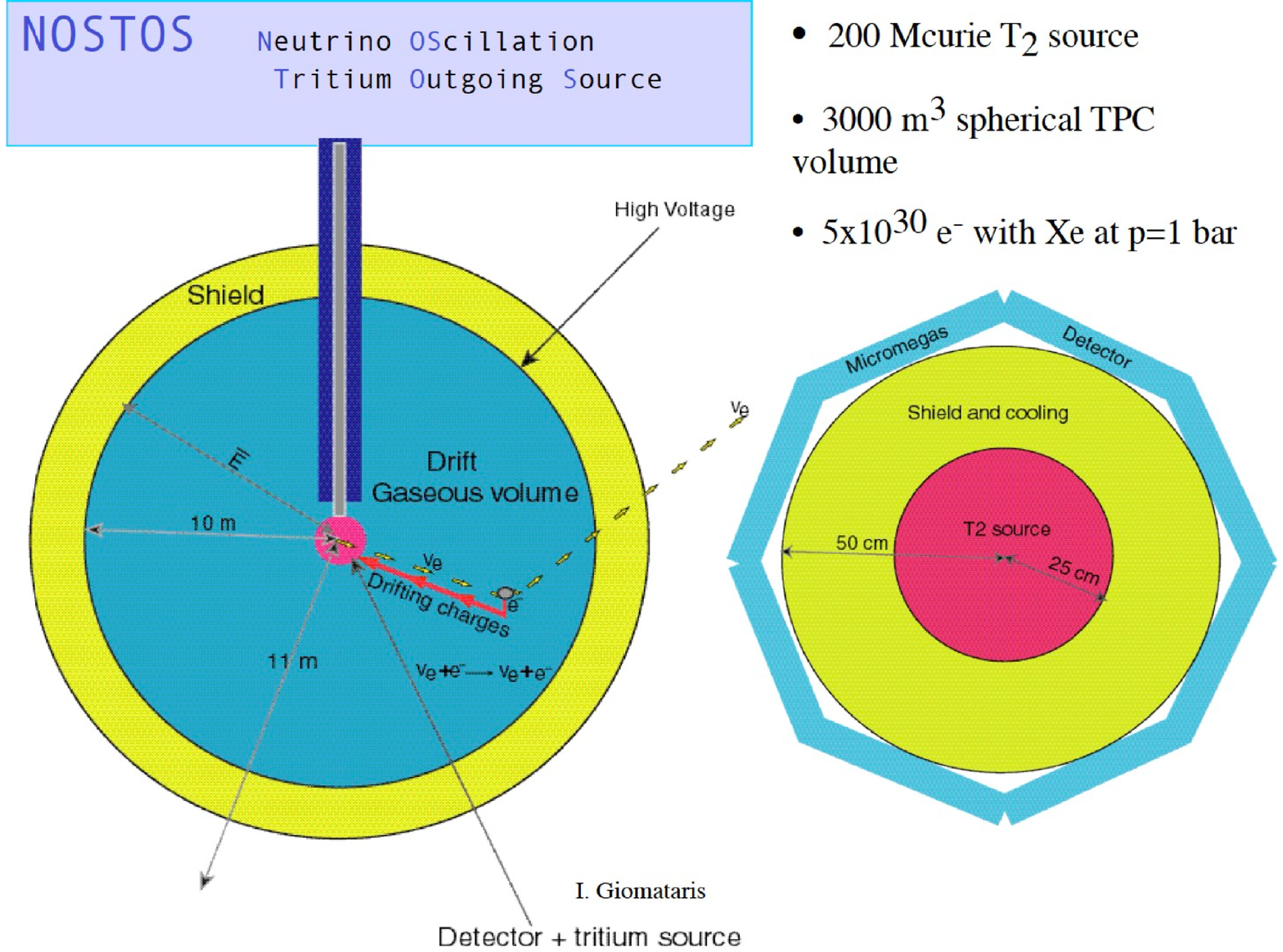}
\caption{Conceptual design of the NOSTOS experiment. The image on the right shows a close-up of the `Detector + tritium source' at the centre of the left image. Figure by I.~Giomataris.
\label{fig:nostos}}
\vspace{-0.45cm}
\end{figure}

Against this backdrop, Ioannis Giomataris and Ioannis Vergados proposed in 2003 the development of a spherical gaseous TPC of about 10 meter in radius with a strong tritium source at the centre~\cite{Giomataris:2003bp}. The proposed experiment was named ``Neutrino OScillation Tritium Outgoing Source'' (NOSTOS), and its conceptual design is presented in Fig.~\ref{fig:nostos}. This would enable $4\pi$ coverage for the study of low energy neutrinos from the decay:
\ce{^3_1H\to^3_2He + e- + \bar{\nu}_e}, through their scattering off electrons that would subsequently recoil with a maximum kinetic energy of 1.27 keV. 
The oscillation length of said neutrinos is similar to the detector radius and, thus, the modulation of neutrino flavour as a function of distance would be contained and directly observed within the spherical TPC. The position of the interaction would be estimated by the diffusion of the ionisation electrons that would directly affect the pulse rise-time, and thus the oscillation parameters would be obtained by a single experiment, suppressing several sources of systematic uncertainties. As a result, NOSTOS would provide sensitivity of a few percent in the measurement of $\sin^2 2\theta_{13}$, opening the possibility for the observation of CP violation in the leptonic sector. 
Furthermore, energy detection thresholds of approximately 100 -- 200 \eV\ were expected. This threshold could potentially be reduced  down to the detection of single ionisation electrons, as suggested by the high pressure operation studies conducted in the context of HELLAZ  which demonstrated high electron multiplication factors (gas gain), of order $10^6$, at 20 bar pressure~\cite{Gorodetzky:1999ty, Dolbeau:2005kt}. Such low energy detection thresholds would further enhance sensitivity for the measurement of neutrino magnetic moment, potentially down to $10^{-12}\mu_B$. A broad physics programme was developed for NOSTOS~\cite{Giomataris:2003bp,Vergados:2009ei, Dedes:2009bk, Vergados:2011gia,Vergados:2011na}, including -- among other topics -- searches for new physics in neutrino oscillometry and radiative electron neutrino scattering, measurements of the weak mixing angle with small effects from radiative corrections. Moreover, once the detector is constructed, a number of other uses may be envisioned, for example a dedicated world-wide network of detectors for neutrino detection from supernova explosions~\cite{Giomataris:2005fx}.

The original design concept of NOSTOS envisioned to equip the experiment with  Micromegas detectors~\cite{Giomataris:1995fq} at the centre, which were already considered for low-background applications~\cite{Collar:2000jq}. This could be achieved either by constructing spherical Micromegas detectors, following a comprehensive research and development effort, or, alternatively, by approximating a sphere by means of several flat Micromegas elements. 
With the decommissioning of the CERN Large Electron Positron collider~\cite{Myers:1990sk}, the opportunity came up to use some of its spherical copper RF cavities
with a diameter ($\varnothing$) of $1.3\;\si{\meter}$ and thickness of $6\;\si{\milli\meter}$ to construct a prototype spherical detector. In anticipation of an ``ideal'' solution for the amplification structure, it was decided to begin the tests using a small spherical anode as a proportional counter at the centre of the detector~\cite{Giomataris:2005bb}. Tests in a variety of gas pressures demonstrated that such a simple amplification element provided large gas gain and stable detector operation. Thus, it would enable fast progress in understanding the spherical TPC properties and in accomplishing part of the immediate detector development program. 

\begin{figure}[h]
\centering
\vspace{-0.2cm}
\subfigure[\label{fig:pulseShape}]{\includegraphics[width=0.52\linewidth]{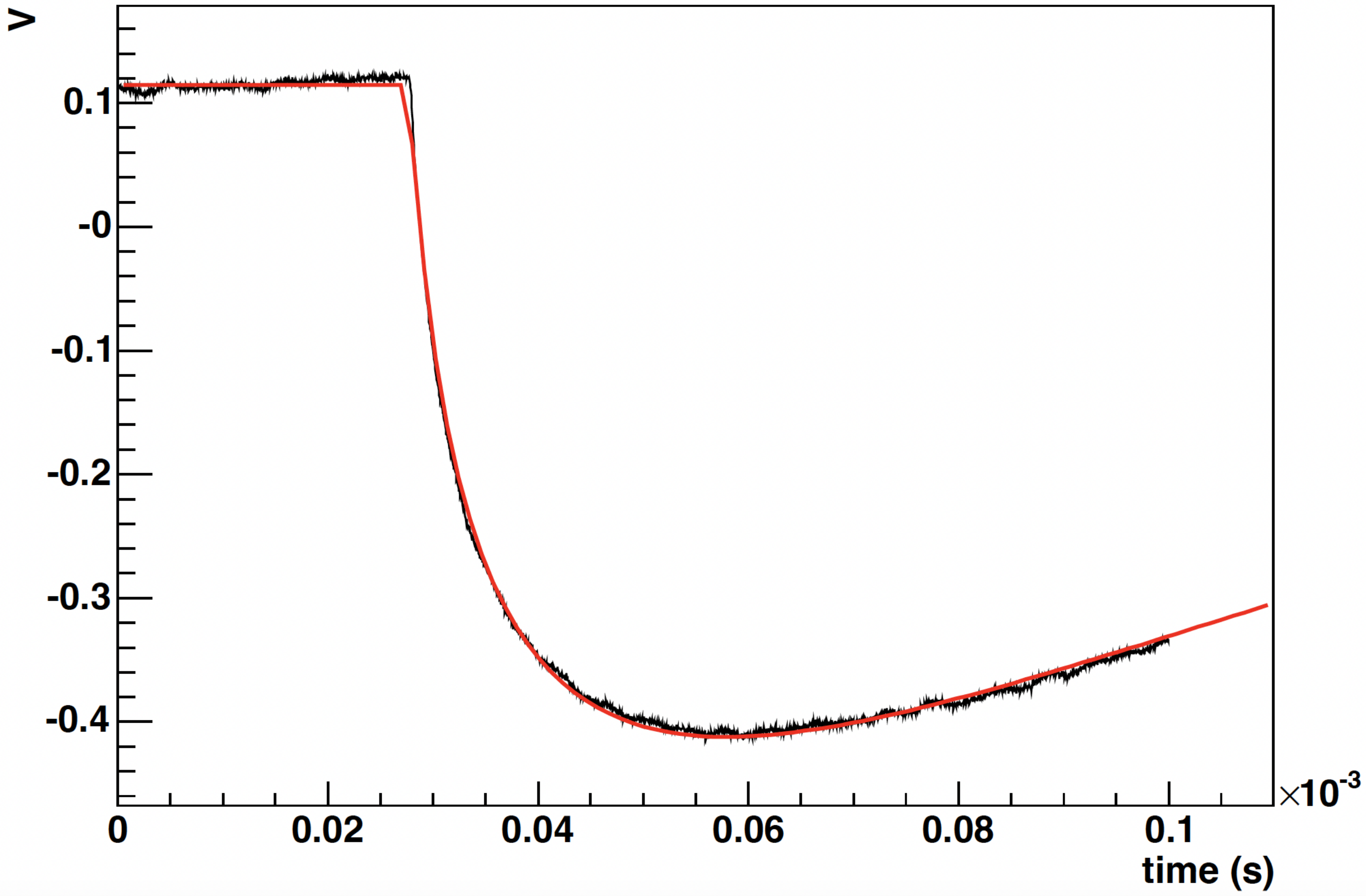}}
\subfigure[\label{fig:deconvolutionDistance}]{\includegraphics[width=0.365\linewidth]{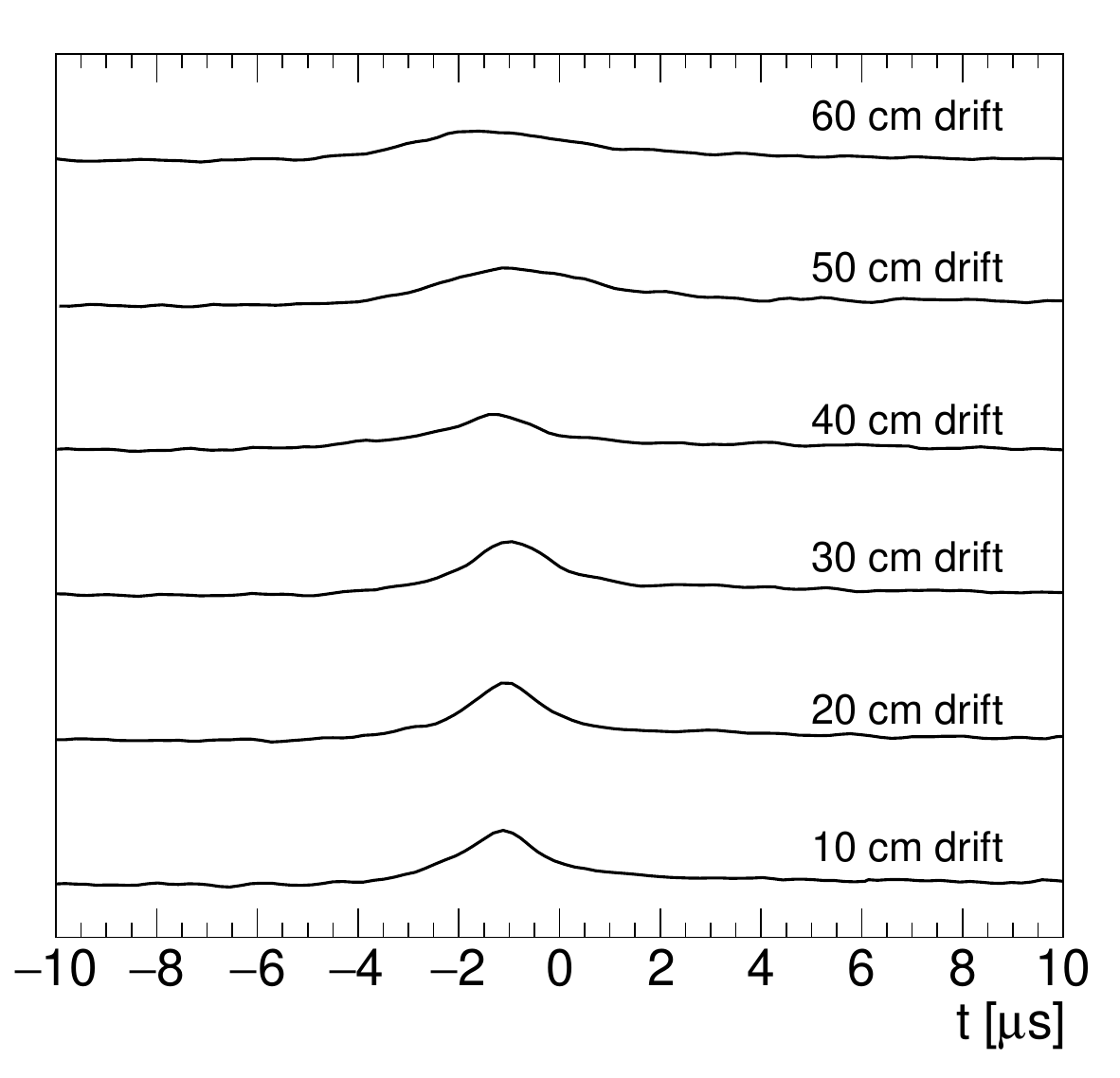}}
\vspace{-0.5cm}
\caption{Performance of a LEP RF cavity-based $\varnothing 1.3\;\si{\meter}$ spherical proportional counter with a $\varnothing 10\;\si{\milli\meter}$ anode: \subref{fig:pulseShape} Example of a pulse induced by an \ce{^{55}Fe} X-ray source, along with the expected pulse shape based on the detector and electronics response~\cite{Giomataris:2008ap}; 
\subref{fig:deconvolutionDistance} Effect of the diffusion on the deconvolved pulses. Each line corresponds to pulses taken with the \ce{^{55}Fe} source at a different
distance from the anode. Adapted from Ref.~\cite{Aune:2005is}.}
\vspace{-0.2cm}
\end{figure}

A number of investigations were performed with this early implementation~\cite{Giomataris:2005bb,Aune:2004bc,Aune:2005is,Aune:2005hv}, confirming several performance characteristics for the NOSTOS experiment. In Fig.~\ref{fig:pulseShape} the pulse induced by an \ce{^{55}Fe} X-ray source is presented, while in Fig.~\ref{fig:deconvolutionDistance} the pre-amplifier and ion-response have been deconvolved from the pulse, leaving the electron signal. 
The effect of the different radius of the initial interaction, which results to different drift distance for the ionisation electrons, is seen as an increased diffusion of the electrons, and thus a broader signal. The benefits of this design, namely a very low
level of electronic noise thanks to the very small detector capacitance, became apparent. As a result investigations focused on the spherical anode, which also permitted substantial flexibility on the size of the detector. 
Following these detailed studies, in 2008 the first investigation dedicated to the spherical proportional counter as a detector concept, independent of any specific experiment, was published~\cite{Giomataris:2008ap}, marking its offical date of birth. In the same publication, the topic of correcting for the distortions of the electric field induced by the anode support structure through the addition of an additional electrode, which was touched upon in an earlier publication, is being addressed at length.

\begin{wrapfigure}{R}{0.55\linewidth}
\centering
\vspace{-0.65cm}
\subfigure[\label{fig:schematic}]{\includegraphics[width=0.43\linewidth]{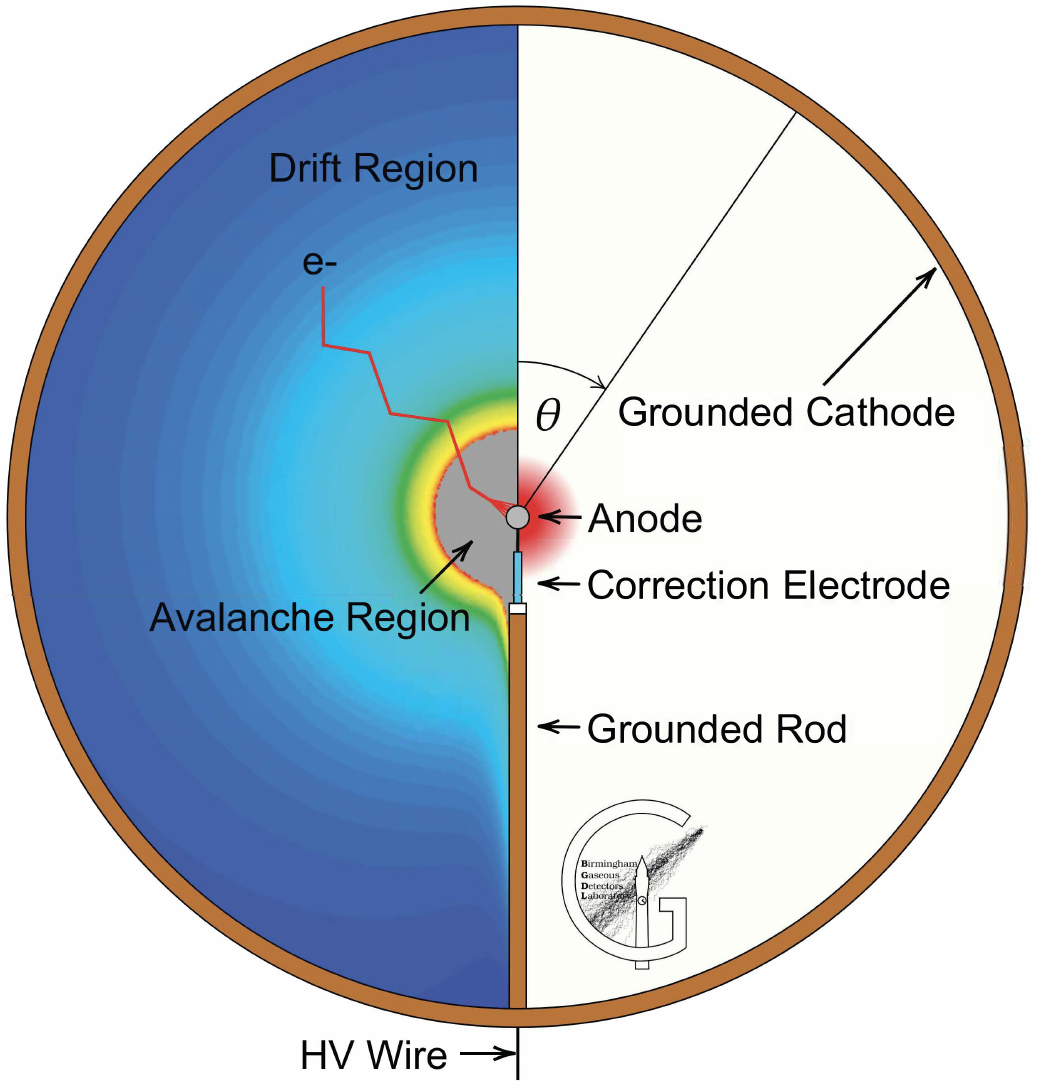}}
\subfigure[\label{fig:earlySensor}]{\includegraphics[width=0.56\linewidth]{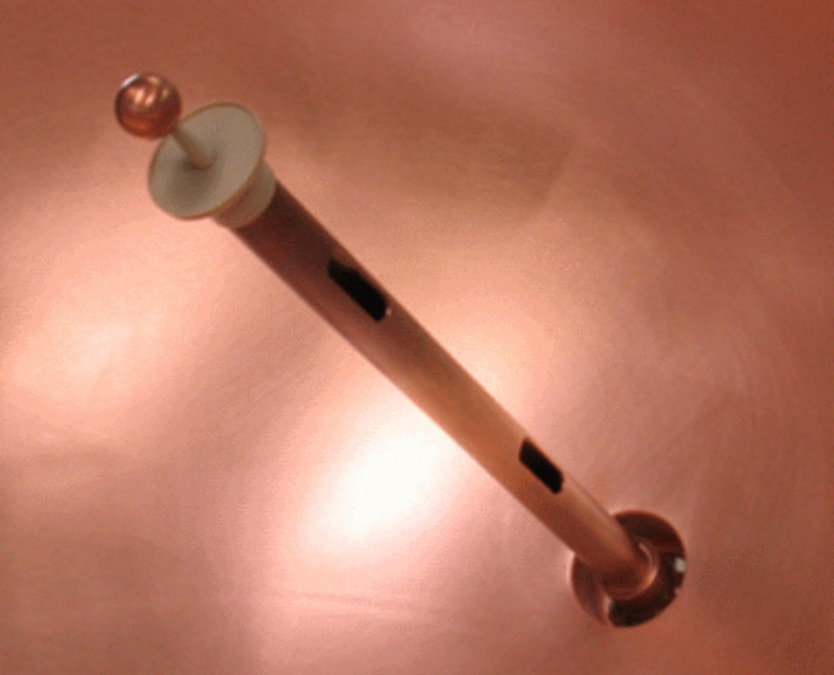}}
\vspace{-0.5cm}
\caption{\subref{fig:schematic} Schematic diagram and principle of operation of a spherical proportional counter, with electric field map overlaid.
\subref{fig:earlySensor} Early read-out structure with an $\varnothing 8\;\si{\milli\meter}$ anode.
}
\vspace{-1.05cm}
\end{wrapfigure}

In the baseline design, the detector, presented schematically in Fig.~\ref{fig:schematic},  consists of a grounded spherical shell, the cathode, and a small spherical co-centric anode
with a diameter of $\mathcal{O}$($1-10\;\si{\milli\meter}$), the sensor. An early realisation of this structure is shown in Fig.~\ref{fig:earlySensor}. Ideally, the electric field strength in the detector is:
\begin{equation}
\label{eq:field}
    \vec{E}(r) = \frac{V_0}{r^2} \frac{r_{a} r_{c}}{r_{c}-r_{a}} \hat{r}\underset{r_c\gg r_a}{\approx} \frac{V}{r^2} r_{a} \hat{r}\,.
\end{equation}
A sufficiently small central sphere radius $r_{a}$ biased at a voltage $V_0$ could enable electric fields near the anode to reach $\num{e4}$-$\num{e5}\;\si{\volt\per\centi\meter}$, as required based on the gas mixture used. This enables electrons drifted to within $10-100\;\si{\micro\meter}$ of the anode to produce further ionisation, inducing avalanches and, thus, signal amplification. As a result, the gas volume is naturally divided in a region where ionised electrons
drift to the anode and one, near the anode, where they multiply creating an avalanche.
The capacitance of the ideal detector is:
\begin{equation}
\label{eq:capacitance}
C = 4\pi\epsilon\epsilon_0\frac{r_ar_c}{r_c-r_a}\underset{r_c\gg r_a}{\approx} 4\pi\epsilon\epsilon_0r_a\,,
\end{equation}
where $\epsilon_0$ and $\epsilon$ are the electric permittivity of the vacuum and the relative electric permittivity of the gas used, respectively. From Eq.~\ref{eq:capacitance} it follows that the capacitance of the detector only depends on the radius of the central anode, and not on the radius of the cathode shell, and thus it is independent of the size of the gas volume. This means that large size detectors may be constructed without deterioration of the electronic noise.
In reality, the sensor requires a support structure and a sensor wire to carry the high voltage for the detector operation and the read-out of the signal. As a result, a significant contribution to the electronic noise arises from the length of the sensor wire. Nevertheless, this could be suppressed, for example by using a coaxial cable where the inner core is the sensor wire while the concentric shield is set to the same voltage. In this case, the induced charge to the sensor wire due to the capacitance would be minimised. 

The avalanche takes place near the small anode. The electrons rapidly reach the anode, while the positive ions drift towards the grounded shell. The signal in the detector is induced by the moving charges. For integration times typically used, from tens to hundreds of $\si{\micro\second}$, it predominantly originates from the moving ions. 
For the ideal detector geometry, the magnitude induced charged at time $t$ from a single electron avalanche with total charge $q$ created at $t=0$ is:
\begin{equation}
\label{eq:voltagePulse}
Q\left(t\right)= \frac{q}{\left(1/r_a-1/r_c\right)}\left[\frac{1}{\sqrt[3]{\frac{3\mu V_0 t}{1/r_a-1/r_c}+r^3_a}}-\frac{1}{r_c}\right]
\end{equation}
From Eq.~\ref{eq:voltagePulse}, the expected shape of the induced signal can be derived by accounting for the 
shaping constant of the preamplifier. An example is shown in Fig.~\ref{fig:pulseShape}, where a low-noise charge amplifier with large time constant  was used to pick-up the induced signal. Convolving the prediction of  Eq.~\ref{eq:voltagePulse} with the transfer function of the amplifier, the pulse 
observed from an \ce{^{55}Fe} source can be reproduced well.

While the $1/r^{2}$ electric field enables amplification of drifted electrons near the anode, it also means that the electric field becomes increasingly weak towards the outer detector volume. This results in long ionisation electron drift times in large volume detectors, and makes the detector susceptible to electron attachment by electronegative impurities in the gas.
This was already discussed in Ref.~\cite{Giomataris:2008ap}, where it is noted that the possibility of using Micromegas structures in a spherical arrangement at the centre of the detector, although more complex, has the major advantage  -- in addition to the well understood properties of the Micromegas in terms of energy and time resolution -- that the amplification and drift fields are decoupled, and the geometrical requirements are therefore less stringent. The detector read-out is a consideration that will be revisited again in Section~\ref{sec:multianode}.

\section{The spherical proportional counter as light dark matter detector}	
\label{sec:spcDM}
\begin{wrapfigure}{L}{0.45\linewidth}
  \centering
    \vspace{-0.6cm}
 \includegraphics[width=0.99\linewidth]{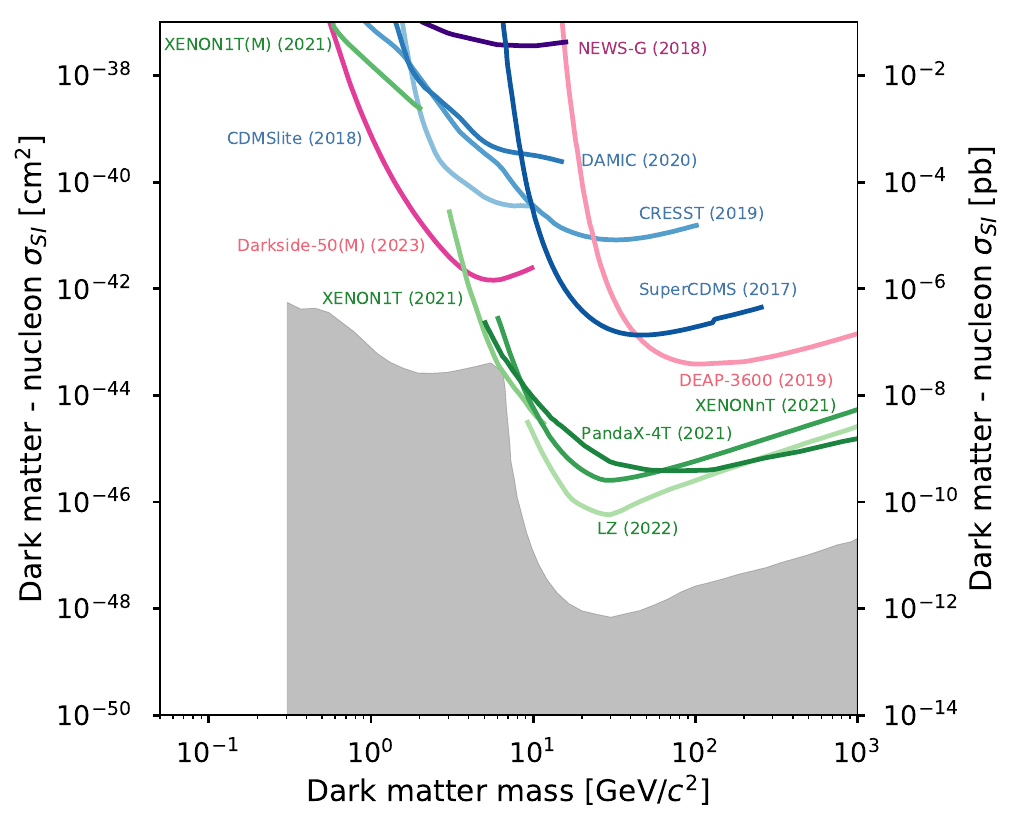}
  \vspace{-0.42cm}
\caption{Spin independent DM-nucleon cross section constraints as a function of DM candidate mass. Figure by L. Millins. \label{fig:pdgDMSI}}
\vspace{-0.5cm}
\end{wrapfigure}

The experimental constraints on
particle DM, as reviewed in Ref.~\cite{Billard:2021uyg}, are presented in Fig.~\ref{fig:pdgDMSI}, where a
significant disparity in the sensitivity to DM candidates with mass
above and below, approximately, 5~\gev\ is observed. In the latter
case, low mass DM candidates are currently, practically,
unconstrained.
However, as discussed in Section~\ref{sec:introduction}, 
DM candidates with masses below a few~\gev\ are predicted by several models, including:
\begin{inparaenum}[a)]
\item dark (or hidden) sectors containing light stable particles with \mev\ to \gev\ mass, communicating with the SM through a light mediator~\cite{Essig:2013lka};
\item asymmetric DM, aiming to explain the similarity between dark and ordinary matter abundances as an excess of DM particles over antiparticles which is dynamically related to ordinary matter-antimatter asymmetry~\cite{Petraki:2013wwa,Zurek:2013wia}; and
\item self-interacting DM, motivated by observations of the galactic structure~\cite{Tulin:2017ara}. 
\end{inparaenum}
Also, such candidates fit well in generic effective descriptions~\cite{Fitzpatrick:2012ix,Cerdeno:2018bty}.

For the spherical proportional counter to be used for light particle
DM searches~\cite{Gerbier:2014jwa,Nikolopoulos:2020vma}, a demonstration of its capability to detect low energy recoils is crucial. This was shown~\cite{Bougamont:2010mj} by exposing the spherical proportional counter prototype of $\varnothing 130\;\si{\centi\meter}$ filled with Ar:CH$_4$ (2\%) to
single electrons extracted from the copper cathode shell using a UV lamp. Furthermore, the response of the spherical proportional counter to  low energy nuclear recoils~\cite{Savvidis:2016wei} was demonstrated using a neutron source.

\begin{figure}[h]
  \centering
  \vspace{-0.5cm}
 \subfigure[\label{fig:Recoil}]{\includegraphics[width=0.330\linewidth]{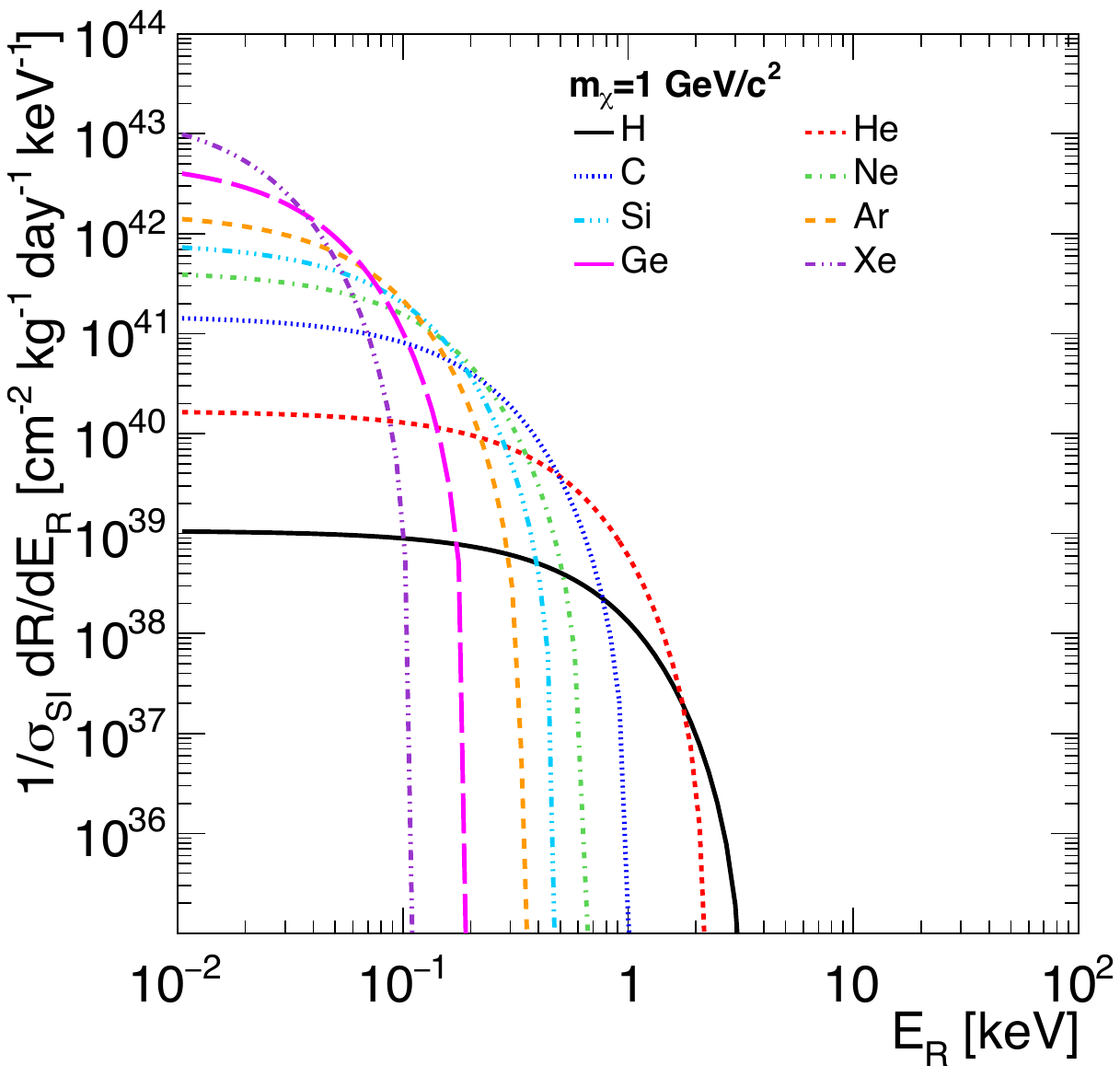}}
 \subfigure[\label{fig:risevsamplabel}]{\includegraphics[width=0.330\linewidth]{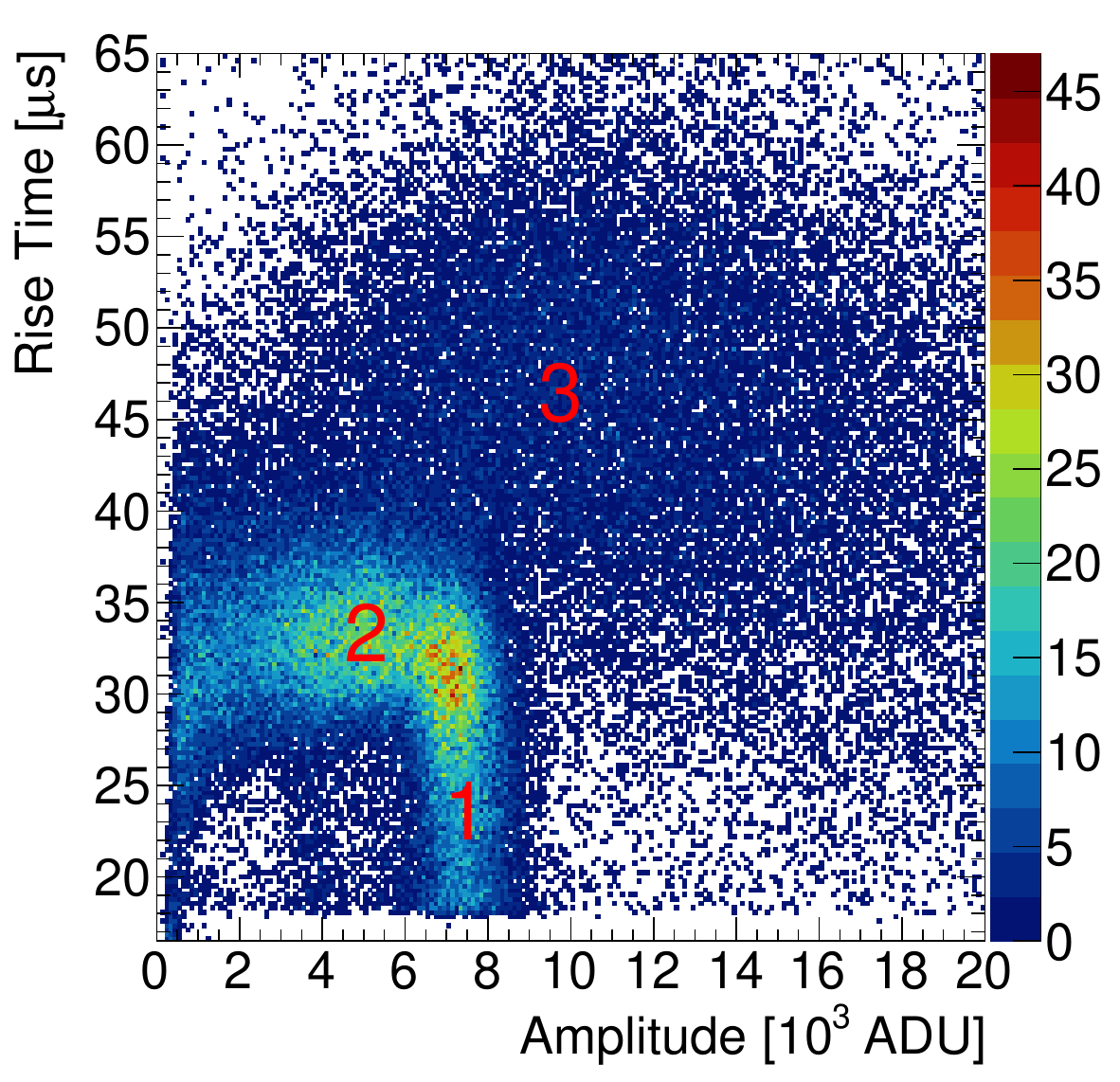}}
 \vspace{-0.5cm}
\caption{\subref{fig:Recoil} Recoil energy for 1~GeV mass DM candidate on different targets.
\subref{fig:risevsamplabel} Pulse rise time versus amplitude in an $\varnothing 30\;\si{\centi\meter}$ spherical proportional counter operated with $1.3\;\si{\bar}$ He:Ar:CH$_{4}$ (51.7\%:46\%:2.3\%)~\cite{Katsioulas:2022cqe}. The annotated populations are (1) 5.9~keV photons from an $^{55}$Fe source interacting in the gas volume, (2) interactions near the cathode, and (3) cosmic muon interactions.
}
\vspace{-0.4cm}
\end{figure}

\begin{wrapfigure}{R}{0.25\textwidth}
  \centering
  \vspace{-0.5cm}
  \includegraphics[width=0.24\textwidth]{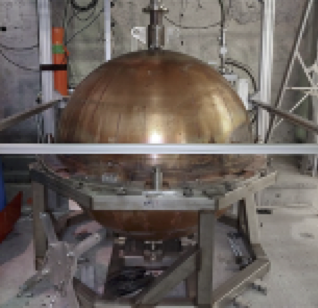}
  \vspace{-0.25cm}
  \caption{The $\varnothing$1.4~m NEWS\=/G detector with electroplated inner surface~\cite{Balogh:2020nmo}.\label{fig:snoglobe}}
  \vspace{-0.5cm}
\end{wrapfigure}
Beyond the attractive features of the spherical proportional counter already discussed in Section~\ref{sec:invention}, advantages directly related to light DM searches, include:
\begin{inparaenum}[a)]
\item {kinematic light DM candidate-target matching}, operating the detector with light gas mixtures naturally maximises the recoil energy, as shown in Fig.~\ref{fig:Recoil};
\item {extremely low energy threshold}, down to a single ionisation electron, thanks to
small detector capacitance independently of the detector volume; 
\item {naturally scalable}, through construction of larger vessels and operation at high gas pressures;
\item {flexible target swapping}, which enables background estimation and confirmation of potential signals; 
\item {small number of readout channels}, which facilitates read-out schemes; 
\item {background rejection and detector fiducialisation}, through pulse shape analysis~\cite{Arnaud:2017bjh}, as shown in Fig~\ref{fig:risevsamplabel}, and track reconstruction; and 
\item {favourable ionisation quenching factor}, as will be discussed in Section~\ref{sec:quenching}. 
\end{inparaenum}
Moreover, a spherical detector has the minimum surface area for a given volume. 
The majority of background in direct DM detection experiments originates from the detector materials in contact with the active volume, thus, a spherical detector maximizes the signal to noise ratio.

This research programme is realised within the international NEWS\=/G collaboration,
which, at the time of writing, comprises approximately 40 members from 10 institutes in 6 countries
NEWS\=/G extended DM sensitivity for the first time down to 0.5~\gev\ with a $\varnothing 60\;\si{\centi\meter}$ spherical proportional counter, {\it SEDINE}, operating at the Laboratoire Souterrain de Modane (LSM), France~\cite{Arnaud:2017bjh}.
Currently, {\it S140}, a $\varnothing 140\;\si{\centi\meter}$ spherical proportional counter constructed of 99.99\% pure copper shown in Fig.~\ref{fig:snoglobe}, is taking data in SNOLAB, Canada.
A 500\,\si{\micro\meter} thick layer of ultra-radiopure copper,
electrodeposited on the inner surface, acts as an internal shield~\cite{Balogh:2020nmo}. This is the largest electroformation of a single piece deep underground~\cite{Knights:2019tmx}, opening
the way for detectors fully electroformed underground.

Looking to the future, the proposed {\it DarkSPHERE}~\cite{NEWS-G:2023qwh} detector, a $\varnothing$3~m spherical proportional counter, will be constructed entirely from underground electroformed copper. To benefit fully from the significant radiopurity improvement in the copper, a water shielding is envisaged. The currently available space in the Boulby Underground Laboratory, UK, is a potential host for {\it DarkSPHERE} as set out in a recently published conceptual design~\cite{NEWS-G:2023qwh}.

\section{Simulation of detector response}
\label{sec:simulation}

\textsc{Geant4}~\cite{GEANT4:2002zbu, Allison:2016lfl} is the established simulation framework in particle physics, used in fundamental research, including accelerator-driven physics, rare-event searches, astronomy, and astrophysics, medical imaging and treatment,  biology, and industry, e.g. energy production, shielding, national security. However, \textsc{Geant4} is
detector agnostic: it only provides energy depositions of incoming particles in various detector parts. Thus, detector specific toolkits are used for detailed modelling and characterisation.

As far as the modelling of gaseous detectors is concerned, a major challenge is the multiple length and energy scales involved: Detectors have physical dimensions from $0.01$ to $1\;\si{\metre}$, while detector physics takes place at much shorter distance scales, below $10\;\si{\micro\metre}$. This, typically, forces a trade-off between the modelling detail and execution speed. The advent of micro-pattern gaseous detectors~\cite{Giomataris:1995fq, Bouclier:1996im} in the second half of the 1990s made this schism even wider.

\begin{wrapfigure}{L}{0.60\linewidth}
\centering
\vspace{-0.7cm}
    \subfigure[\label{fig:diffusion}]{\includegraphics[width=0.49\linewidth]{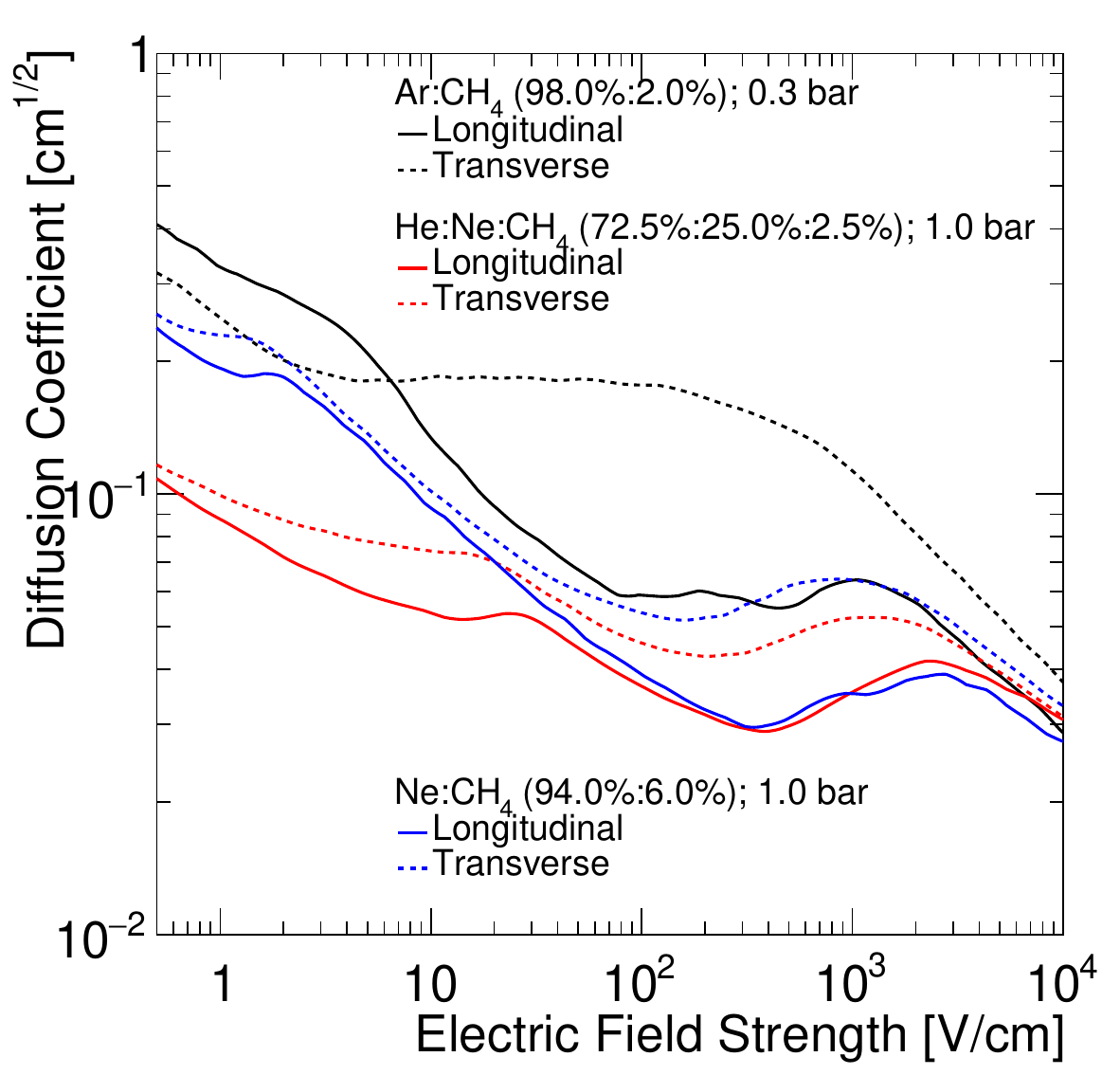}}%
    \subfigure[\label{fig:driftv}]{\includegraphics[width=0.49\linewidth]{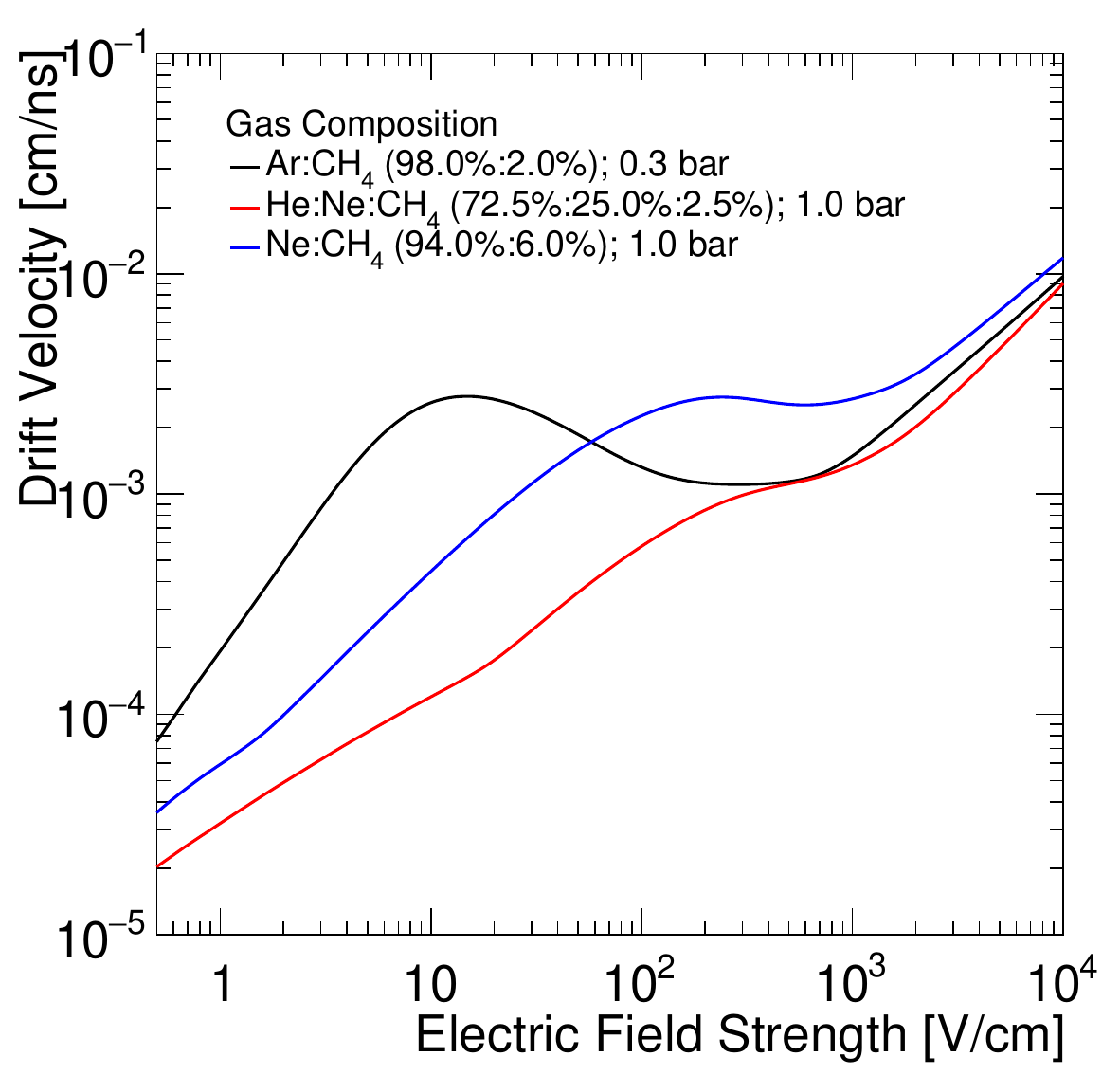}}
\vspace{-0.5cm}
    \caption{Electron transport parameters as a function of electric field strength, calculated by \textsc{Magboltz}: \subref{fig:diffusion}~longitudinal and transverse diffusion coefficients and \subref{fig:driftv}~drift velocity~\cite{Katsioulas:2019sui}.
    \label{fig:transportParameters}}
\vspace{-0.4cm}
\end{wrapfigure}

\textsc{Garfield/Garfield++}~\cite{Veenhof:1993hz,Veenhof:1998tt, garfield} is the
established toolkit for the modelling of gaseous detectors. It interfaces a variety of modules, for example \textsc{Heed}~\cite{Smirnov:2005yi} for modelling particle interactions in the gas volume and \textsc{Magboltz}~\cite{Biagi:1999nwa} for modelling electron transport parameters in gases. 
For traditional gaseous detectors, the simulation of detector response was based on the estimation of electron and ion drift lines based on macroscopic transport coefficients. 
An example of the estimated electron transport parameters for a variety of gas mixtures of interest as a function of the electric field magnitude is given in Fig.~\ref{fig:transportParameters}. The electric field in \textsc{Garfield++} can be described either analytically or with the use of finite element method software like \textsc{ANSYS} or \textsc{gmsh}~\cite{gmsh}. 
However, this approach is inadequate for micro-pattern gaseous detectors, where the precise description of the detector micro-physics for electron transport is critical for the  accurate modelling of the detector response. Thus, in the late 2000s, \textsc{Garfield/Garfield++} was enhanced with microscopic tracking algorithms , where 
semi-classical  Monte Carlo techniques utilising the cross-sections of the atomic processes in the detector are used. Essentially, these are the algorithms utilised within \textsc{Magboltz} to derive the macroscopic electron transport parameters. More details on these algorithms and examples of early successes in describing micro-pattern gaseous detectors are provided in Refs.~\cite{Nikolopoulos:2011zza, Schindler:2012wta}.

\begin{figure}[h]
\centering
    \vspace{-0.5cm}
    \subfigure[\label{fig:signalCurrent}]{\includegraphics[width=0.3\linewidth]{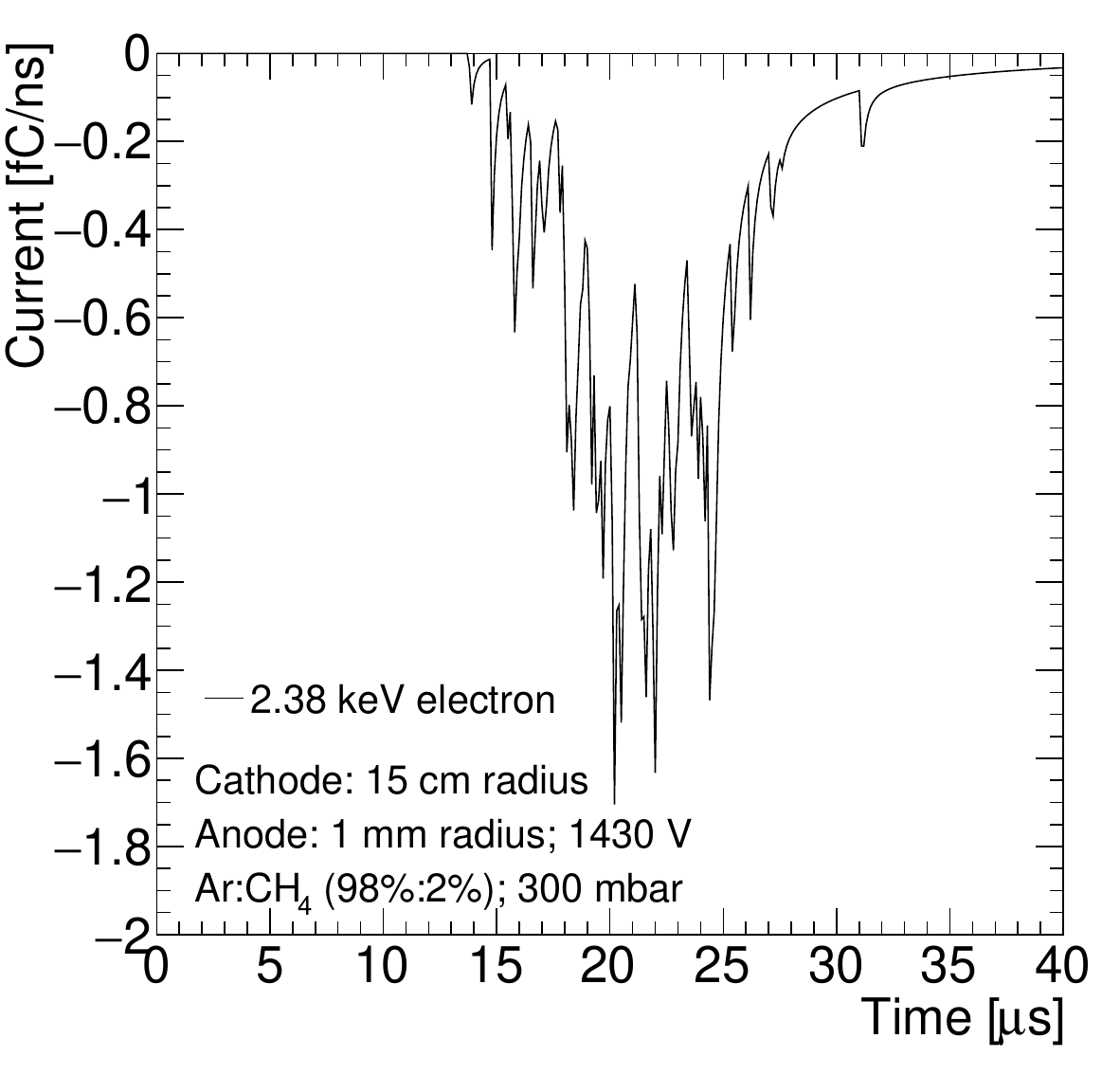}}%
    \subfigure[\label{fig:pulse}]{\includegraphics[width=0.3\linewidth]{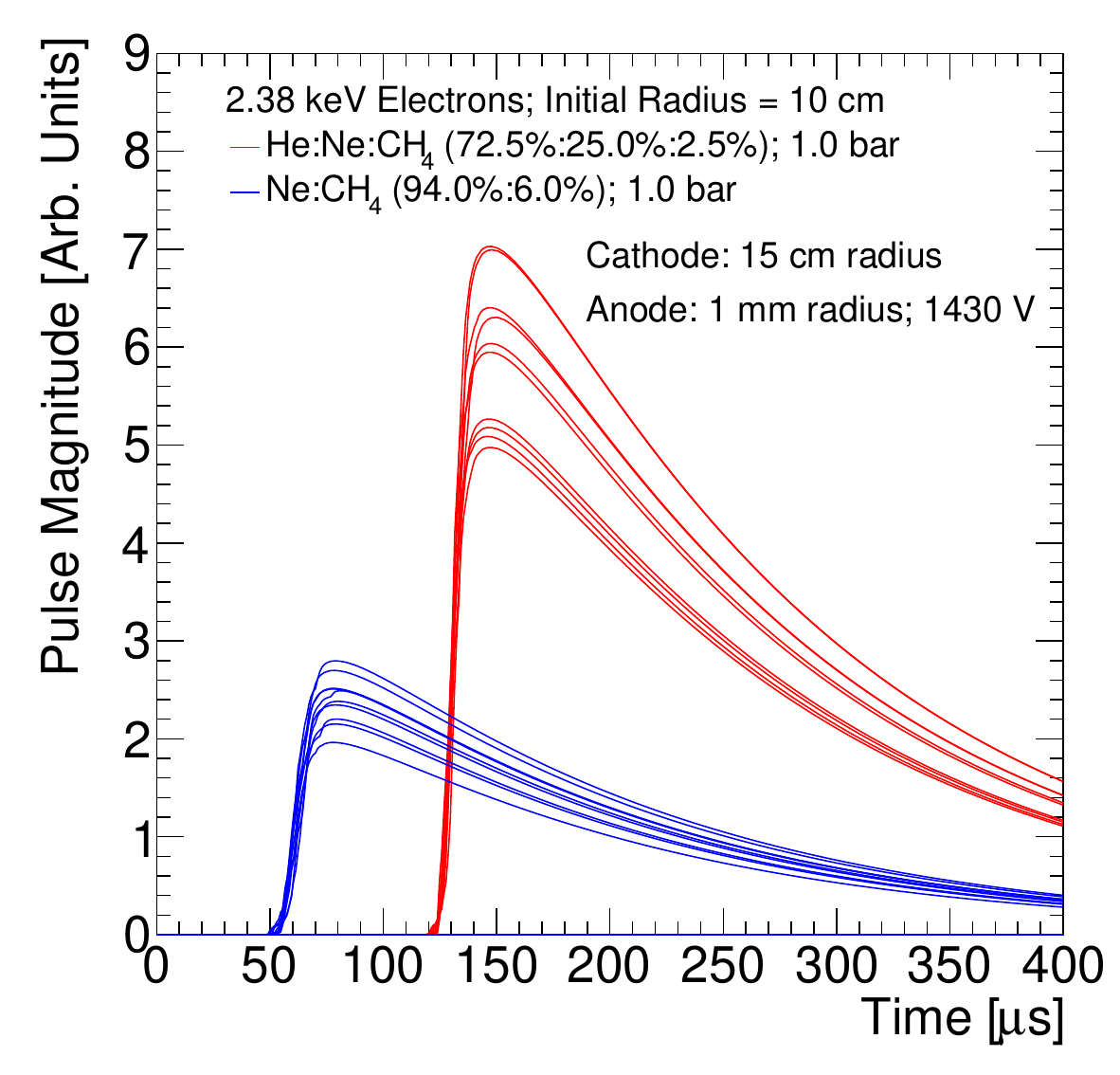}}
        \vspace{-0.5cm}
    \caption{$2.38$\;keV electron interacting in the spherical proportional counter from an initial radius of $10$\;cm and $\theta = 0$: \subref{fig:signalCurrent} the electric current induced in $300\;\si{\milli\bar}$ Ar:CH$_{4}~(98\%:2\%)$, and \subref{fig:pulse} readout pulses following integration produced in $1.0\;\si{\bar}$ He:Ne:CH$_{4}~(72.5\%:25.0\%:2.5\%)$ (red) and $1.0\;\si{\bar}$ Ne:CH$_{4}~(94\%:6\%)$ (blue)~\cite{Katsioulas:2019sui}.\label{figure:pulse}}
    \vspace{-0.4cm}
\end{figure}

As part of the research and development on spherical proportional counters, the development of a predictive framework is pursued~\cite{Katsioulas:2019sui}.
The framework combines \textsc{Geant4} and \textsc{Garfield++}, along with finite element software, thus is equipped with all the elements required to provide physics output. A proof-of-principle demonstration of the combination of the above toolkits was presented in Ref.~\cite{Pfeiffer:2018yam}. The simulation workflow is divided into thee main segments:
\begin{inparaenum}[a)]
\item primary ionisation;
\item electron transport and multiplication in the gas; and
\item signal formation.
\end{inparaenum}
\textsc{Garfield++} is used within \textsc{Geant4} both for primary ionisation, electron transport, and amplification, through \textsc{Geant4}'s physics parameterisation feature, and subsequently for calculation of the induced signal at the end of each event.  
This framework has been employed in various applications of the spherical proportional counter, including detector R\&D~\cite{Herd:2023hmu, Bouet:2020lbp} and neutron spectroscopy with a nitrogen filled spherical proportional counter~\cite{Giomataris:2022bvz}.

A drawback of microscopic tracking algorithms is that they are significantly more CPU-intensive with respect to the traditional approaches to gaseous detector simulation. As a result, there have been several efforts to  speed-up simulations, for example through parametrised simulations~\cite{Amoroso:2020llb}, but also through advanced computational techniques such as hardware acceleration methods relying on graphical processing units (GPUs). Typically, such attempts resulted in the production of new software packages with the desired features, e.g. OuroborosBEM~\cite{Quemener:2021bzz}, but with limited up-take by the community. Recently, a new version of \textsc{Garfield++} has been developed and tested that incorporated GPU-based acceleration in the code-base of \textsc{Garfield++}~\cite{neep-gpu}. This development comes with the additional advantage that this is directly available to all users of \textsc{Garfield++}.

\section{Instrumentation developments}
\label{sec:sensor}
\begin{figure}[htbp]
\vspace{-0.3cm}
\centering
    \subfigure[\label{fig:idealfield}]{\includegraphics[width=0.24\linewidth]{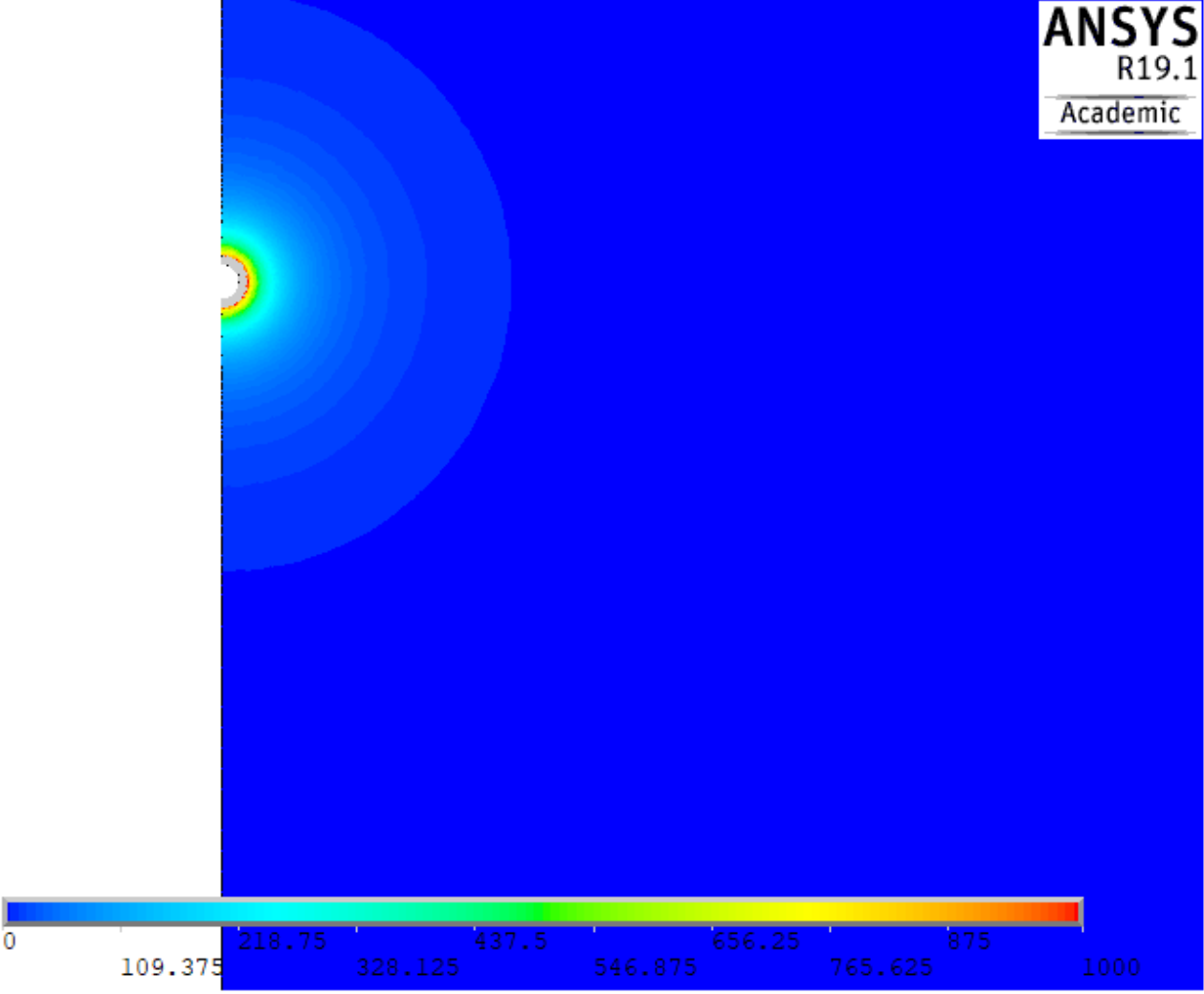}}%
    \subfigure[\label{fig:wirefield}]{\includegraphics[width=0.24\linewidth]{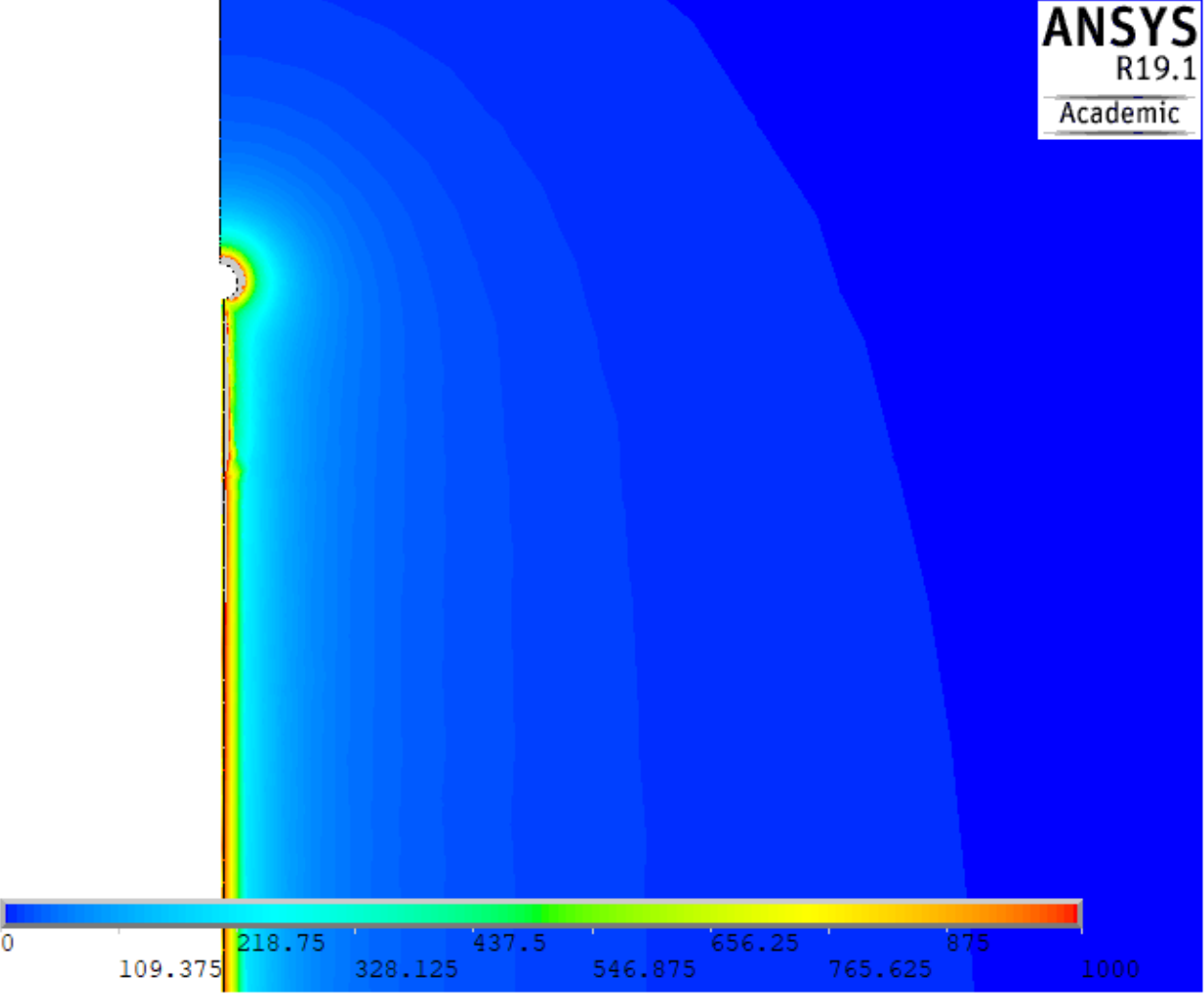}}%
    \subfigure[\label{fig:rodfield}]{\includegraphics[width=0.24\linewidth]{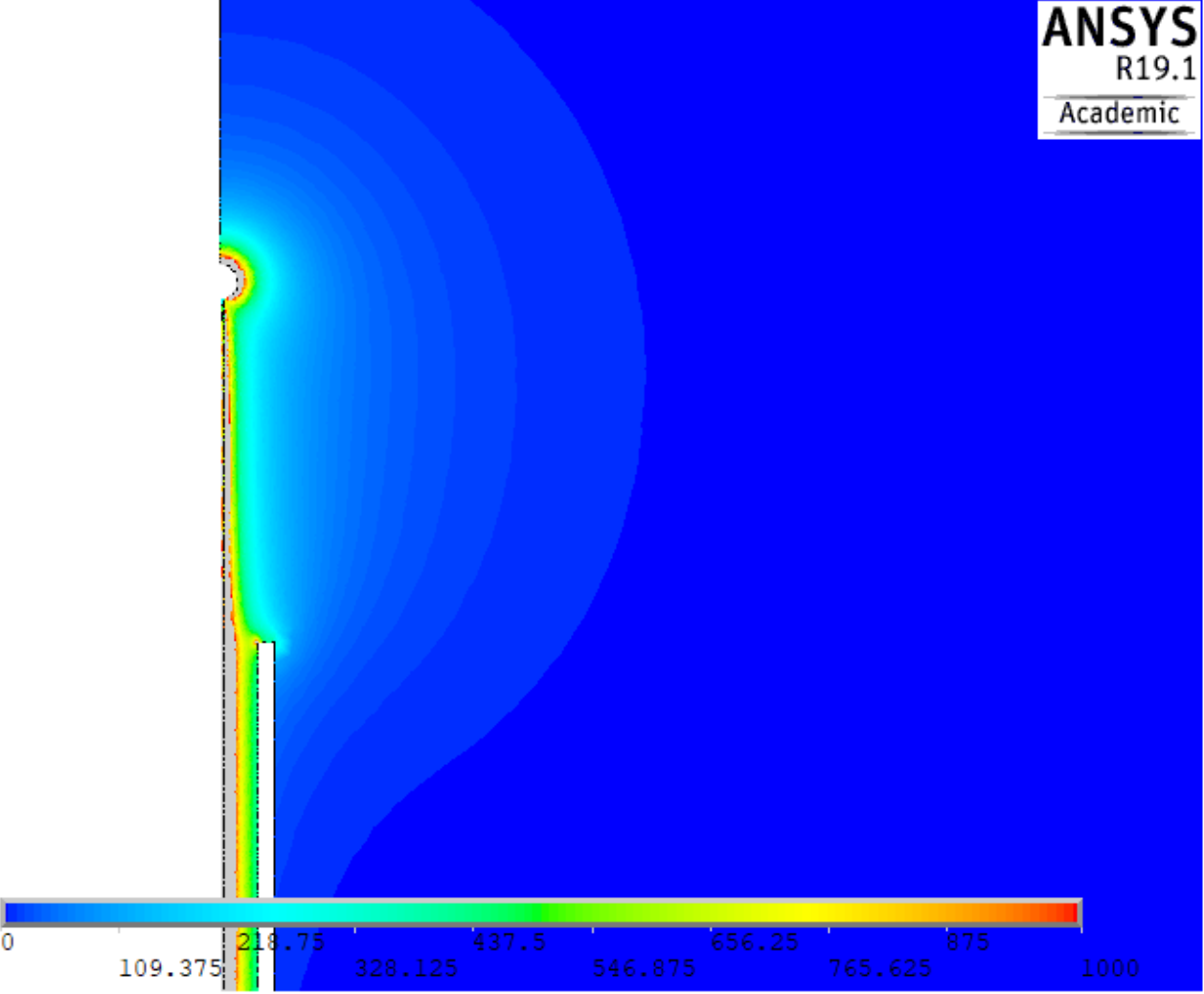}}%
    \subfigure[\label{fig:umbrellafield}]{\includegraphics[width=0.24\linewidth]{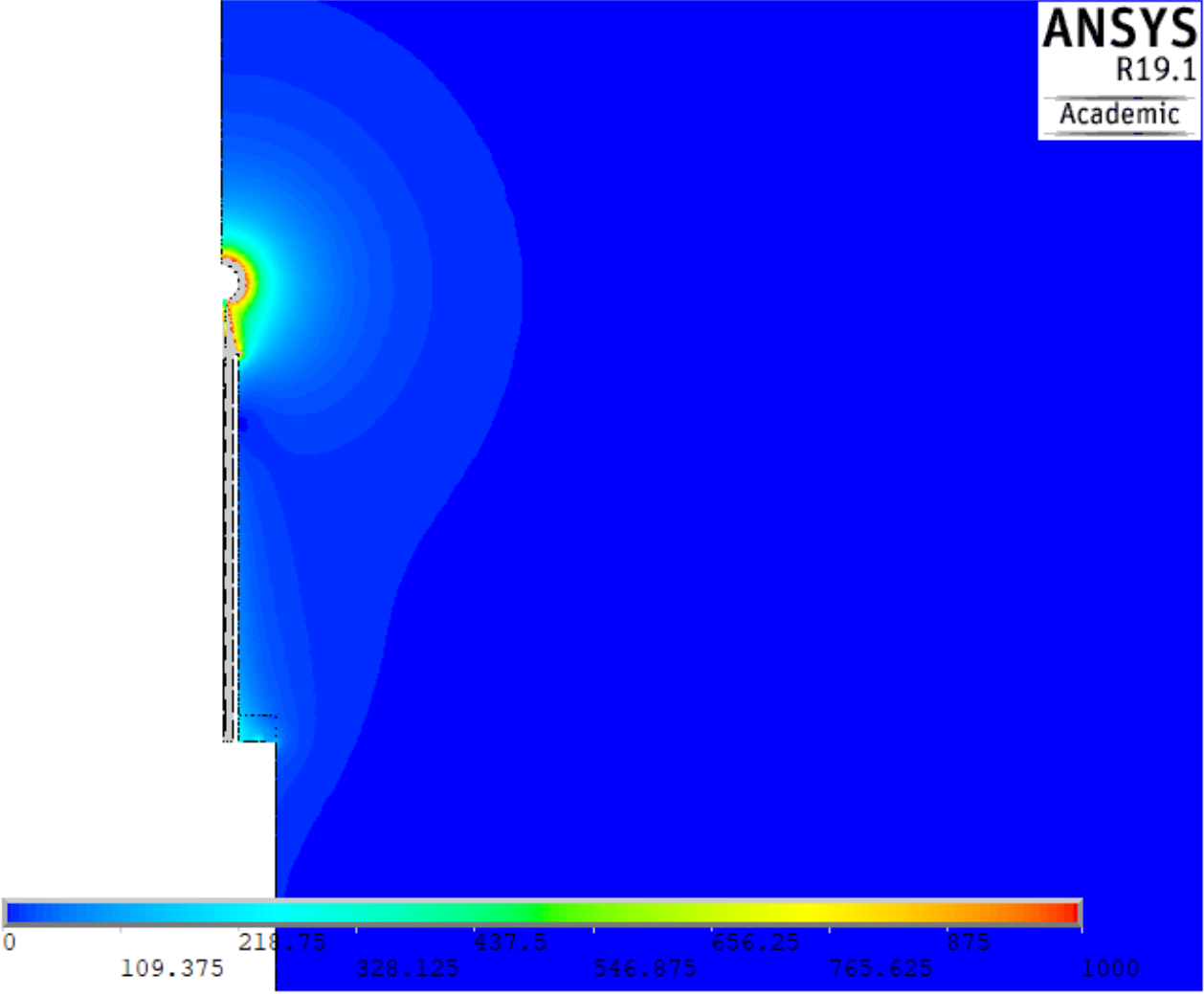}}
    \vspace{-0.5cm}
\caption[]{Electric potential near the $r_a = 1\;\si{\milli\meter}$ anode for four cases each including a new element: \subref{fig:idealfield}, anode only;\subref{fig:wirefield},including the sensor wire; \subref{fig:rodfield} including the grounded support rod; and \subref{fig:umbrellafield} including a cylindrical field corrector electrode~\cite{Katsioulas:2018pyh}.\label{fig:fieldPlots}} 
\vspace{-0.7cm}
\end{figure}

\begin{wrapfigure}{R}{0.30\linewidth}
  \centering
  \vspace{-0.5cm}
  \includegraphics[width=0.99\linewidth]{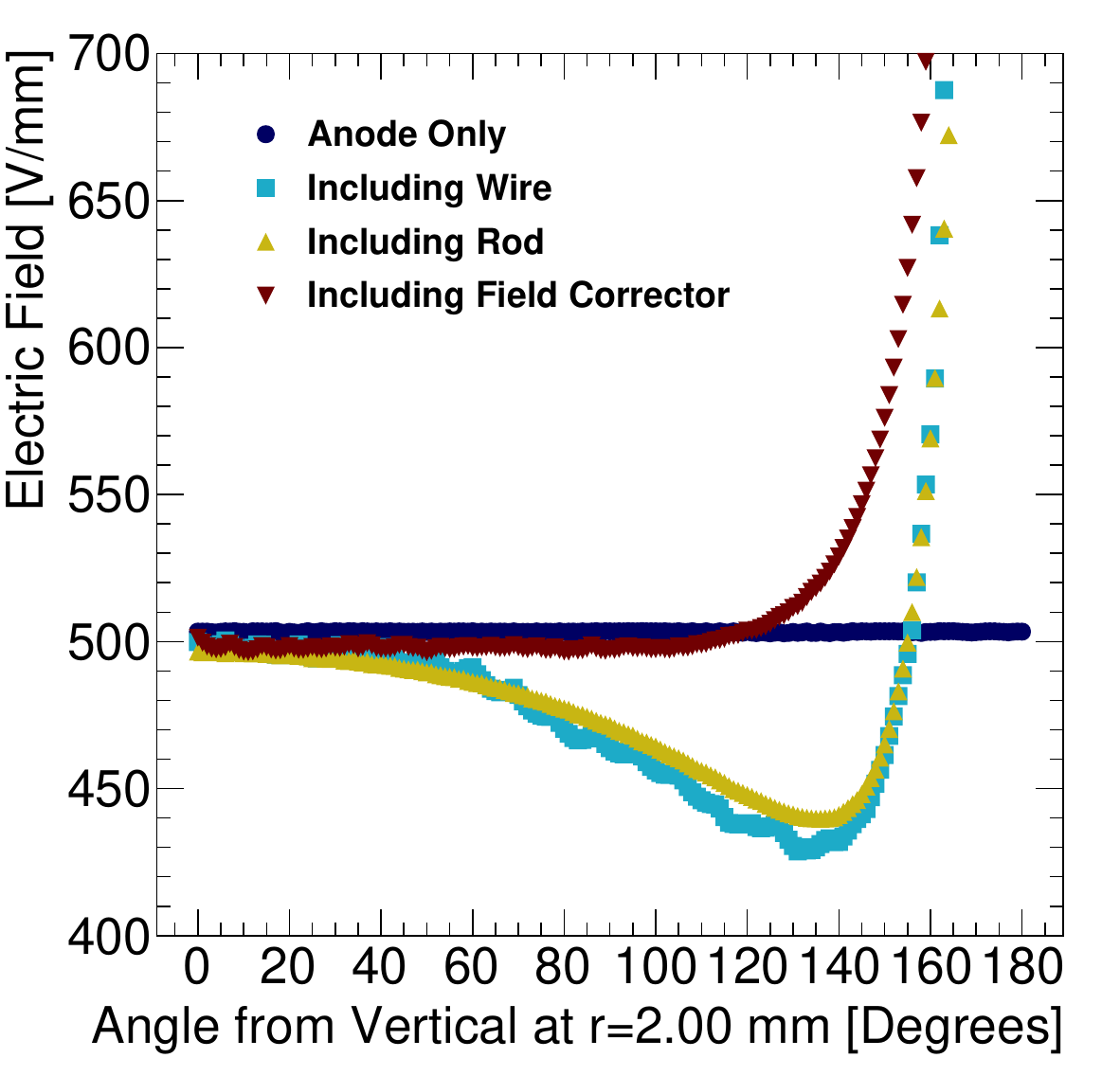}
  \vspace{-0.2cm}
   \caption{Electric field strength at $r=2\;\si{\milli\meter}$ around 
   an $r_a = 1\;\si{\milli\meter}$ anode for four anode support configurations~\cite{Katsioulas:2018pyh}.\label{fig:fieldComparison}}
  \vspace{-0.3cm}
\end{wrapfigure}

As discussed in Section~\ref{sec:invention}, initial detector prototypes used for the read-out sensor a small spherical anode held in place by a metallic rod and an insulated read-out wire. The metallic rod provided the necessary mechanical support and shielded the detector volume from the wire’s electric field. However, the distortion to the ideal electric field caused by the support structures resulted in poor energy resolution. Furthermore, the proximity of the anode to the support rod resulted in instabilities during operation.
The ideal electric field in the vicinity of the anode, shown in Fig.~\ref{fig:idealfield}, is contrasted to the field as modified by the presence of the read-out wire (Fig.~\ref{fig:wirefield}) and the the grounded rod in Fig.~\ref{fig:rodfield}.
For this reason, a field correction electrode was devised and mounted at the end of the support rod near the anode, enabling tuning of the electric field~\cite{Giomataris:2008ap}. The corresponding improvement of the electric field around the anode is shown in Fig.~\ref{fig:umbrellafield}.  
The uniformity in the gas gain achieved with the correction electrode is shown quantitatively in Fig.~\ref{fig:fieldComparison}, comparing the electric field magnitude at a radius of $2\;\si{\milli\meter}$ around the anode for different configurations.

\subsection{Single-anode sensor}
\label{sec:singleanode}

\begin{figure}[h]
\centering
\includegraphics[width=0.65\linewidth]{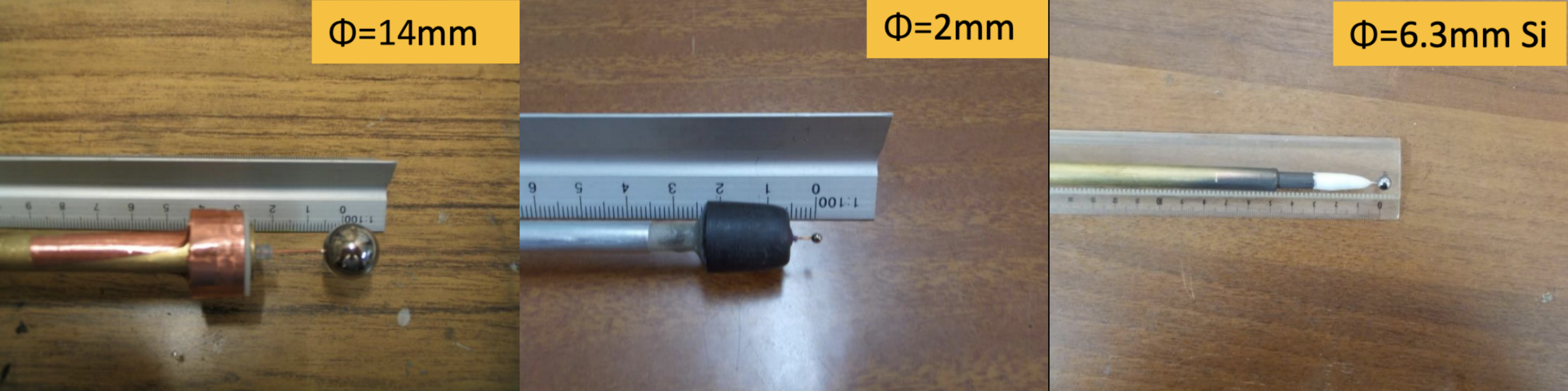}%
    \caption[]{A selection of design iterations for the spherical proportional read-out.Credit: I.~Giomataris and I.~Savvidis.\label{fig:sensorEvolution}} 
\end{figure}

Figure~\ref{fig:sensorEvolution} shows the field correction electrode design evolution through the years, as discussed in Ref.~\cite{Savvidis:2010zz}. Initial designs were implemented using coiled wire around a plastic structure. However, it was found that insulating materials used were subject to charging up, which led to time-varying gain dependencies and negatively impacted energy resolution. Similarly, directly using exposed conductors for the electrode resulted in reduced stability due to the potential for discharges between the anode and the electrode. Instead, resistive materials were found to present the best solution, enabling the biasing of the electrode, while suppressing charging up and sparking.

\begin{figure}[htbp]
\centering
\vspace{-0.5cm}
    \subfigure[\label{fig:sensorDesign}]{\includegraphics[width=0.24\linewidth]{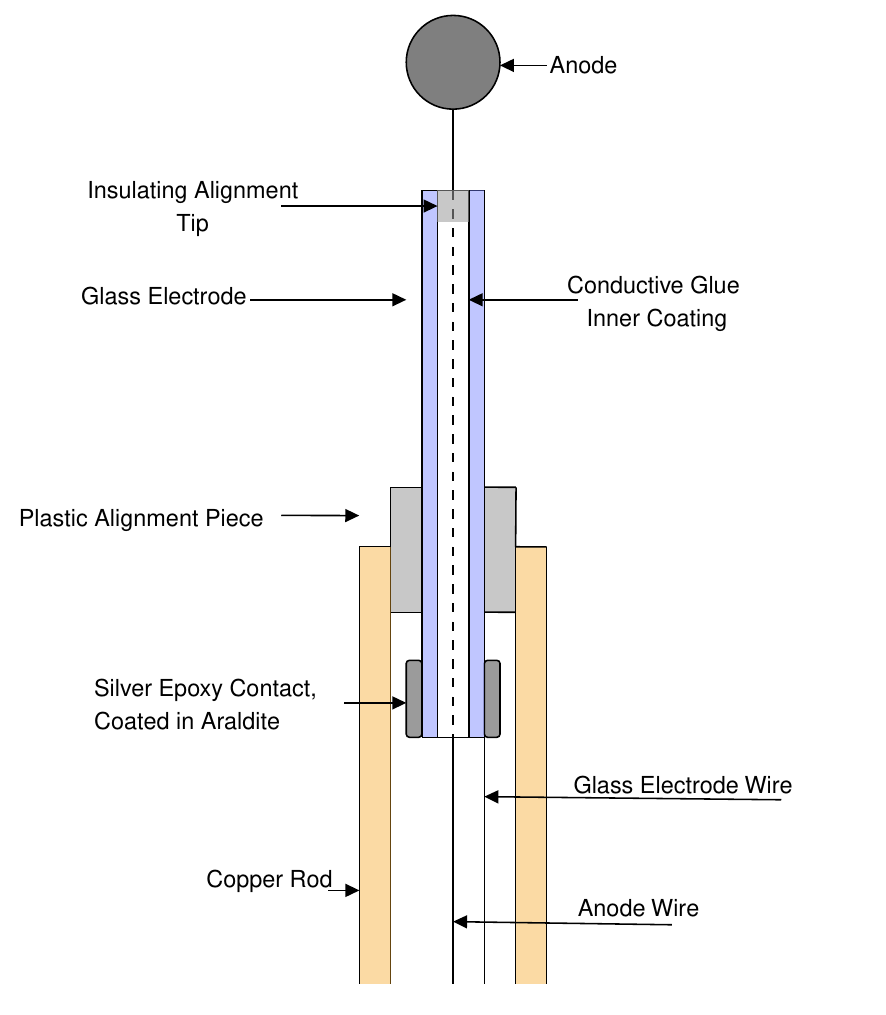}}%
    \subfigure[\label{fig:sensor}]{\includegraphics[width=0.24\linewidth]{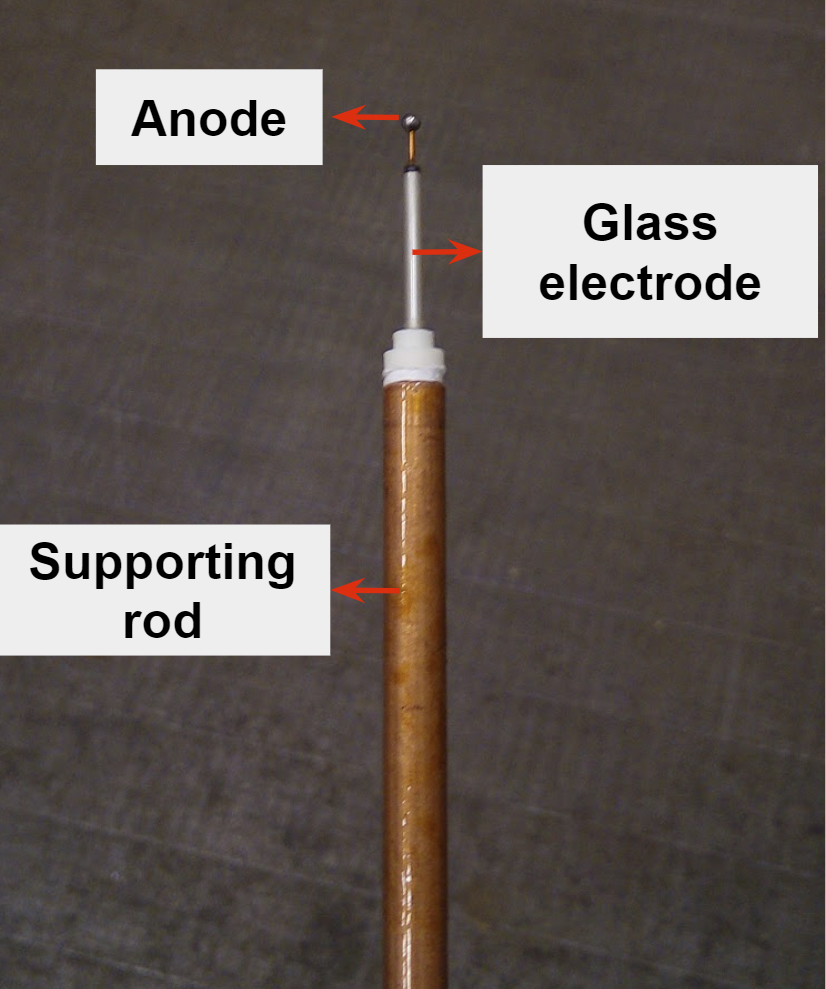}}
    \vspace{-0.5cm}
   \caption[]{
   Design of the single-anode support structure incorporating a resistive, glass electrode, and photograph of a sensor~\cite{Katsioulas:2018pyh}. 
   The support structure primarily features a $1.2\;\si{\milli\meter}$ in diameter soda-lime glass tube, which is connected to a ceramic-coated copper wire by a silver expoxy contact. 
   The inner surface of the glass tube was coated with a conductive epoxy layer to improve electrical contact with the glass. To prevent discharges to the anode, this layer did not extend the full length of the tube, and the end of the tube is sealed with an insulating epoxy. Purpose made, 3D-printed alignment pieces were produced to electrically isolate the glass from the grounded rod and to secure it in place.\label{fig:singleAnodeSensor}}
\vspace{-0.4cm}
\end{figure}

Several different resistive materials were tested for the electrode. 
While Bakelite was found to have favourable properties, such as a resistivity of $\mathcal{O}\left(\num{1e10}\;\si{\ohm\meter}\right)$ and easily machinable, 
its resistivity was found to be sensitive to environmental changes, resulting in temporal variation to the sensor properties. Soda-lime glass was adopted as 
the electrode material, thanks to its commercial availability in cylinders of $\mathcal{O}\left(\si{\milli\meter}\right)$ diameter 
and its resistivity of $\num{5e10}\;\si{\ohm\cm}$~\cite{Katsioulas:2018pyh}. 
In addition, the geometry of the electrode also evolved. Extensive simulations suggested that longer cylinders could provide a better electric field shape.
Figure~\ref{fig:singleAnodeSensor} shows a schematic and a  photograph of an example sensor using a glass correction electrode. 
Through simulation studies it was found that increasing the length of the glass electrode would further improve energy resolution. However, experimentally, it was found that the resistivity of the glass could lead to a poor electrical connection between its two ends, deteriorating  detector performance. For this reason, a metalisation coating of silver epoxy was applied to the inside of the glass, to provide better contact.

\begin{figure}[htbp]
\centering
    \subfigure[\label{fig:singleAnodeStability}]{\includegraphics[width=0.32\linewidth]{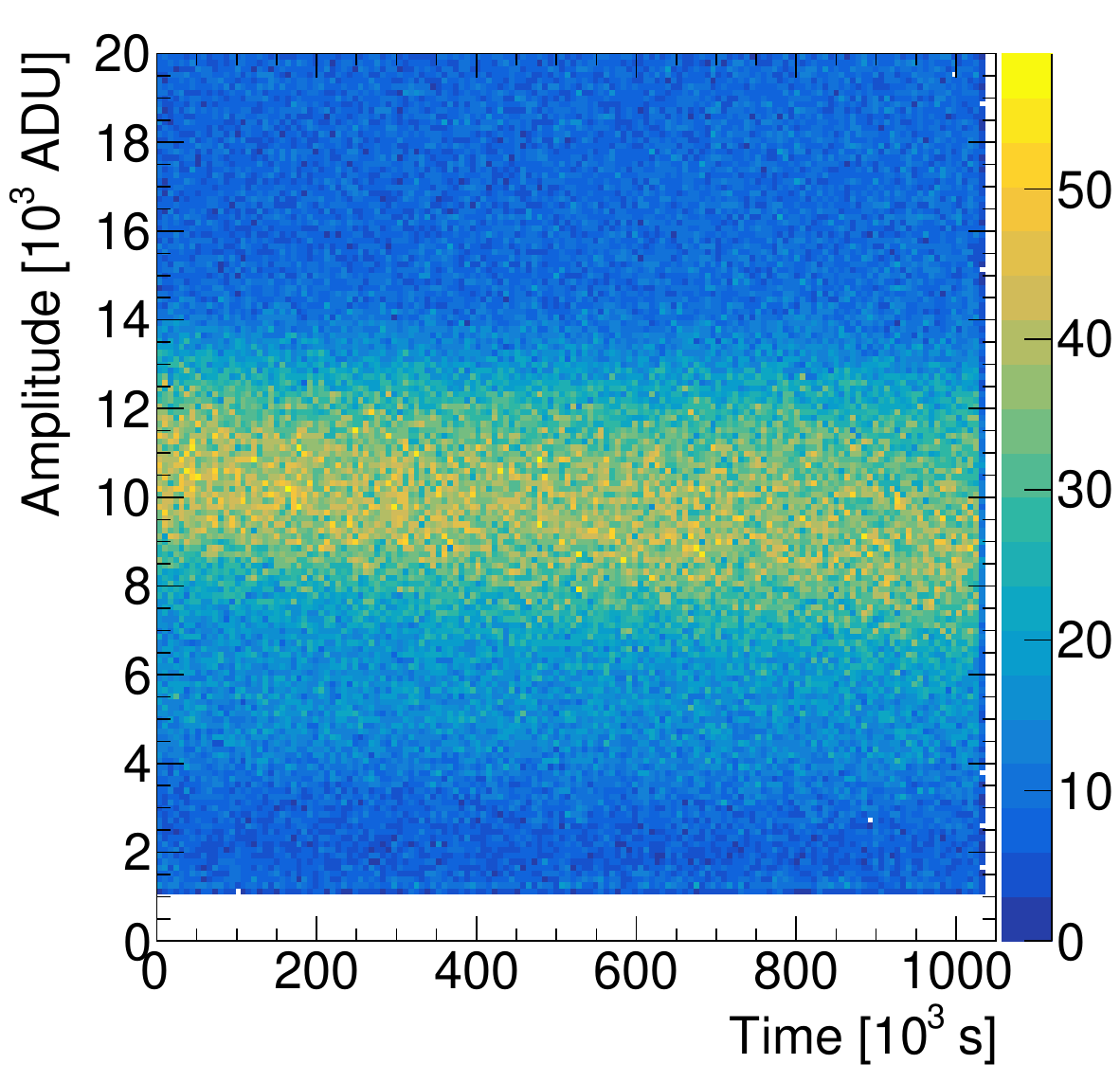}}%
    \subfigure[\label{fig:singleAnodeResponse}]{\includegraphics[width=0.32\linewidth]{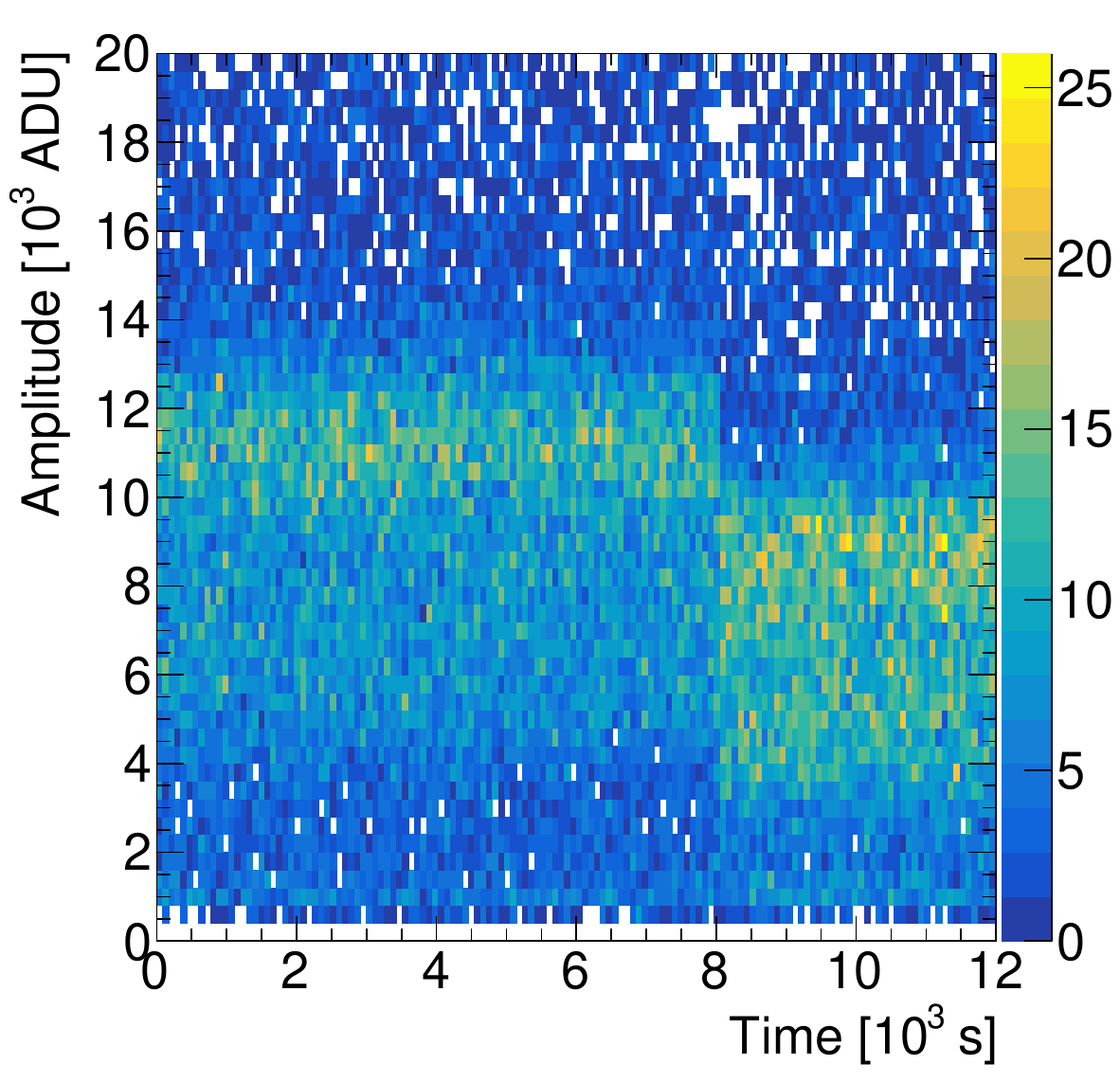}}%
    \subfigure[\label{fig:singleAnodeAmplitude}]{\includegraphics[width=0.32\linewidth]{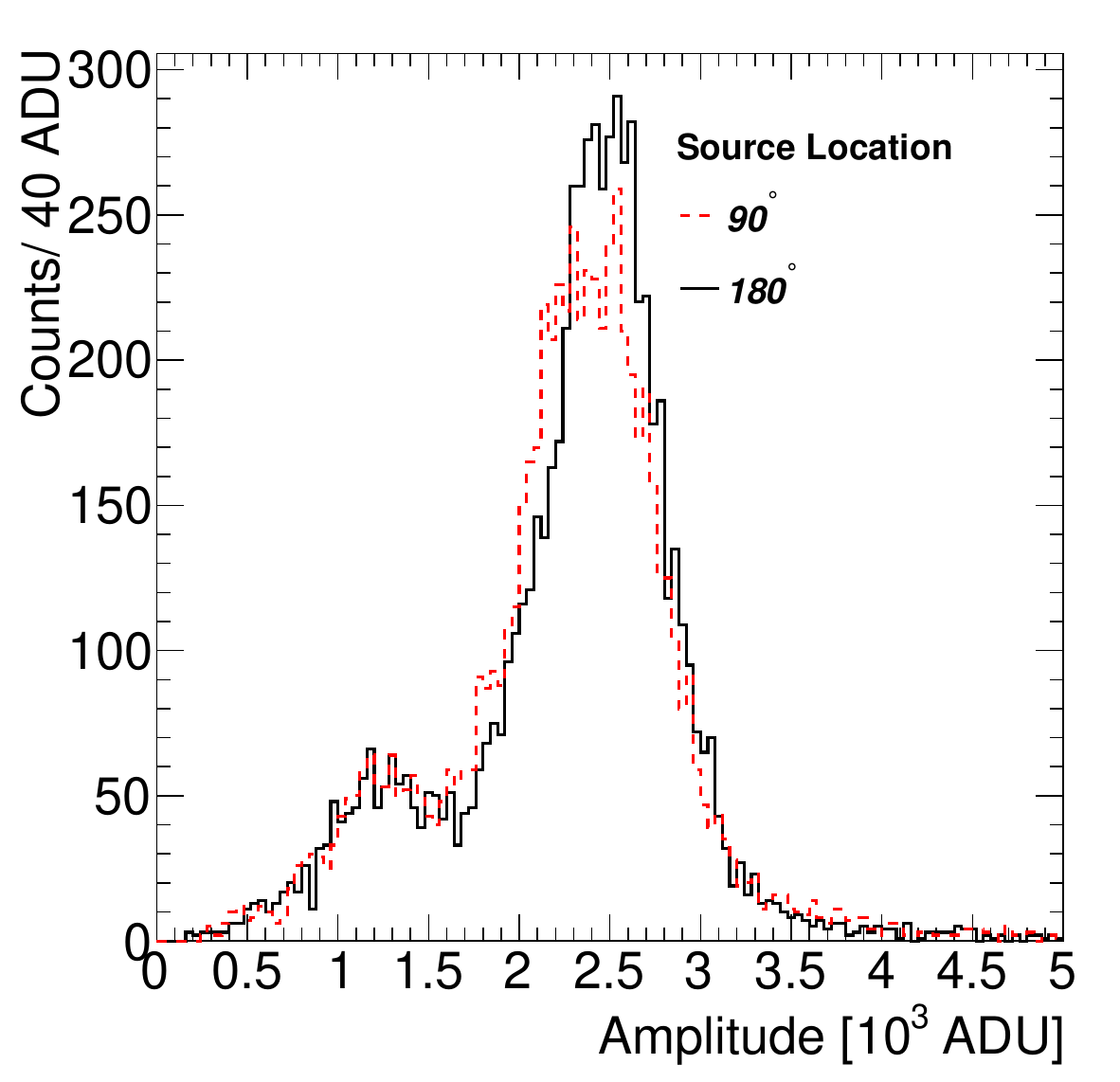}}\\
    \vspace{-0.5cm}
        \caption[]{Data collected with a single, $\varnothing 2\;\si{\milli\meter}$  anode with a glass correction electrode inside a $\varnothing 30\;\si{\centi\meter}$ spherical proportional counter with an $^{55}$Fe source mounted on the inner surface: \subref{fig:singleAnodeStability}  the pulse amplitude as a function of time, demonstrating detector stability, with a small, gradual decrease in amplitude due to gas quality degredation across the 13\,day measurement; \subref{fig:singleAnodeResponse} pulse amplitude versus time with an abrupt change of the voltage applied on the correction electrode  at $t=8000\;\si{\second}$ from $100\;\si{\volt}$ to $200\;\si{\volt}$; and \subref{fig:singleAnodeAmplitude} the pulse amplitude distributions for the source placed opposite the wire, corresponding to $180^{\circ}$ in the plot, and perpendicular to the wire, corresponding to $90^{\circ}$~\cite{Katsioulas:2018pyh}.\label{fig:singleAnodeSensorResults}.}
         \vspace{-0.5cm}
\end{figure}

Studies were performed to evaluate the stability and energy resolution that could be achieved with the glass correction electrode. Data was collected with the sensor mounted inside a $\varnothing 30\;\si{\centi\meter}$ spherical proportional counter with an $^{55}$Fe source inside. The position of the source on the cathode inner surface was remotely adjustable. The $^{55}$Fe decay chain gives $5.9\;\si{\kilo\eV}$ X-rays. Figure~\ref{fig:singleAnodeStability} shows the recorded amplitude of events in the detector as a function of time, showing  operational stability without discharges observed over approximately 12 days of continuous operation. The signal amplitude  as a function of time is shown in Fig.~\ref{fig:singleAnodeResponse}. At $t=8000\;\si{\second}$ the bias voltage applied to the correction electrode was abruptly modified from $100\;\si{\volt}$ to $200\;\si{\volt}$. The response of the detector is immediate, demonstrating the good electrical connection to the electrode. The energy resolution of the detector shown in Fig.~\ref{fig:singleAnodeAmplitude} was assessed by placing the source in two positions:
\begin{inparaenum}[a)]
\item opposite the wire, where the field is expected to be least affected by the anode support structure; and 
\item perpendicular to the wire. 
\end{inparaenum}
It is shown that the detector exhibits a homogeneous response to events in theses two regions.

\begin{wrapfigure}{R}{0.17\linewidth}
  \vspace{-0.35cm}
  \centering
  \includegraphics[width=0.99\linewidth]{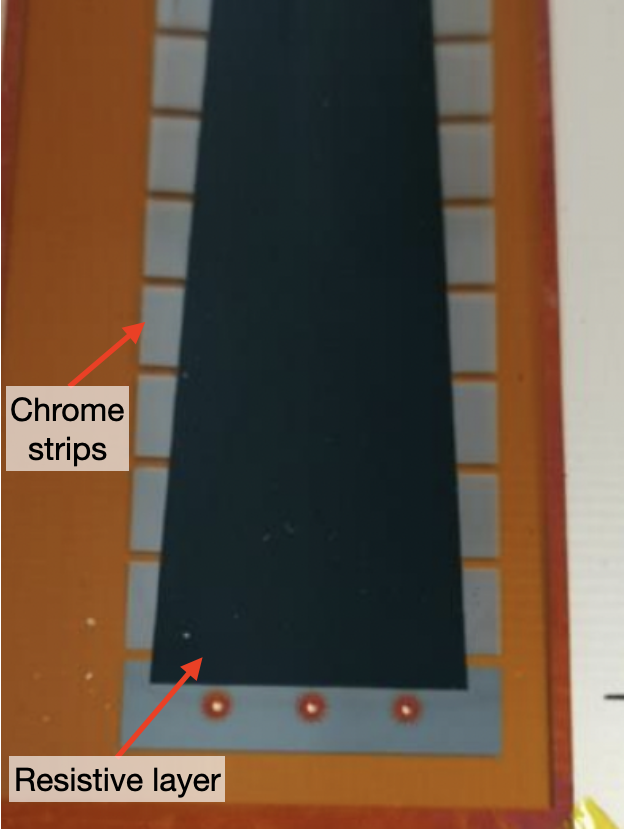}
  \vspace{-0.4cm}
  \caption{Continuous voltage electrode, 
  with a resistive layer applied to kapton-backed chrome strips~\cite{giomatarisrd512020}. Figure by I.~Giomataris. \label{fig:designDegrader}}
  \vspace{-0.5cm}
\end{wrapfigure}

While the correction electrode alleviates the electric field distortions in the vicinity of the anode, the grounded rod causes the electric field to deviate from the ideal in the drift region. 
While technically challenging to implement, a support rod with an applied voltage that changes as $1/r$ along its length would correct for this distortion. A method for implementing an approximation of this is shown in Fig.~\ref{fig:designDegrader}, using restive paste or diamond-like-carbon (DLC) applied over chrome strips on a kapton backing.  This structure is wrapped around the grounded rod to provide an electrode. The shape of the resistance layer ensures the correct   voltage reduction between descending strips. Initial tests with resistive paste suffered from damage when wrapping the structure around the rod, however, tests with DLC coating were promising~\cite{giomatarisrd512020}.

\subsection{Multi-anode sensor}
\label{sec:multianode}
Several of the spherical proportional counter applications require increasingly larger target mass, which scales proportionally to the gas volume and pressure, $P$. However, detector operation
depends on $E/P$, and -- as seen also for Eq.~\ref{eq:field} -- unacceptably low values are obtained for large radii and high pressures. 
A commensurate increase in the anode radius and its voltage would be required to maintain the electric field near the cathode of  a larger detector, while maintaining the same gas gain. This could lead to detector instability and discharges. 

The invention of the
multi-anode sensor ACHINOS~\cite{Giganon:2017isb, Katsioulas:2022cqe} is a breakthrough in this direction. 
It relies on multiple anodes located at a fixed distance $r_A$ from the detector centre.
The electric field at large radii is comprised by the contributions of the ensemble of anodes, while at small radii each anode forms an independent avalanche region, with the local electric field  determined by the size and voltage of that anode.
As a result, the electric
field strengths in the drift and avalanche regions are decoupled, allowing large enhancement at large radii -- scaling with the number of anodes and their distance from the detector centre, without
detector instabilities. The increase in electric field is shown in Fig.~\ref{fig:DarkSPHERE_sensorField}.

\begin{figure}[b]
\centering 
\vspace{-0.5cm}
    \subfigure[\label{fig:bakeliteAchinos}]{\includegraphics[width=0.33\linewidth]{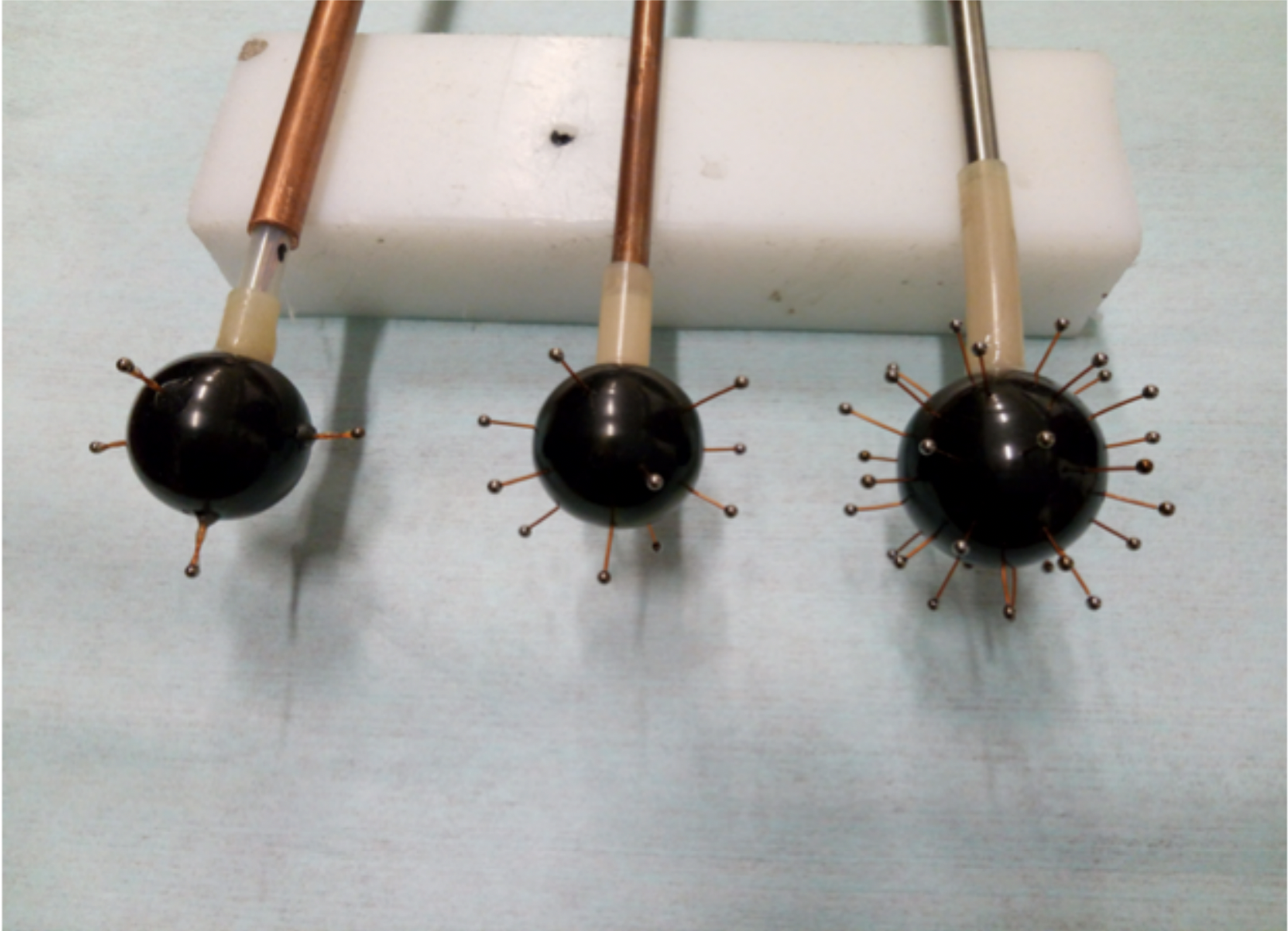}}%
   \hspace{0.02\linewidth}
     \subfigure[\label{fig:uob_achinos}]{\includegraphics[width=0.32\linewidth]{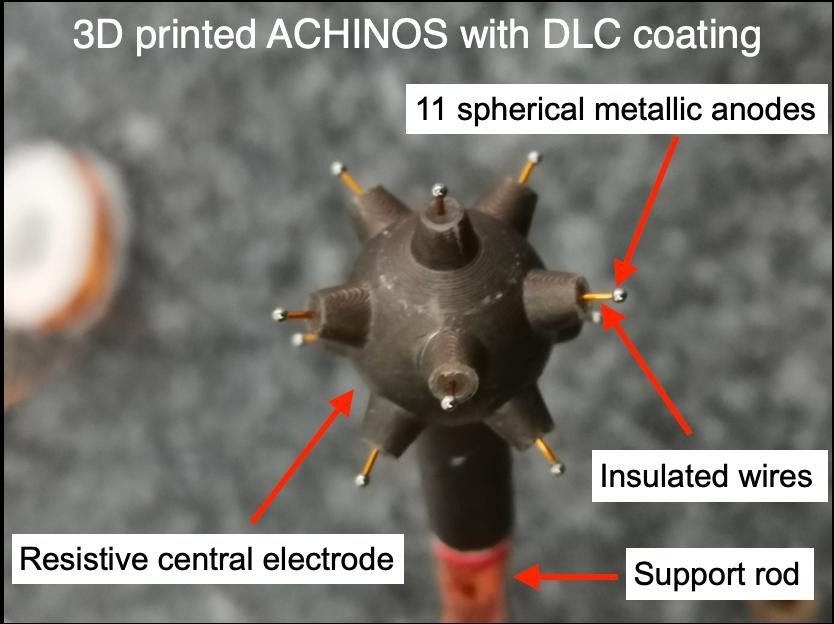}}%
    \vspace{-0.5cm}
    \caption[]{\label{fig:achniosPhotos} \subref{fig:bakeliteAchinos} Early  (6,11, and 33 anodes). Figure by I.~Giomataris. \subref{fig:uob_achinos} current (11-anode) ACHINOS implementations. The former employ central electrodes made of Bakelite, while the latter use DLC-coated, 3D printed, central electrode.} 
    \vspace{-0.2cm}
\end{figure}

The anodes of ACHINOS are supported by a central support structure, which also acts as a correction electrode. Similarly to single-anode sensors, early realisations of ACHINOS utilised a bakelite central correction electrode, as shown in Fig.~\ref{fig:bakeliteAchinos}. To improve the construction reproducibility, additive manufacturing techniques were employed to construct the central support structure, which was then coated in resistive material to form the central electrode. Resistive epoxy mixtures, such as copper-epoxy, were used. However, they were found to be susceptible to damage from discharges. The state-of-the-art sensors, currently employed in NEWS-G and in spherical proportional counter applications more generally, utilise a 3D-printed central support structure coated in a diamond-like carbon (DLC) layer~\cite{Giomataris:2020rna}. An example is shown in Fig~\ref{fig:uob_achinos}. 

Current ACHINOS technologies use 11-anodes located at the vertices of an icosahedron, chosen to ensure uniform arrangement of the anodes, with the twelfth vertex being occupied by the grounded support rod. For  larger detectors, more anodes are required. This has motivated the developement of a 60-anode ACHINOS, where the anodes are located at the vertices of a truncated icosahedron. The increase in the drift electric field is shown in Fig.~\ref{fig:DarkSPHERE_sensorField}.

\begin{figure}[h]
\centering
    \vspace{-0.3cm}
\subfigure[\label{fig:achinosAnglePlotDV0V}]{\includegraphics[width=0.32\linewidth]{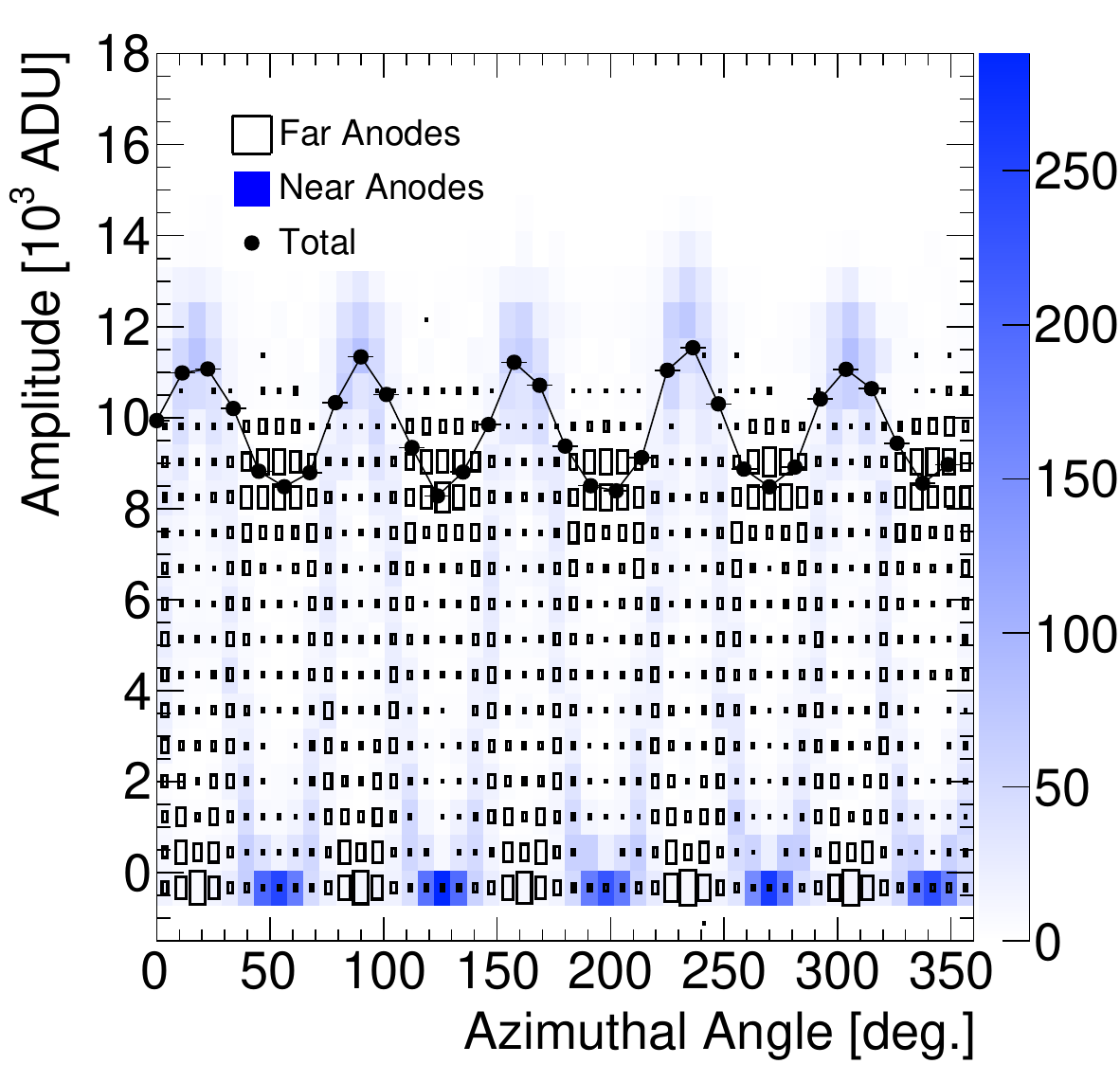}}%
     \subfigure[\label{fig:achinosAnglePlotDiffV}]{\includegraphics[width=0.32\linewidth]{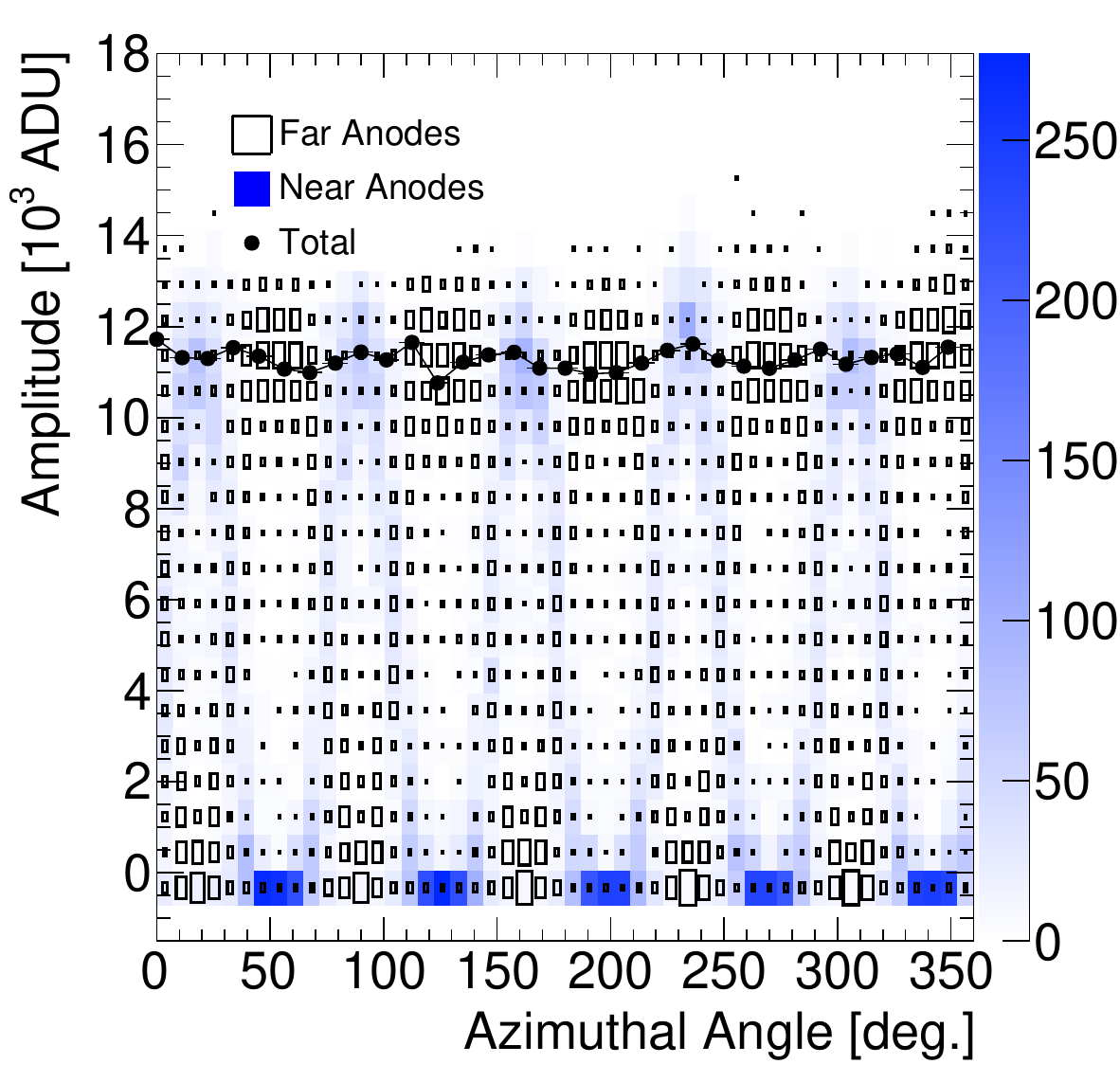}}%
    \vspace{-0.5cm}
        \caption[]{\label{fig:achinosAmplitudeAngle} Simulated pulse amplitude in the Near and Far anodes as a function of azimuthal angle of an $^{55}$Fe source in the detector: \subref{fig:achinosAnglePlotDV0V} when the same voltage is applied to all anodes, and \subref{fig:achinosAnglePlotDiffV} when $1.4\%$ higher voltage is applied to the Far anodes~\cite{Giomataris:2020rna}.} 
    \vspace{-0.35cm}
\end{figure}

Typical implementations of 11-anode ACHINOS, are read-out in two channels, one electrode consisting of the 6 anodes farthest from the grounded rod (Far), and another consisting of the 5 anodes nearer the grounded rod (Near). However, due to the proximity of the grounded rod to the Near anodes, they exhibit higher electric fields, and, thus, higher gains. 
Figure~\ref{fig:achinosAmplitudeAngle} shows the amplitude recorded as a function of azimuthal angle of a simulated $^{55}$Fe source on the inside of the cathode. The open boxes (blue squares) show the magnitude of the signal on the Far (Near) electrode, 
while the points show the total recorded signal. The open boxes (blue squares) show the magnitude of the signal on the Far (Near) electrode, while the points show the total recorded signal. In Fig.~\ref{fig:achinosAnglePlotDV0V}, the difference in the Near and Far gains results in a sinusoidal total amplitude as the source is moved around the detector in the azimuthal angle.
This effect is corrected for by increasing the voltage applied to the Far anodes by $1.4\%$, as shown in Fig.~\ref{fig:achinosAnglePlotDiffV}. This correction resolves the main gain variation expected for an ACHINOS operated in 2-channel read-out.
\begin{figure}[h]
     \centering
  \includegraphics[width=0.65\linewidth]{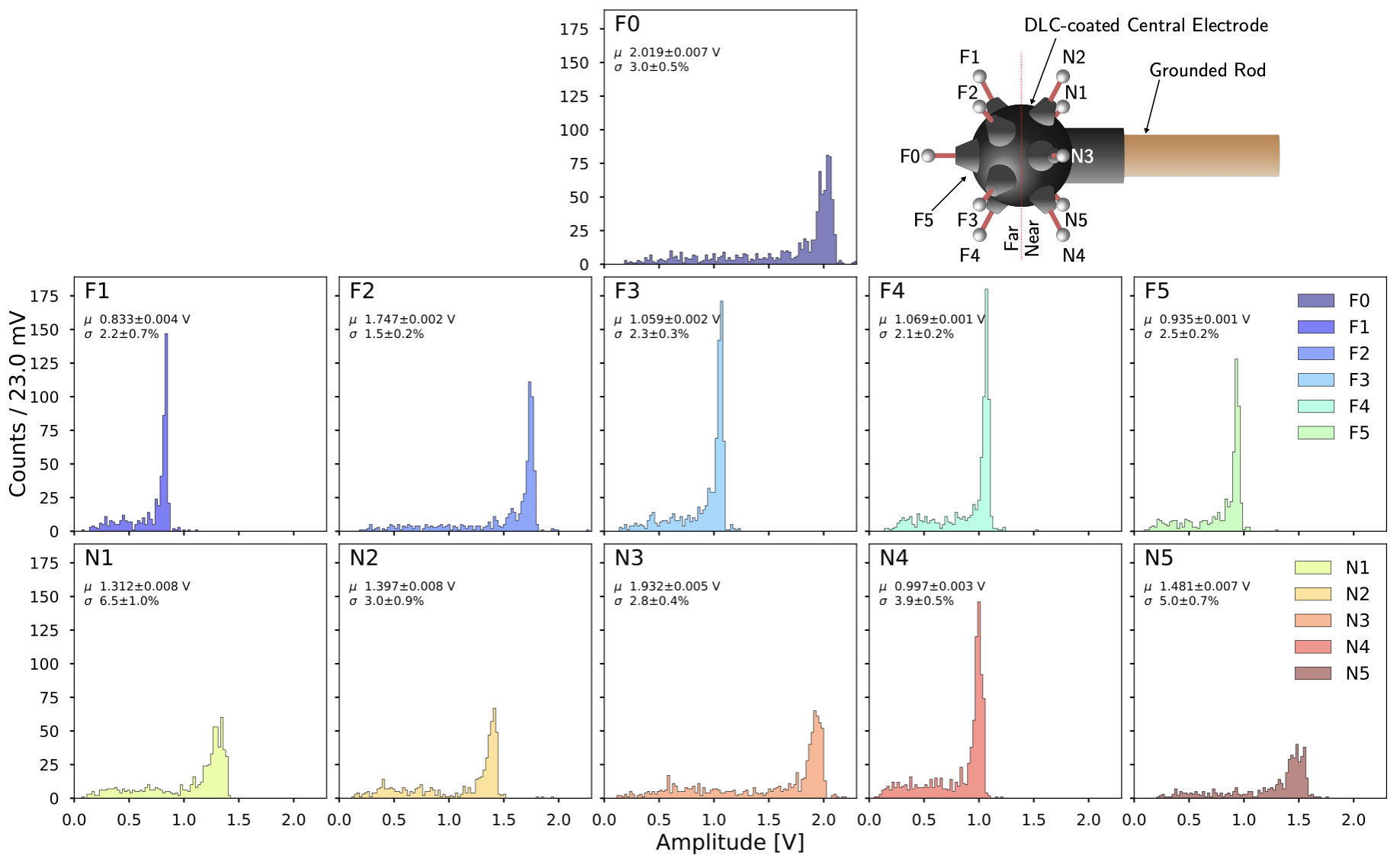}
    \vspace{-0.3cm}
     \caption{Recorded amplitude for an individually read-out ACHINOS when an $^{55}$Fe source was directed at each anode, indicated in each subfigure. The layout of the read-out channels is shown in the diagram. Adjusted from Ref.~\cite{Herd:2023hmu}. \label{fig:indivAnode}}
    \vspace{-0.2cm}
\end{figure}

\begin{wrapfigure}{L}{0.30\linewidth}
  \centering
  \vspace{-0.6cm}
  \includegraphics[width=0.99\linewidth]{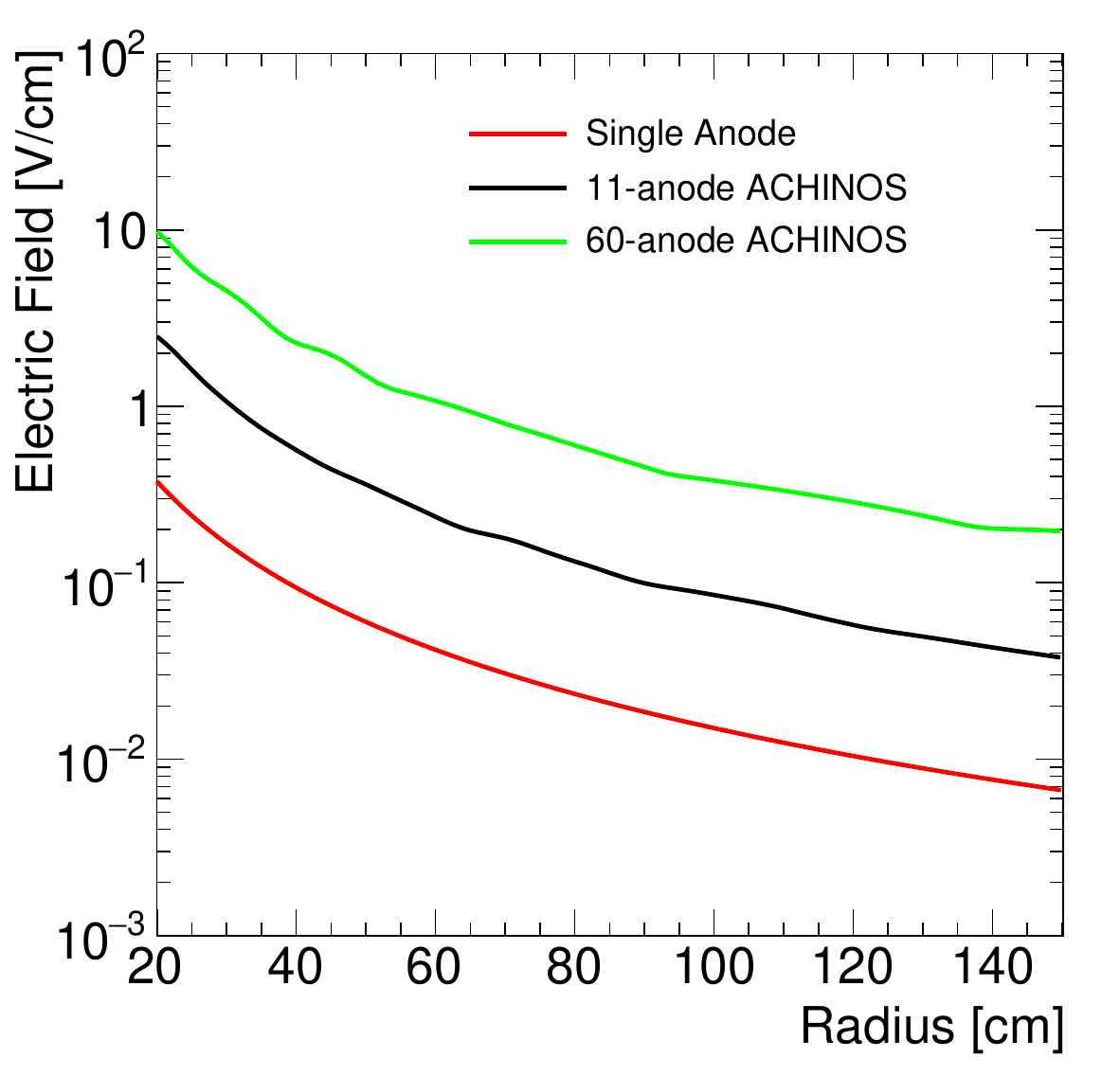}
  \vspace{-0.8cm}
  \caption{Electric field magnitude in the drift region 
  for different ACHINOS configurations~\cite{NEWS-G:2023qwh}.\label{fig:DarkSPHERE_sensorField}}
  \vspace{-0.8cm}
  \end{wrapfigure}
  While the Near versus Far electric field difference induces the dominant gain difference, imperfections in the ACHINOS construction or in the anodes used can induce further gain differences between anodes. Recent work has gone into the development of the necessary electronics and acquisition system to individually read-out the anodes of ACHINOS, allowing for individual anode calibration~\cite{Herd:2023hmu}. Figure~\ref{fig:indivAnode} shows the amplitude spectrum recorded when an $^{55}$Fe source was directly aligned with each anode, indicated in each subfigure, when a selection was applied to the data to only keep events when the signal from all other anodes was opposite in polarity to the anode of interest. This first work in this direction has successfully demonstrated a $32\%$ improvement in the energy resolution achieved for the ACHINOS.

\subsection{Event localisation and track reconstruction}

\begin{figure}
     \centering
     \vspace{-0.45cm}
   \subfigure[\label{fig:risetimeVsRadiusSimulation}]{\includegraphics[width=0.30\linewidth]{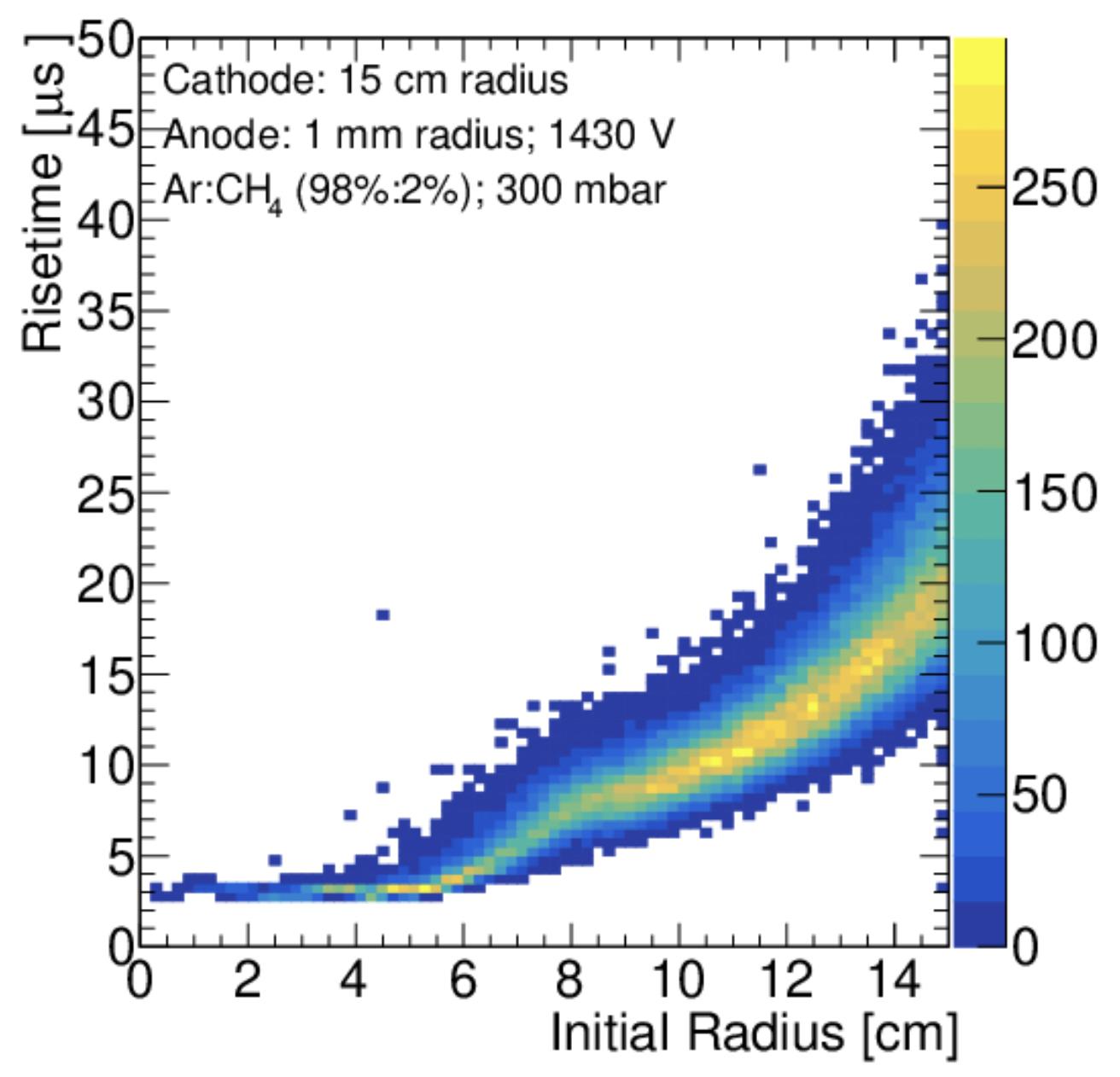}}
   \subfigure[\label{fig:weightingField}]{\includegraphics[width=0.63\linewidth]{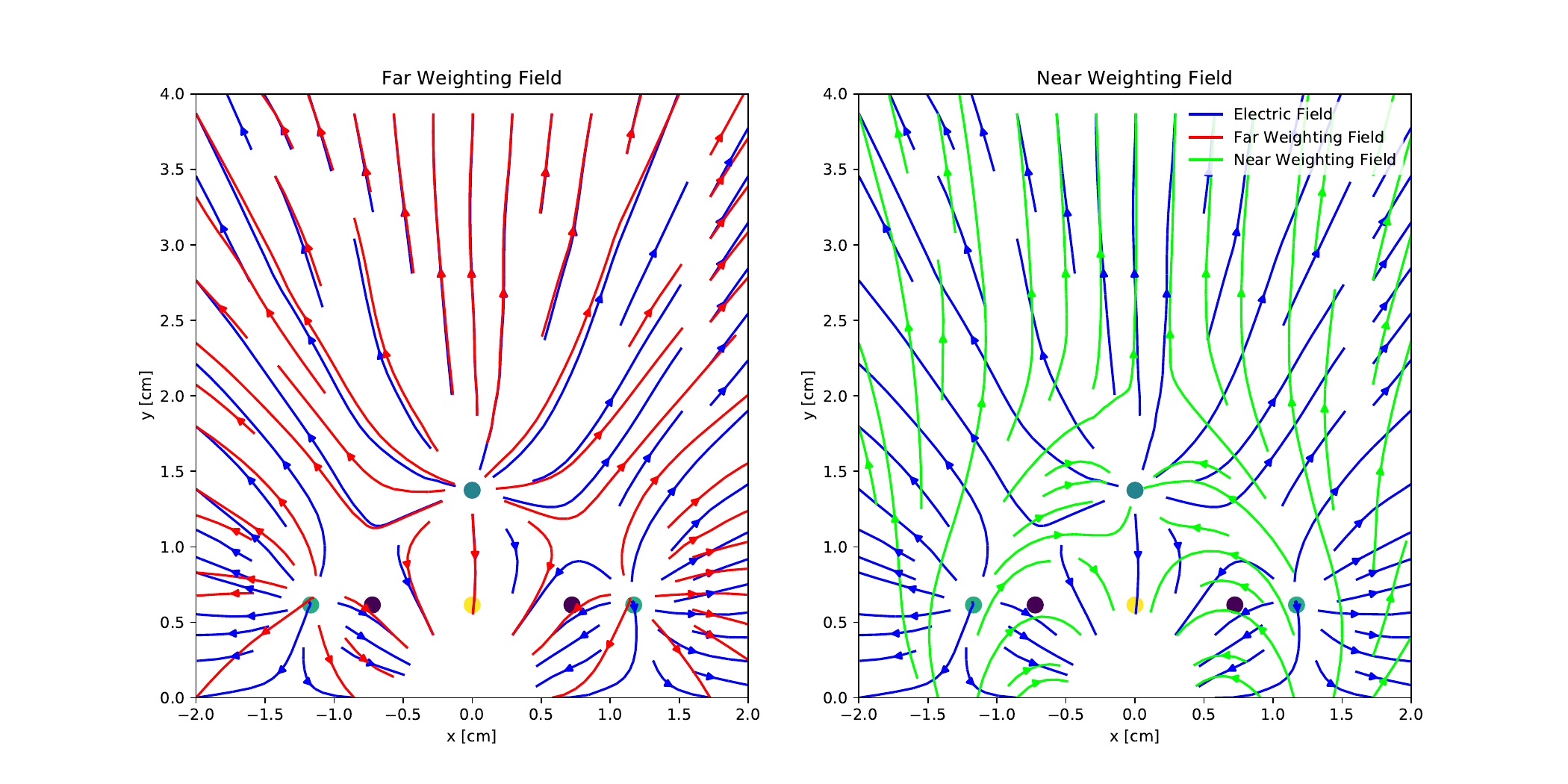}}
   \vspace{-0.3cm}
     \caption{\subref{fig:risetimeVsRadiusSimulation} The pulse risetime as a function of interaction radius for $2.38\;\si{\kilo\eV}$ electrons simulated uniformly in a $\varnothing 30\;\si{centi\meter}$  detector with a single-anode sensor~\cite{Katsioulas:2019sui}. \subref{fig:weightingField} The weighting field~~\cite{Katsioulas:2022cqe}.}
\vspace{-0.4cm}
\end{figure}

The detector can also provide spatial information for events. For example, electrons generated at larger radii in the detector undergo more diffusion as they travel to the anode. This translates to an increased pulse risetime, defined as the time difference between the 10\% to 90\% of pulse amplitude in the rising edge. 
This is shown in Fig.~\ref{fig:risetimeVsRadiusSimulation}. Furthermore, the risetime provides a handle to distinguish `point-like' and `track-like' energy deposits, where the latter generally have large risetime due to the spatial spread for the produced ionisation electrons.

ACHINOS not only provides an increase in the electric field magnitude at large radii, but by having multiple anodes, it can provide improved spatial information for interactions in the detector volume through charge sharing techniques. The weighting field for the two-channel ACHINOS electrodes in the vicinity of the Far anodes (coloured circles) are shown in Fig.~\ref{fig:weightingField}. At the surface of a given Far anode, the weighting fields for the Near and Far channels are anti-parallel. Thus, the induced signal on the two channels by an electron arriving to an anode, given by the Shockley-Ramo theorem~\cite{shockley,ramo}, will have opposite polarity. 

\begin{wrapfigure}{Rh}{0.30\linewidth}
     \centering
     \vspace{-0.5cm}
   \includegraphics[width=0.85\linewidth]{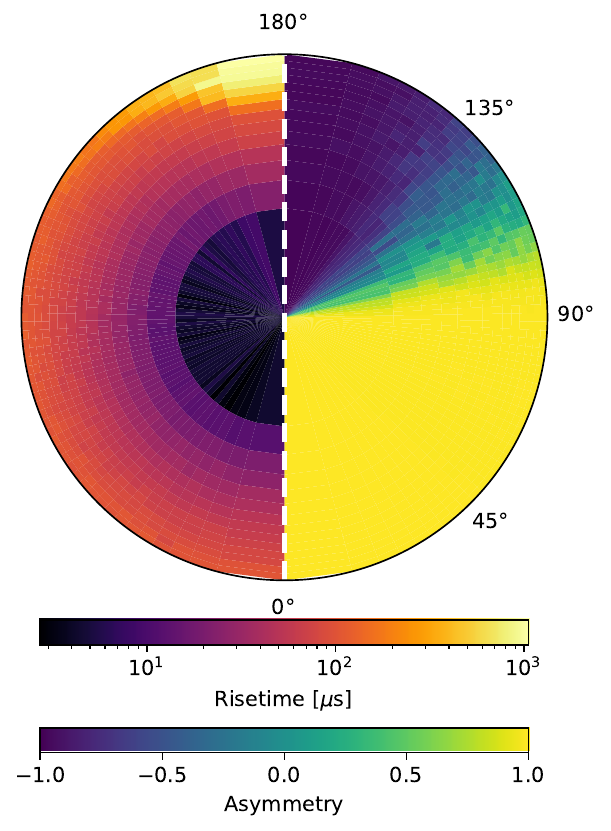}
   \vspace{-0.2cm}
     \caption{Pulse risetime (left) and the Near-Far signal asymmetry (right), as a function of the interaction point for simulated $^{37}$Ar decays in a 11-anode/two-channel read-out spherical proportional counter~\cite{NEWS-G:2023qwh}. \label{fig:achinosAsymmetry}}
        \vspace{-0.75cm}
\end{wrapfigure}

Figure~\ref{fig:achinosAsymmetry} shows the  interaction location for $^{37}$Ar decays reconstructed in simulation using an 11-anode multi-anode sensor. The read-out electrodes are grouped in two channels, denoted as ``Near'' and ``Far''.  On the left part of the figure, the pulse risetime  versus the interaction point is shown. On the right part, the signal asymmetry, defined as $(\text{Far}-\text{Near})/(\text{Far}+\text{Near})$, is shown. 
The former shows sensitivity to the radial position
and the latter to the zenith-angle of a point-like interaction.  These elements combined offer  detector fiducialisation.

\begin{figure}[t!] 
  \centering
  \subfigure[\label{fig:achinosMuon_b}]{\includegraphics[width=0.45\linewidth]{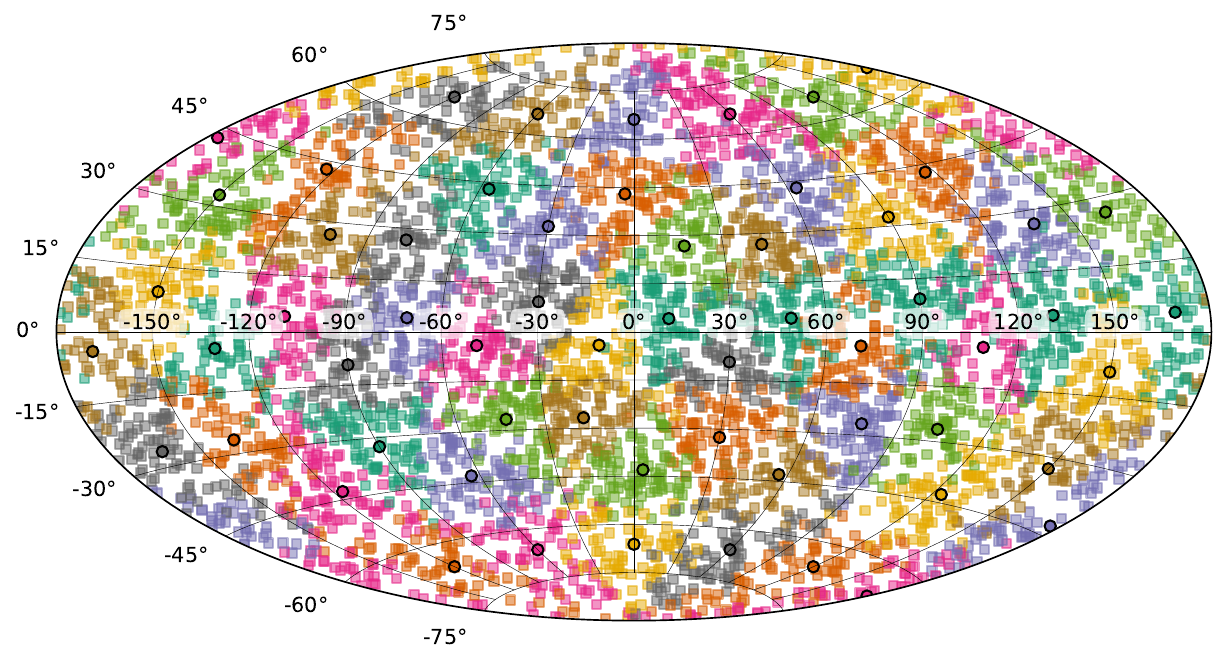}}
\subfigure[\label{fig:achinosMuon_a}]{\includegraphics[width=0.45\linewidth]{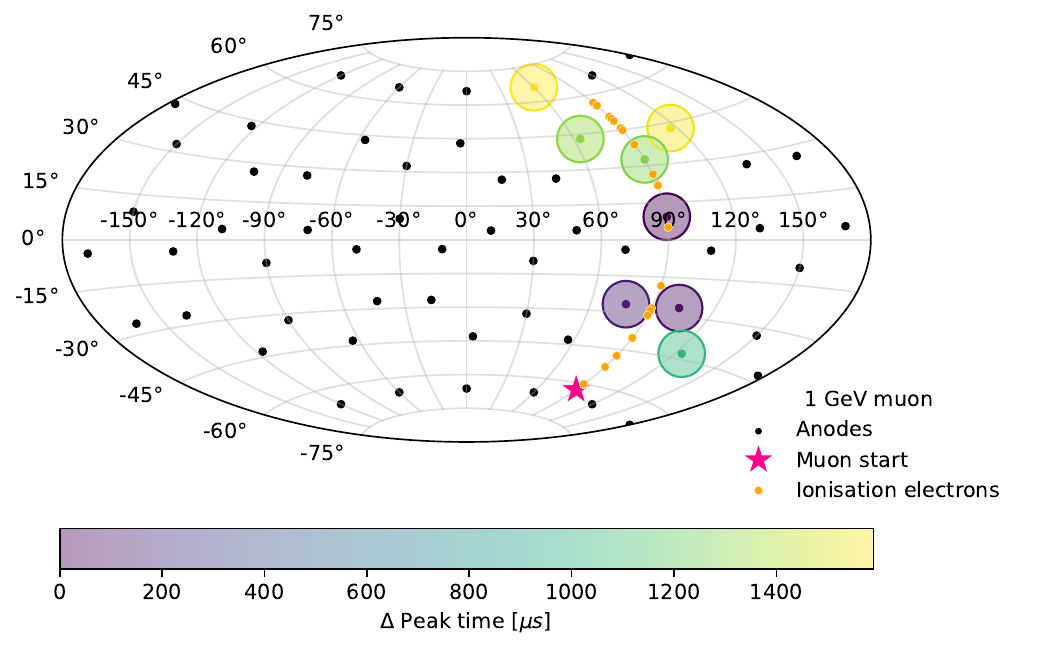}}
  \caption{Simulation of DarkSPHERE filled with isobutane and equipped with a 60-anode ACHINOS. \subref{fig:achinosMuon_b} shows the simulations of individual electrons uniformly distributed inside DarkSPHERE. \subref{fig:achinosMuon_a} shows a $1\;\si{\giga\eV}$ muon traversing the detector, with the muon starting position denoted by with a star. Ionisation electrons produced by the passage of the muon are shown with the orange points. The black points indicate the anode locations, with the larger coloured circle indicating the relative signal peak times for signals recorded on each anode above a set threshold~\cite{NEWS-G:2023qwh}. \label{fig:achinosMuon}}
\end{figure}

The natural next step, towards enhanced localisation information, is the individual read-out of each anode in the ACHINOS. In this case, generalised versions of the Near versus Far signal asymmetry can be derived, e.g. centre-of-gravity techniques, to obtain the position.
Figure~\ref{fig:achinosMuon_b} shows individual electrons simulated  uniformly in detector with their initial positions coloured corresponding to the anode to which they arrive after drifting and diffusing in the gas volume. 
In Fig.~\ref{fig:achinosMuon_a} a muon traversing a spherical proportional counter equipped with a 60-anode ACHINOS is simulated.
Highlighted anodes recorded a signal above a set threshold, and are coloured based on the onset time of the signal relative to the onset time of the earliest recorded anode signal. 
This information could lead to full three-dimensional track reconstruction.

\subsection{Scintillation read-out}
In particle interactions in the gas, along with the ionisation charge, photons are also produced from de-excitation of atoms/molecules. Additionally, photons are also generated during the electron amplification at the anode. This could provide a prompt signal, compared to the relatively slow drift of electrons, and thus timing information to enable detector operation as a Time Projection Chamber (TPC). A first proof of principle application of simultaneous charge and light read-out has been achieved as part of the R2D2 research programme~\cite{Meregaglia:2017nhx, Katsioulas:2021usd}, exploring the application of the spherical proportional counter to neutrinoless double $\beta$-decay searches~\cite{Bouet:2022kav}. 

The R2D2 project aims to use a spherical proportional counter filled with high-pressure xenon to search for neutrinoless double $\beta$-decay of $^{136}$Xe, utilising the many strengths of the spherical proportional counter, not least its energy resolution~\cite{Bouet:2020lbp}, required to distinguish the shape of the double $\beta$-decay spectrum. 

\begin{figure}[htbp]
\centering
\vspace{-0.5cm}
    \subfigure[\label{fig:r2d2setup}]{\includegraphics[width=0.21\linewidth]{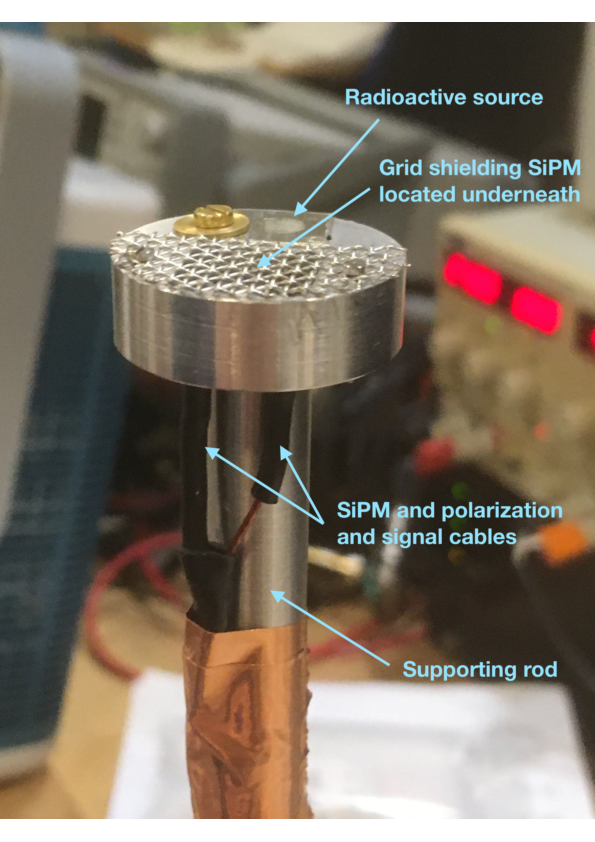}}%
    \hspace{0.02\linewidth}
     \subfigure[\label{fig:r2d2result}]{\includegraphics[width=0.435\linewidth]{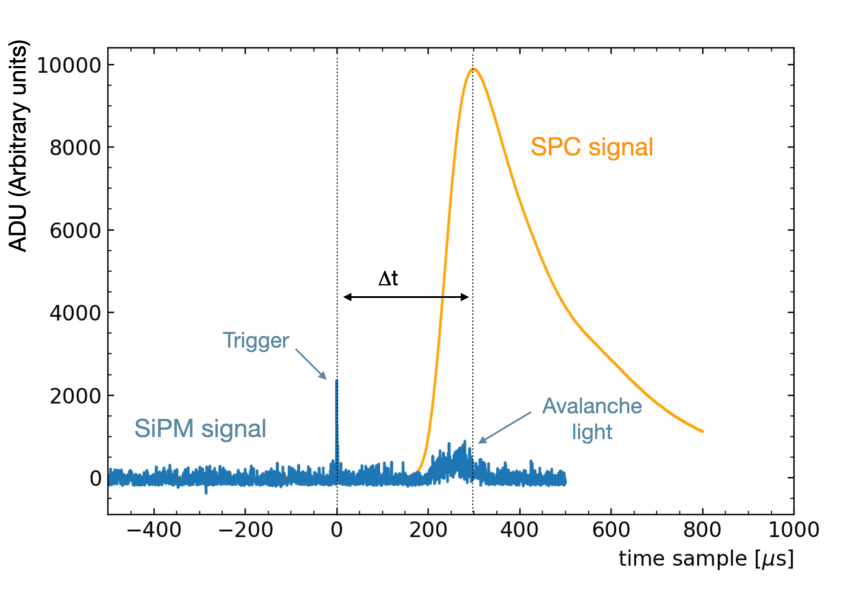}}%
     \vspace{-0.3cm}
    \caption[]{\label{fig:r2d2} \subref{fig:r2d2setup} The silicon photomultiplier and $^{210}$Po source holder, which was arrange to sit at an opening in the cathode of the spherical proportional counter. \subref{fig:r2d2result} Amplitude recorded by the silicon photomultiplier and the spherical proportional counter anode as a function of time. From Ref.~\cite{Bouet:2022kav}}. 
    \vspace{-0.3cm}
\end{figure}

Light read-out was achieved using the set-up shown in Fig.~\ref{fig:r2d2setup}, where a Hamamatsu S13370 $6\times6\;\si{\milli\meter\squared}$ silicon photomultiplier is mounted to a holder along with a $^{210}$Po source. The holder was inserted through an opening in the SPC cathode such that the holder was flush with the inner cathode surface. This allowed both the light and charge, read-out by a single-anode sensor, to be read-out simultaneously, with an example shown in Fig.~\ref{fig:r2d2result}. The spike in amplitude labelled "Trigger" corresponds to the initial interaction, where the time between this and the charge signal onset corresponds to the electron drift time, which was found to agree with simulations. Additionally, the light produced in the avalanche is seen, showing the arrival of groups of electrons. 

\section{Ionisation Quenching Factor}
\label{sec:quenching}

As mentioned in Section~\ref{sec:introduction}, direct detection experiments search for signatures of scattering of DM particles off a target nucleus. The momentum transfer gives rise to a nuclear recoil, which will subsequently dissipate its kinetic energy in the medium through inelastic Coulomb interactions with atomic electrons -- electronic energy losses: ionisation electrons and excitation of atomic and quasi-molecular states -- and elastic scattering in the screened electric field of the nuclei -- nuclear energy losses. The secondary recoil atoms and electrons may, in turn, undergo further scattering, transferring energy to other particles. For each type of detector technology, it is important to maximise the recoil kinetic energy dissipated through the sensitive channel --  ionisation, scintillation, or heat -- for the specific system, i.e. the ``visible'' energy, while a precise knowledge of this fraction is of major importance to reconstruct the recoil energy from the observed signals.  

The concept of quenching factor, i.e. the ratio of energy given to the electronic energy losses over the ion kinetic energy, was analysed by Lindhard et al.~\cite{Lindhard1963} and is characteristic of the material, ion species and energy. 
Broadly speaking, electronic energy losses dominate for fast ions, while nuclear energy losses are increasingly important for decreasing ion kinetic energy and become dominant
for ion velocities smaller than the electron orbital velocity. 
However, as noted in Ref.~\cite{Katsioulas:2021pux}, multiple definitions of the quenching factor may be found in the literature:
\begin{enumerate}
\item the fraction of the ion kinetic energy that is dissipated in a medium in the form of electronic energy losses, which coincides with that of  Lindhard et al.~\cite{Lindhard1963};
\item the ratio of the ``visible'' energy in a detector to the kinetic energy of the recoil; and
\item the conversion factor between the kinetic energy of an ion and an electron yielding the same ``visible'' energy in the detector.
\end{enumerate}

For the measurement of ionisation quenching factors in a material, ions with precisely known kinetic energy are required. In the case of ionisation measurement, typically, the amount of ionisation induced by an ion species is measured as a function of its kinetic energy and compared to the corresponding ionisation by electrons of the same kinetic energy. Given the long history of direct searches for DM with detectors using crystalline target materials~\cite{Ahlen:1987mn} and liquid noble elements~\cite{Aprile:2018dbl, Akerib:2016vxi, lux2016, zeplin2012},  a number of investigations of the corresponding ionisation quenching effects have been performed~\cite{Joo:2018hom, xenon_qf, silicon_qf, germanium_qf}. However, measurements of the ionisation quenching factor in gases relevant for DM searches with the spherical proportional counter, remain relatively sparse. In the following, the different approaches pursued in this direction will be summarised.

The COMIMAC facility at LPSC Grenoble features a compact Electron Cyclotron Resonance source~\cite{Muraz:2016upt} which is used to produce and accelerate electrons and ions. Their energy can be selected by means of a tunable extraction potential. This is coupled by means of a narrow -- $1\;\si{\micro\meter}$ in diameter -- opening to a gaseous detector in the volume in which the electrons and ions interact and dissipate their energy. The use of both Micromegas and spherical proportional counters has been explored in this context. 
The capacity to accelerate both electrons and ions through the same equipment enables a direct comparison of the measured signal between the species at the same initial kinetic energy, as per definition 3 above. Thus, any observed differences can be, in principle, attributed to the ionisation quenching factor, and be used directly for their quantification. Measurements of 
ionisation quenching factor for
\ce{He^+} in \ce{He}:\ce{C3H8} mixtures~\cite{santos2008ionization}, and protons in i-\ce{C4H10}:\ce{CHF3} mixtures~\cite{Tampon:2017mpm} and \ce{CH4}~\cite{NEWS-G:2022fym} have been performed. 

Another approach is to measure the ionisation quenching factor by generating recoils through nuclear scattering. Neutrons from a narrow in size and energy beam produced in an accelerator facility scatter off nuclei in the detector gas volume and are subsequently registered in a backing detector. This arrangements enables the estimation of the scattering angle, and, thus, the kinetic energy of the nuclear recoil. By adjusting the relative position of the backing detector ring to the spherical proportional counter, a range of nuclear recoil kinetic energies can be probed. By comparing the nuclear recoil signal measured in the detector to the signal generated by a calibration source, such as $^{55}$Fe, the ionisation quenching factor can be measured, following definition 2 above.
A measurement of the quenching factor for neon in a \ce{Ne}:\ce{CH4} gas mixture
has been performed~\cite{NEWS-G:2021mhf} at the Tandem Laboratory of TUNL, USA. Protons accelerated by a tandem Van de Graaf accelerator to $20\;\si{\mega\eV}$ are directed on a lithium fluoride target. As a result,  (quasi-)monochromatic neutrons are produced from the reaction \ce{^{7}Li + p \to ^{7}Be + n} reaction.

\begin{figure}[h]
  \centering
    \vspace{-0.2cm}
 \includegraphics[width=0.75\linewidth]{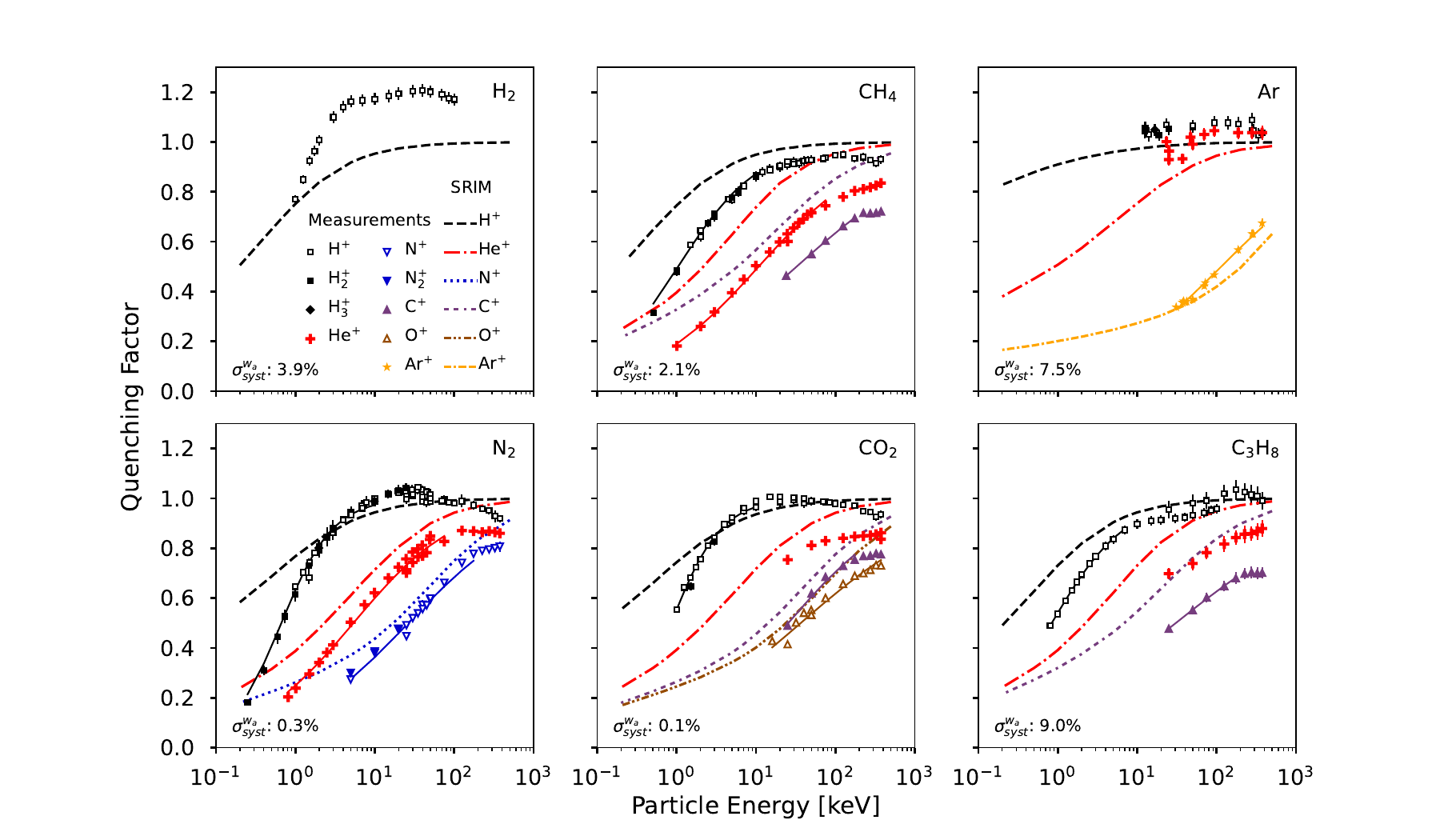}
  \vspace{-0.2cm}
\caption{Ionisation quenching factors estimated from measurements of the $W$-value from various authors, as discussed in Ref.~\cite{Katsioulas:2021pux}. Data are fit (solid lines) with a functional form $q (E) = \frac{E^{\alpha}}{\beta+E^{\alpha}}$, where $q$ is the ionisation quenching factor, $E$ is the particle energy, and $\alpha, \beta$ are constants. Results form SRIM simulations are shown for comparison.\label{fig:WvaluequenchingFactor}}
\vspace{-0.3cm}
\end{figure}

A more recent approach utilises measurements of the $W$-value, i.e. the mean energy required for the creation of an electron-ion pair, of electrons and ions to estimate the ionisation quenching factor in pure gases~\cite{Katsioulas:2021pux}. Historically, $W$-values, particularly in common gases, tissue-equivalent gases, and its constituents, have been measured in the context of dosimetry. Typically, these measurements are performed with ionisation chambers, and experiments are carefully designed to control and mitigate any detector-specific effects that could influence the measured $W$-value. By comparing measurements of the $W$-value of electrons to those of ions the ionisation quenching factor is estimated, corresponding to either as definition 2, in the case where the asymptotic electron $W$-value is estimated, or definition 3 when the energy-dependent electron $W$-value is used. Figure~\ref{fig:WvaluequenchingFactor} shows the ionisation quenching factor in several gases estimated using this method. A discrepancy between ionisation quenching factors estimated using SRIM~\cite{ZIEGLER20101818}, also reported elsewhere~\cite{santos2008ionization}, is apparent at low energies. It can be seen that the ionisation quenching factor for protons in some gases exceeds unity in a narrow energy band. While seemingly counter intuitive, this is explained by a minimum in the $W$-value for protons, caused by charge exchange reactions involving protons and the molecule (atom) of the gas~\cite{VDNguyen_1980,EWaibel_1992}. 

While these three methods are promising avenues to elucidate the ionisation quenching factor, varying degrees of tension have been observed with other estimates, such as SRIM~\cite{ZIEGLER20101818} calculations. 
The COMIMAC method has been used to perform measurements down to sub-keV quenched ion energies, however, stable operation with ion beams of kinetic energies below $1\;\si{\kilo\eV}$ remains a challenge. Furthermore, while significant progress in understanding this method has been made, including those enabled by the development of the spherical proportional counter instrumentation read-out, open questions remain, including understanding the observed pressure dependence of some measurements~\cite{santos2008ionization}. For the neutron scattering method, the calibration of the detector over a larger range of energies would enhance the measurement technique, which shows significant potential to achieve low ion energies. The $W$-value estimates can be further extended to other gases as measurements become available, and can also be applied to gas mixtures. 
Direct comparison of the results obtained by these methods for the same gas mixtures could provide further insights.

\section{Radiopurity considerations}
\label{sec:radiopurity}

The suppression of experimental backgrounds originating from radioisotopes is of critical importance to rare-event searches. Nuisance radioisotopes, primarily from the $^{238}$U and $^{232}$Th decay chains, can be found in the materials of the laboratory, the detector materials used to construct the detector, or the gas with which it is filled. Much of the effort in applying spherical proportional counters to DM searches has been in suppressing these backgrounds, by improving the detector shielding, using of radiopure construction materials, and by developing dedicated gas purifiers. 
\subsection{Detector Materials}
Copper is an attractive material from rare-event search experiments, thanks to its relatively low cost, favourable mechanical and electrical properties, and its commercial availability in high purity. SEDINE was constructed from commercial NOSV copper from Aurubis. While copper has no long-lived radioisotopes of its own, it can become contaminated during manufacturing, activated by fast neutrons produced by cosmic ray interactions, or have contamination remaining from processing of the ore. A particular risk is contamination with $^{222}$Rn, a progeny of $^{238}$U, present in the manufacturing environment, whose subsequent decay chain includes the long-lived isotope $^{210}$Pb. This breaks the secular equilibrium in the $^{238}$U chain, meaning that sensitive radioassay techniques measuring the uranium content are insensitive to this contamination. The decay of $^{210}$Pb and its daughters can present a significant long-term background to rare-event searches. However, copper's favourable electrochemical properties mean that it can be purified through additive-free potentiostatic electroformation~\cite{Abgrall:2013rze}, which has been shown to have favourable properties~\cite{Hoppe2009JRADIOANALNUCLCHEM,Overman2012TECHREPmajoranaCopper,Overman2015}.   

Electroformation is performed using an electrolytic cell with an electrolyte separating two electrodes. An electric current supplies electrons to the cathode which are used to reduce ions from the electrolyte and form atoms on its surface through the reaction
\begin{equation*}
    A^{z+} + ze^{-} \rightarrow A\,,
\end{equation*}
where $A$ is the species and $z$ is its ionic charge. Other reactions, where the ionic species is not completely electrically neutralised may also occur. Opposite reactions, where an atomic species is oxidised and forms ions, occur at the anode. The possible cathode and anode reactions  are each known as the half-cell reaction. The tendency of a given ionic species to reduce in a given half-cell reaction is given by its reduction potential $E^{0}$, where higher reduction potentials correspond to more easily reduced species. Reduction potentials for several relevant ions are provided in Table~\ref{tab:reductionPotentials}. The combined reaction in the cell is characterised by the standard cell potential defined as the difference of the cathode and anode half-cell reduction potentials,
\begin{equation*}
    E_{\text{cell}} = E^{0}_{\text{C}} - E^{0}_{\text{A}}\,.
\end{equation*}
When $E_{\text{cell}}\geq0$, the reaction is spontaneous, or in chemical equilibrium for the case of equality. However, when $E_{\text{cell}}<0$, additional energy is required for the reaction. Given the relatively high reduction potential of copper, many radiocontaminants are unable to reduce, meaning that copper can be purified in this way. However, this simplified model neglects that a small electrical potential is required to drive the reaction, which must be carefully controlled to prevent deposition of contaminants. Moreover other factors, e.g.  mass transport of contaminant ions, may result in species with lower reduction potentials to be deposited with the copper in small amounts.
\begin{table}[!h]
\centering
\caption{\label{tab:reductionPotentials}Reduction potential $E^{0}$ of copper and common radioactive contaminants.}
\vspace{0.5em}
\begin{tabular}{lllll}
\hline
Reductants &  & Oxidants & $E^{0}\;$[$\si{\volt}$] & \\ \hline
$\text{Cu}^{2+} + 2 e^{-}$  & $\leftrightharpoons$  	&$\text{Cu}$   	& $ +0.34$&\cite{bard1985standard} \\
$\text{Pb}^{2+}+2e^{-}$	&$\leftrightharpoons$	&$\text{Pb}$	& $-0.13$&\cite{atkins1997crc}     \\    
$\text{U}^{3+}+3e^{-}$	&$\leftrightharpoons$	&$\text{U}$	& $-1.80$&\cite{lide2006crc}     \\     	
$\text{Th}^{4+}+4e^{-}$	&$\leftrightharpoons$	&$\text{Th}$	& $-1.90$&\cite{lide2006crc}    \\    	
=$\text{K}^{+}+e^{-}$    	&$\leftrightharpoons$	&$\text{K}	$       & $-2.93$&\cite{haynes2011crc}     \\   \hline  
\end{tabular}
\end{table}
Several experiments have employed this technique to produce a variety of detector components. NEWS-G has used it to deposit a $500\;\si{\micro\meter}$ thick ultra-pure copper layer to the internal surface of its current detector, acting as an internal shield to the underlying commercial $99.99\%$ pure copper~\cite{Knights:2019tmx, Balogh:2020nmo}. The electroplating was performed at LSM, suppressing cosmogenic activation of the copper while the process was underway. The electroforming rate achieved was around $1\;\si{\milli\meter\per month}$. 

Next-generation NEWS-G detectors, ECuME and DarkSPHERE, will be constructed entirely from electroformed copper, intact, with minimal additional machining or welding required. Furthermore, they will be constructed directly in the underground laboratories where they will be operated. This will suppress the cosmogenic activation of the produced copper, which is expected to be the dominant background with \textit{S140}, after the copper intrinsic background. 

Depending on the electroformed vessel's requirements, there are several challenges that must be overcome. One of the major ones is the malleability of copper, which can make it deform at large scales or under pressure. A proposed method to overcome this is the use of ultra-pure alloys, e.g. an alloy of copper and chromium~\cite{Vitale:2021xrm}. 

\subsection{Environmental backgrounds and detector shielding}
The rock and support structures around the underground experimental area where a direct DM experiment is situated have their own intrinsic radioactivity. Isotopes of concern include $^{40}$K, which can decay via electron capture and emit high-energy photons, and the $^{238}$U and $^{232}$Th decays chains, which contain a variety photon-emitting decays that can provide a background to an experiment. Furthermore, gaseous radon isotopes from these chains are present in the laboratory air in varying activities depending on the laboratory and its geology. Additionally, cosmic muon spallation  can produce fast neutrons, which can mimic a DM particle.

To suppress these contributions, experiments employ a shield around the sensitive detector to stop or reduce the flux of incoming photons and neutrons. 
NEWS\=/G's first detector, SEDINE, employed a cubic shield of $8\;\si{\centi\meter}$ of copper, $15\;\si{\centi\meter}$ of lead and $30\;\si{\centi\meter}$ of polyethylene, from inside to outside. The copper and lead shield the detector from photons, while the polyethylene thermalises, scatters, and captures neutrons. Copper was placed at the inner layer to shield the detector from the lead, which itself contains radioisotopes that could induce a background. The intervening space was continuously flushed with reduced-radon air, to displace the laboratory air, which is typically radon-burdened.

\begin{wrapfigure}{R}{0.34\linewidth}
  \centering
    \vspace{-0.65cm}
    \includegraphics[width=0.99\linewidth]{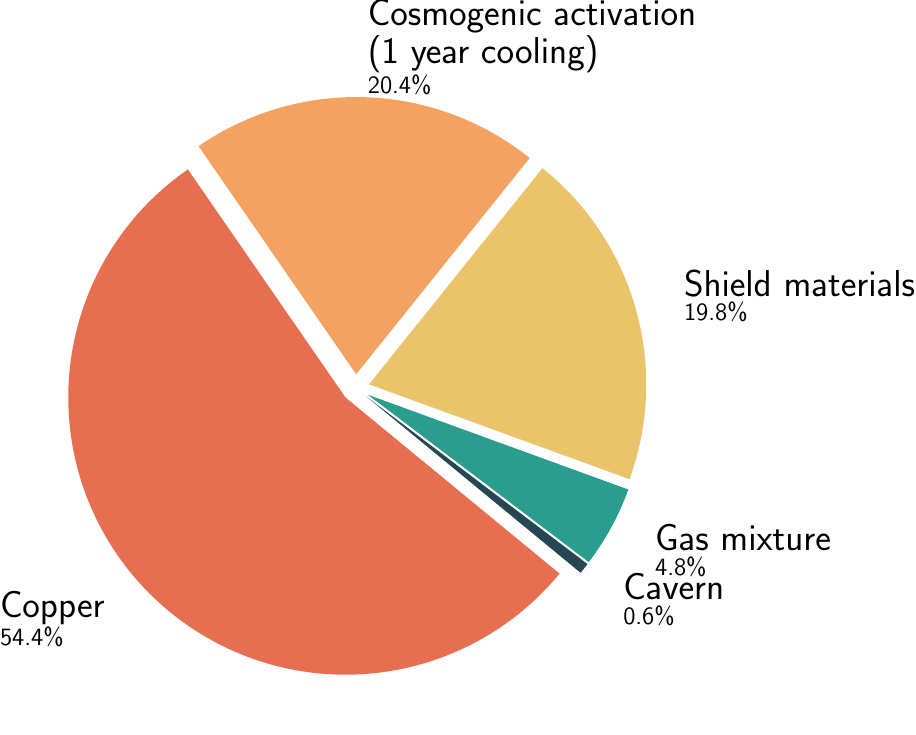}
  \vspace{-0.4cm}
    \caption{Expected \textit{S140} background~\cite{NEWS-G:2022kon}.\label{figure:snoglobeBackgrounds}}
  \vspace{-0.6cm}
\end{wrapfigure} 
The current NEWS-G detector, S140, uses an compact shielding with reduced background. From inside to out, it comprises a spherical shell of $3\;\si{\centi\meter}$ archaeological lead and $22\;\si{\centi\meter}$ of commercial low-activity lead, housed inside an octogonal neutron shield of $40\;\si{\centi\meter}$-thick high density polyethylene. The shielding was partially sealed with foam between the joints, and the intervening space was flushing with nitrogen gas. Similarly to the copper in SEDINE's shielding, the archaeological lead acts as a shield to the low-activity lead. The expected background contributions are summarised in Fig.~\ref{figure:snoglobeBackgrounds}. 

To take advantage of the exceptional radiopurity of a fully electroformed detector, the shielding for DarkSPHERE needs to change from that used by \textit{S140}; Figure~\ref{figure:snoglobeBackgrounds} shows that the shielding material would make up the vast majority of the remaining experimental background, because the copper's intrinsic background, including cosmogenic activation would be suppressed. Instead, a modular pure-water design is envisaged. Simulation studies have shown~\cite{NEWS-G:2023qwh} that a water shielding of $2.5\;\si{\meter}$ thick walls could suppress environmental background from the laboratory to below the target background level. The dominant background component remains the cosmic muons, however, it is expected that the majority of these can be suppressed through the use of an active veto. This could be implemented by either instrumenting the water shield, or including a veto scintillator plane above the detector and shield.

\subsection{Radiopurity of the gas mixture}
The progress achieved or anticipated given the earlier discussion, suggests that the radiopurity of the gas mixture may become relevant. Typically, radioactivity of the gas mixture arises either due radioisotopes of constituents, contaminants, thus requiring high grade of purity, or from cosmogenic activation. For example, hydrocarbons such as methane or isobutane should be sourced from natural gas deposits, where cosmogenically produced isotopes are suppressed. In the future, measurements of the radioactive contamination in the gas will be performed prior to use in detectors.

In addition to radiopurity, the contamination of the gas with electronegative impurities can impact detector performance. While the $1/r^{2}$ electric field enables amplification of drifted electrons near the anode, it also means that the electric field becomes increasingly weak towards the outer detector volume. This makes the detector susceptible to electron attachment by electronegative impurities in the gas. To overcome this, commercial purifiers have been employed to remove O$_{2}$ and H$_{2}$O contamination from a gas as it was introduced to the detector~\cite{Knights:2019tmx}. While this greatly improved detector performance, it was found that the purifiers emanated $^{222}$Rn, which is an additional background for spherical proportional counter employed for rare direct DM searches. 

Given the proprietary nature of such purifiers, it was challenging to
identify the specific components resulting in $^{222}$Rn
emanation. Instead, custom purifiers, based on CU-0226 (Q-5) for \ce{H2O}
adsorption, $3\;\si{\angstrom}$ and $4\;\si{\angstrom}$ molecular
sieves have been developed~\cite{Mavrokoridis:2011wv,Knights:2021}. These have
demonstrated a factor of 20 reduction in $^{222}$Rn emanation rate
compared to a commercial purifier. They were also not found to
preferentially remove one gas component over another, when a
Ne:i-C$_{4}$H$_{10}$ gas mixture was circulated through the purifier
for several weeks with the respective fractions of Ne and
i-C$_{4}$H$_{10}$ monitored by a binary gas analyser. In parallel,
efforts to develop $^{222}$Rn removal systems to be used in
conjunction with this purifier are being developed by
NEWS-G~\cite{pobrienMastersThesis}.

\section{Physics Opportunities}
\label{sec:opportunities}

The spherical proportional counter provides a large number of physics
opportunities, which is discussed in this section.  The
sensitivity of spherical proportional counters to search for light
DM through DM-nucleon and DM-electron scattering is presented, as
well as capability to constraint more exotic scenarios, such as
strongly-interacting DM candidates with a mass near the Planck
scale. The physics potential of the derector in neutrino physics,
through coherent neutrino-nucleus scattering, and in neutrinoless
double-beta decay is also briefly discussed.

In the following, except if noted otherwise, the \textit{DarkSPHERE} detector
-- a $3\;\si{\meter}$ in diameter fully electroformed underground
spherical proportional counter with a $2.5\,\si{\meter}$-thick pure
water shield at Boulby, operated with a He:i-C$_{4}$H$_{10}$
(90\%:10\%) gas mixture at a pressure of $5\,\si{\bar}$ -- is
considered, as discussed in Section~\ref{sec:spcDM} and in Ref.~\cite{NEWS-G:2023qwh}.  A running time
of $300\;\si{days}$ and a flat background rate of $0.01\;\si{dru}$
below $1\;\keV$ is assumed~\cite{NEWS-G:2023qwh}.
The sensitivity is quantified as the $90\%$ confidence level (CL)
upper limit on the DM interaction cross-section, estimated
using a profile likelihood ratio
test statistic~\cite{Cowan:2010js} in the energy range from
$14\;\si{\eV}$ to $1\;\si{\kilo\eV}$.

\subsection{Direct DM searches through DM-nucleus scattering and DM-electron scattering}
\label{sec:PhysicsNR}
\begin{figure}[h] 
\centering
\vspace{-0.75cm}
\subfigure[\label{fig:limitsSI}]{\includegraphics[width=0.32\columnwidth]{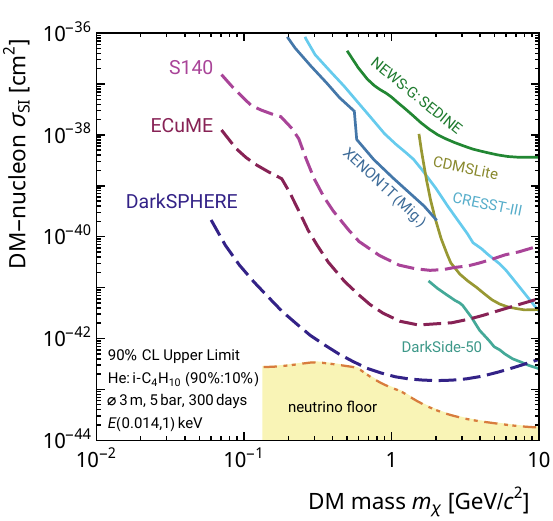}}
\subfigure[\label{fig:limitsSDp}]{\includegraphics[width=0.32\columnwidth]{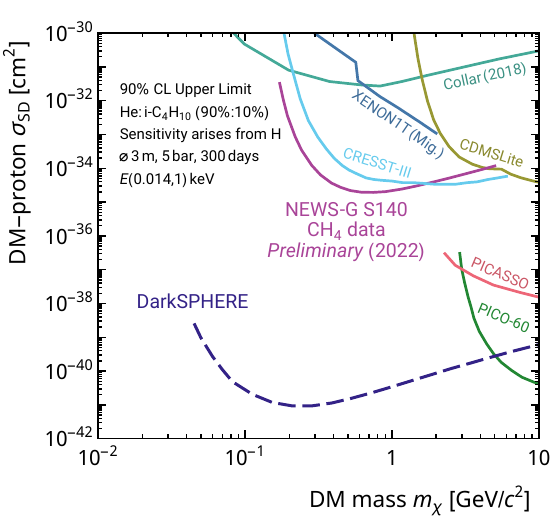}}
\subfigure[\label{fig:limitsSDn}]{\includegraphics[width=0.32\columnwidth]{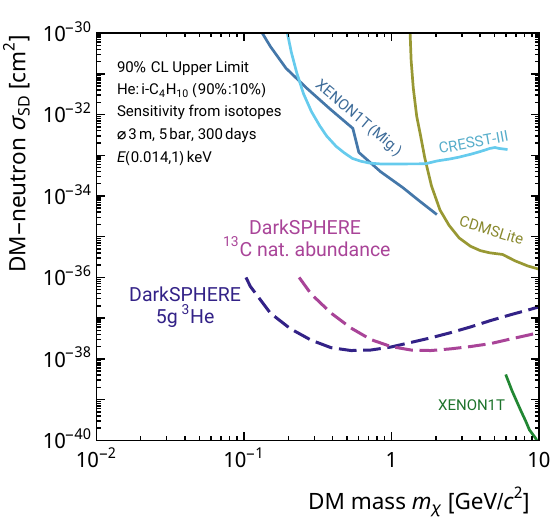}}
  \vspace{-0.3cm}
  \caption{Expected DarkSPHERE sensitivity for \subref{fig:limitsSI}~spin-independent DM-nucleon interaction cross-section, and spin-dependent \subref{fig:limitsSDp}~DM-proton and \subref{fig:limitsSDn}~DM-neutron interaction cross-section. 
  Existing constraints are
  summarised both for spin-independent~\cite{Agnese:2018gze,Abdelhameed:2019hmk,Agnes:2018ves,Arnaud:2017bjh,Aprile:2019jmx} and spin-dependent~\cite{SuperCDMS:2017nns,Collar:2018ydf,CRESST:2022dtl,Behnke:2016lsk,PICO:2019vsc,Aprile:2019jmx} searches. From Ref.~\cite{NEWS-G:2023qwh}.
  \label{fig:limits}
  }
  \vspace{-0.3cm}
\end{figure}
\begin{wrapfigure}{R}{0.30\linewidth}
  \centering
  \vspace{-0.4cm}
  \includegraphics[width=0.99\linewidth]{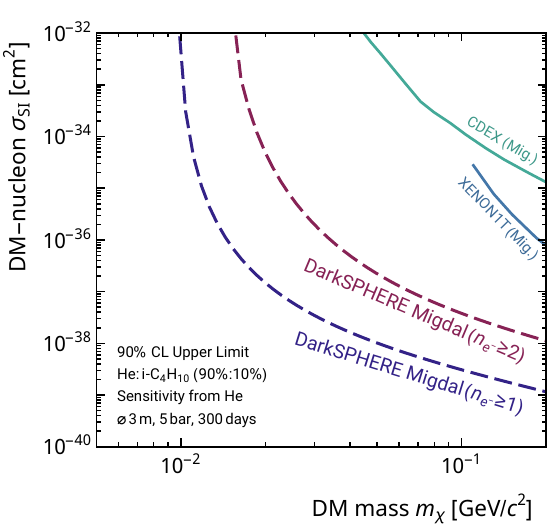}
  \vspace{-0.3cm}
\caption{DarkSPHERE sensitivity for spin-independent DM-nucleon interactions
exploiting the Migdal effect~\cite{NEWS-G:2023qwh},
compared to existing constraints~\cite{CDEX:2021cll,Aprile:2019jmx}. From Ref.~\cite{NEWS-G:2023qwh}.\label{fig:Migdallimits}}
  \vspace{-0.5cm}
\end{wrapfigure}

\textit{DarkSPHERE} is expected to probe extensive new regions of parameter space. The sensitivity to  spin-independent (SI) DM-nucleon cross-section versus the DM candidate mass is shown in Fig.~\ref{fig:limitsSI}. The corresponding reach for \textit{S140} and \textit{ECuME}, estimated using the optimal interval method~\cite{Yellin:2002xd}, is shown for comparison~\cite{NEWS-G:2022kon}, along with other published constraints.  The `neutrino floor' is also shown, calculated for He:i-C$_4$H$_{10}$ (90\%:10\%) based on Refs.~\cite{Baxter:2021pqo} and  Ref.~\cite{Billard:2013qya}. Approximately two events from solar neutrino interactions are expected in the exposure assumed for \textit{DarkSPHERE}.
  
The use of hydrogen -- a spin-$1/2$ proton nucleus -- in the gas
mixture provides sensitivity to interactions that depend on the
spin of the nucleus~\cite{Fitzpatrick:2012ix}. Fig.~\ref{fig:limitsSDp} 
presents \textit{DarkSPHERE}'s sensitivity to the DM-proton spin-dependent (SD) interaction.
The results from {\sc S140} obtained with CH$_{4}$ gas are also shown, which provide the strongest constraint in the $0.2$-$2\,\si{\giga\eV}$ DM mass range~\cite{NEWS-G:2024jms}.  
 
Approximately~1.1\% of natural carbon is $^{13}$C. This provides sensitivity to spin-dependent DM-neutron interaction, shown in Fig.~\ref{fig:limitsSDn}, coming from the carbon found in the i-C$_4$H$_{10}$ component of that gas. Furthermore,  small amounts of $^3$He could be added to the gas, albeit for an additional financial cost, which would result in enhanced sensitivity at even lower candidate masses, as also shown for $5\,\si{\gram}$ of $^3$He in Fig.~\ref{fig:limitsSDn}. Moreover, isotopically separated gases with significantly enhanced fraction of $^{13}C$ can be procured, which would result to a significant improvement in the corresponding sensitivity to spin-dependent DM-neutron interactions. 

As discussed in Section~\ref{sec:introduction}, the Migdal effect is
employed by several collaborations to extend their sensitivity to
light DM particles. This is applicable also to the searches with the
spherical proportional counter, and in the case of DarkSPHERE enables
the exploration of masses down to $100\;\MeV$.  This is shown in
Fig.~\ref{fig:Migdallimits}, where the projected sensitivity of
DarkSPHERE for spin-independent DM-nucleon cross-section using the
Migdal effect is presented\footnote{The Migdal probability for helium
from Ref.~\cite{Cox:2022ekg} is used. Although this  should apply also to molecules~\cite{Lovesey:1982,Blanco:2022pkt}, probabilities for i-C$_4$H$_{10}$ have not been calculated and are, thus, not included.}.

\begin{figure}[t!]
\centering
\subfigure[\label{fig:limits_electron1}]{\includegraphics[width=0.32\columnwidth]{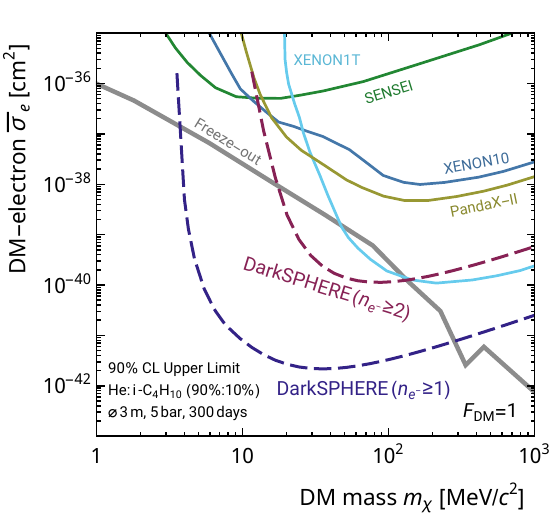}}
\subfigure[\label{fig:limits_electron2}]{\includegraphics[width=0.32\columnwidth]{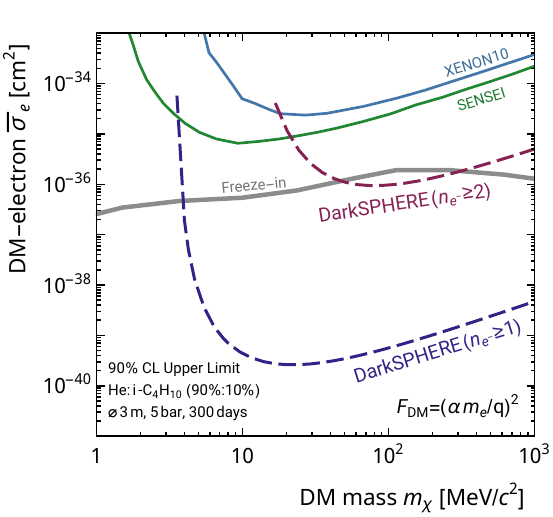}}
  \caption{Expected DarkSPHERE sensitivity for 
  DM-electron interaction cross-section for \subref{fig:limits_electron1} contact
  interactions with DM form factor $F_{\rm{DM}}=1$
  and \subref{fig:limits_electron2} a light mediator with DM form
  factor $F_{\rm{DM}}=(\alpha m_e/q)^2$, where $\alpha$, $m_e$ and $q$
  are the fine-structure constant, mass of electron  and transfer of momentum. The two DarkSPHERE scenarios
  correspond to two different thresholds: ($n_{e}\!\geq 1$) and 
  ($n_{e}\!\geq 2$), where $n_{e}$ is the number of electrons. Solid coloured lines show
  existing  constraints~\cite{SENSEI:2020dpa,Essig:2017kqs,XENON:2019gfn,PandaX-II:2021nsg}. The grey line correspond to the benchmark freeze-out and freeze-in  light DM models~\cite{Essig:2011nj,Essig:2015cda}.   From Ref.~\cite{NEWS-G:2023qwh}.\label{fig:limits_electron}}
\end{figure}

The spherical proportional
counter, thanks to its single ionisation electron threshold, has also the potential to search for
DM-electron interactions~\cite{Hamaide:2021hlp}.  
DM-electron scattering provides a search channel with sensitivity to even lower mass candidates, between 0.005 and 1~GeV~\cite{Essig:2011nj}. DarkSPHERE's sensitivity through DM-electron interaction channel,  following the approach of
Ref.~\cite{Hamaide:2021hlp}, is shown for contact interactions in Fig.~\ref{fig:limits_electron1} and for  interactions through a light mediator in Fig.~\ref{fig:limits_electron2}. 
 
\subsection{Searches for heavy DM candidates}
\begin{figure}[t!] 
\centering
\includegraphics[width=0.32\columnwidth]{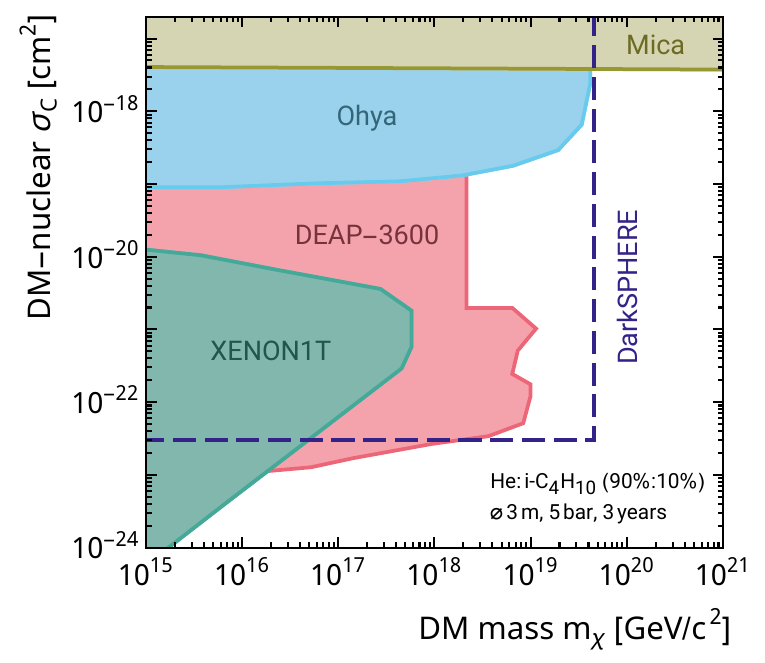}
  \caption{\label{fig:limitsHeavyDM}
  \textit{DarkSPHERE}'s expected sensitivity to the contact DM-nuclear cross-section (dashed line). Available limits are shown in the shaded region, corresponding to the DEAP-3600 search (`Model I' with extrapolation)~\cite{Adhikari:2021fum},
  XENON1T single scattering constraints~\cite{Clark:2020mna}, Ohya etching plastic searches~\cite{Bhoonah:2020fys} and constraints from ancient mica samples~\cite{Acevedo:2021tbl}. %
  From Ref.~\cite{NEWS-G:2023qwh}.
}
\end{figure}

Spherical proportional counters are also able to probe strongly interacting DM candidates with Planck-scale masses. Many massive candidates are provided by several theories, for example grand unification~\cite{Burdin:2014xma} among others~\cite{Raby:1997pb, Hardy:2014mqa, Ponton:2019hux,Lehmann:2019zgt}. The large mass of these candidates suppresses their number density, to satisfy the observed relic density, and also means that they interact much more strongly with matter. Therefore, it is expected that such candidates will interact multiple times in the detector and its overburden. The large  cross-sectional area of the detector means that {\it DarkSPHERE} would be able to provide a factor $3$--$10$ improvement in mass reach per exposure time, with the estimated sensitivity shown in Fig.~\ref{fig:limitsHeavyDM}. Such a search can be performed concurrently with those already discussed above, as multiple scattering events will not be rejected online.

\subsection{Kaluza-Klein axions}
The standard QCD axions, postulated to solve the strong CP problem~\cite{Peccei:1977hh}, can have higher mass excited states in theories with extra dimensions, known as Kaluza-Klein axions~\cite{Dienes:1999gw, Chang:1999si}. These would have significantly shorter lifetimes meaning their decay could be observed. Furthermore, if they are produced in the Sun~\cite{DiLella:2002ea}, they remain gravitationally bound to our solar system, enhancing their observable decay rate in Earth-bound detector. Their decay to two photons is an experimental signature that the spherical proportional counter is well suited to search for, thanks to its relatively low density target and large volume. Using SEDINE, NEWS\=/G has performed a search for Kaluza-Klein axions, looking for two equal energy deposits in the detector~\cite{VazquezdeSolaFernandez:2020xcp,NEWS-G:2021vfh}. The concept was demonstrated with a $^{55}$Fe source, where both the $X$-ray from the de-excitation of the daugther $^{55}$Mn nucleus and the subsequent Ar fluorescence photon are detected in the volume filled with an argon-based gas mixture. The search was performed with a 42-day run with SEDINE filled with a 99.3$\%$ neon and 0.7$\%$ methane gas mixture at 3.1 bar, and a 90\% C.L. upper limit on the axion-photon coupling of $\num{9e-13}\;\si{\per\giga\eV}$ 
was set for the benchmark of a KK axion density on Earth of $\num{4.07e13}\;\si{\per\meter\cubed}$ and two extra dimensions of size $R = 1\;\si{\per\tera\eV}$. 

\subsection{Neutrino physics}
In addition to the rich programme of neutrino physics that can be performed with spherical proportional counters, outlined in the introduction, it is also well suited to the study of coherent elastic neutrino-nucleus scattering (CE$\nu$NS). The first detection of CE$\nu$NS in 2017 by COHERENT~\cite{Akimov:2017ade} opened up a new potential method or searching for physics beyond the standard model, and has  wide-reaching implications beyond high-energy physics, into astrophysics, nuclear physics, and beyond.
Detailed studies of CE$\nu$NS with the spherical proportional counter and its low energy threshold, will allow searches for new physics, for example through non-standard neutrino interactions or searches for sterile neutrinos. Studies have been performed into the application of the spherical proportional counter to CE$\nu$NS measurements~\cite{Katsioulas:2016fet, Vidal:2021tmt} and NEWS-G is planning the construction of a $\varnothing$60-cm detector deicated to this study, as part of the NEWS-G3 project. The detector, made from high-purity materials, will be encased in a shielding inspired by GIOVE~\cite{Heusser:2015ifa} and CONUS~\cite{Buck:2020opf} and including an active muon veto. Commissioning and characterisation of the shielding is underway at Queen's University, Canada, to assess environmental and cosmogenic backgrounds. 

Furthermore, the spherical proportional counter is well suited to other rare-event searches, including neutrinoless double $\beta$-decay searches. This is pursued by the Rare Decays with Radial Detector (R2D2) R\&D effort, aspiring to a tonne-scale $^{136}$Xe-filled spherical proportional counter operated as a Time Projection Chamber~\cite{Meregaglia:2017nhx, Katsioulas:2021usd}. In this context, advancements in energy resolution\cite{Bouet:2020lbp} and light read-out~\cite{Bouet:2022kav}-- which is would then establish the time of initial interaction -- have been demonstrated. 

\subsection{Physics  with a xenon-filled spherical proportional counter}
A further potential application for the spherical proportional counter is for supernova neutrino searches~\cite{Vergados:2005ny, Meregaglia:2017nhx}. 
\begin{wrapfigure}{L}{0.32\linewidth}
  \centering
  \vspace{-0.50cm}
\includegraphics[width=0.99\linewidth]{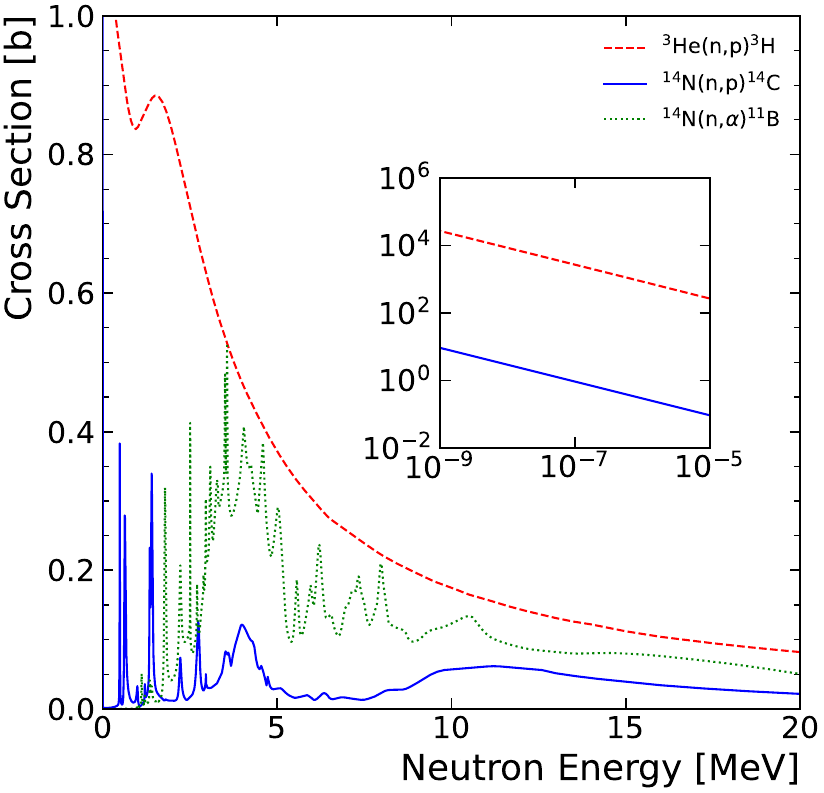}
  \vspace{-0.45cm}
\caption{Neutron capture cross-section  for different elements.\label{fig:neutronCrossSection}}
\vspace{-0.4cm}
\end{wrapfigure}
A $^{136}$Xe filled spherical proportional counter operation at $5\,\si{\bar}$ and benefiting from the same low background and good energy resolution described previously, would be capable of detecting within a short time $5\times (10\,\mathrm{kpc}/d_{\rm{SN}})^2$ neutrino interactions from a supernova explosion of a $27\,M_{\odot}$ progenitor at a distance $d_{\rm{SN}}$~\cite{Lang:2016zhv}.

\section{Measurements of neutron backgrounds}
\label{sec:neutrons}

A notorious background in DM searches arises from neutrons
cosmogenically and environmentally produced in the experimental
cavern.  Neutrons result in practically identical signal events in the
detector as those expected from DM interactions, and thus 
substantial effort is invested in suppressing their contribution
through shielding material and active veto systems. Furthermore,
neutron detection and measurement is important in a broad range of
applications, e.g. non-destructive material inspection, material
studies through neutron activation analysis, nuclear
materials detection, and for dosimetry in medical environments.

However, neutron spectroscopy measurements are still challenging and precise results are limited both within research laboratories and industrial applications~\cite{Brooks:2002oyj,Pietropaolo:2020frm}.
A widely used method is the use of $^3$He-based detectors, which utilise the low photon interaction probability and the large neutron capture cross section of the \ce{^3He + n \to p + ^3H + 765\;\keV} reaction, shown in Fig.~\ref{fig:neutronCrossSection}. 
However, this is not without challenges, e.g. energy measurements for fast neutrons are hampered by the ``wall effect'' -- the loss of a part of the energy of the recoiling nucleus due to its collision with the detector walls -- and the high cost of $^3$He. Other available  techniques, e.g. BF$_3$-based proportional counters, $^{10}$B lined proportional counters, and $^6$Li coated detectors, etc. also face challenges and limitations to their operation~\cite{KOUZES20101035}. 
 
The spherical proportional counter was applied to the measurement of neutrons early on in the development of the detector. An early prototype based on a spherical LEP RF cavity was moved underground at LSM~\cite{Piquemal:2012fs} to measure the flux of neutrons in the experimental cavern~\cite{Savvidis:2010zz}. The detector was equipped with a small, $15\;\si{\milli\meter}$ in diameter, spherical anode made of stainless steel. The  \ce{Ar + 2\% CH4} gas mixture at $275\;\si{\milli\bar}$ was enhanced with $3\;\si{\gram}$ of \ce{^3He}. The latter offers unique sensitivity for neutron detection thanks to the large thermal and fast -- up to several \MeV\ -- neutron capture cross-section, as shown in Fig.~\ref{fig:neutronCrossSection}. The proton and tritium produced through the $\left(n,p\right)$ reaction deposit their energy in the drift volume offering also energy information on the incident neutron. The low background of the detector and the pulse shape discrimination to separate the $\gamma$-ray induced pulses from the nuclear recoil induced pulses increases the sensitivity on neutron detection.
The measured thermal neutron capture rate was $4.8\cdot 10^{-3}$ counts per second, which yielded a thermal neutron flux of $1.9\cdot 10^{-5}\si{\per\square\centi\meter\per\second}$.

\begin{wrapfigure}{R}{0.35\linewidth}
  \centering
  \vspace{-0.50cm}
\includegraphics[width=0.99\linewidth]{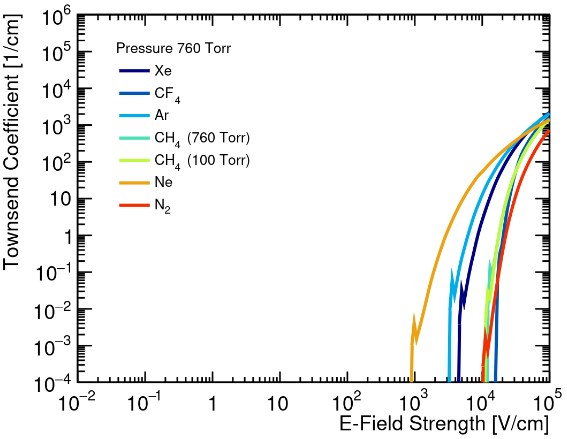}
  \vspace{-0.3cm}
\caption{Townsend coefficient for various gases calculated by \textsc{Magboltz}.\label{fig:townsend_nitrogen}}
\vspace{-0.5cm}
\end{wrapfigure}
Subsequently, operation of the spherical proportional counter with
nitrogen was pursued~\cite{Bougamont:2015jzx}. Nitrogen presents an
interesting opportunity for neutron spectroscopy up to tens of MeV. The
thermal and fast neutrons can be detected via the
\ce{^{14}N + n \to ^{14}C + p + 625.87\;\keV}
reaction. For thermal neutrons the energy of the exothermic reaction
is shared between the \ce{^{14}C} and the proton as kinetic energy of
$41.72\;\keV$ and $584.15\;\keV$, respectively.  Fast neutrons can also be detected by the
\ce{^{14}N + n \to ^{11}B + \alpha  -158\;\keV}
reaction. The cross-section for these reactions is also shown in
Fig.~\ref{fig:neutronCrossSection}.
On the experimental side, the anode diameter was reduced to $8$ or
$3\;\si{\milli\meter}$ and silicon was chosen as the material, to
suppress discharges. These improvements were necessary
to achieve proportional amplification in nitrogen, a process occurring
at extremely high electric fields as shown in
Fig.~\ref{fig:townsend_nitrogen}. On this first attempt, nitrogen
pressures of up to $500\;\si{\milli\bar}$ were achieved. This required an anode voltage of $6.2\;\si{\kilo\volt}$ to obtain adequate signal amplification. Further increase of the anode voltage was prohibited by the sensor technology at the time, and thus it was not possible to explore
higher-pressure operation, making the wall effect a limiting factor to the fast neutron energy reconstruction.

\begin{figure}
\centering
\subfigure[\label{fig:neutron_thermal1p8}]{\includegraphics[width=0.28\linewidth]{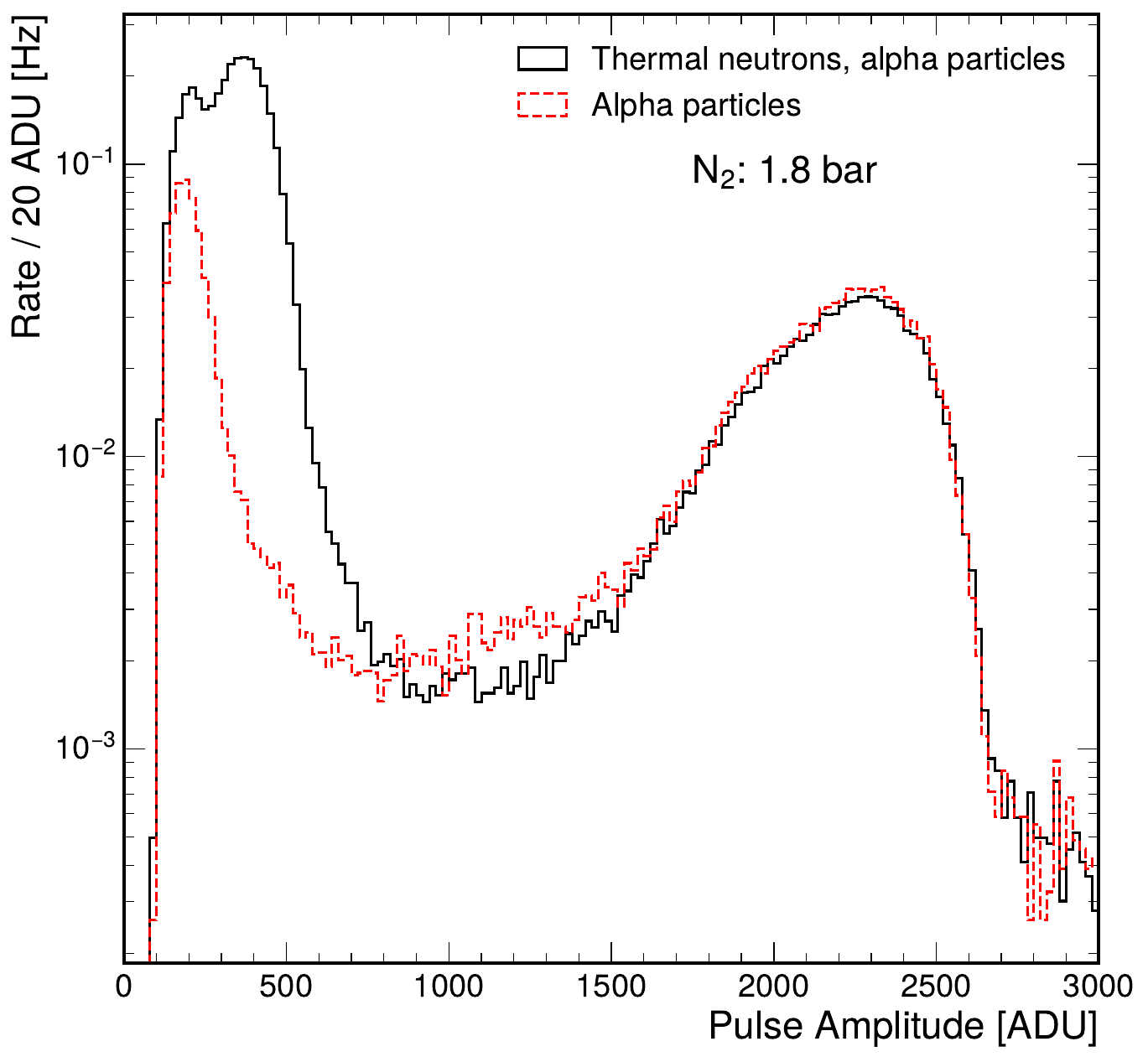}}
\subfigure[\label{fig:neutron_fast1p5}]{\includegraphics[width=0.28\linewidth]{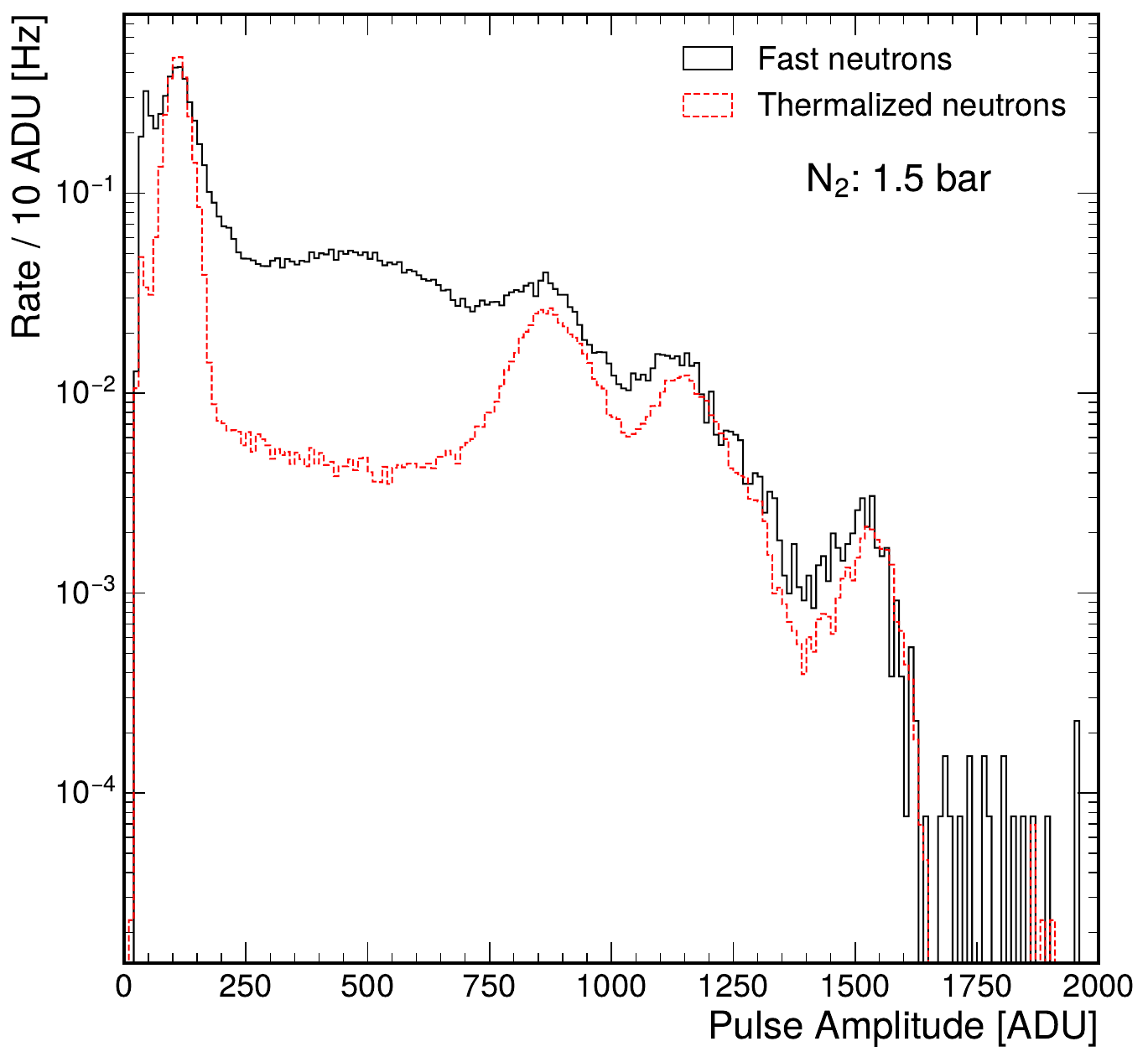}}
  \vspace{-0.45cm}
\caption{\subref{fig:neutron_thermal1p8} Amplitude distribution with (without) thermal neutrons shown as solid (dashed) line with the spherical proportional counter operating with
1.8 bar nitrogen and 5.95 kV anode voltage.
The $\alpha$-particles from \ce{^{210}Po}  are observed as a peak around 2175 ADU. 
\subref{fig:neutron_fast1p5} Amplitude distribution for neutrons from an $^{241}$Am-$^{9}$Be source without (with) neutron thermalisation shown as solid (dashed) line, with the spherical proportional counter  with 1.5 bar nitrogen at $4.5\;\si{\kilo\volt}$ anode voltage. The three rightmost peaks correspond to \ce{^{222}Rn} decay chain. From Ref.~\cite{Giomataris:2022bvz}.}
\label{fig:neutronSpectra}
\vspace{-0.6cm}
\end{figure}

Thanks to the advent of the ACHINOS, previous limitation which prevented the further development of nitrogen-filled spherical proportional counters were overcome. The decoupled drift and avalanche fields allowed for increased anode voltage without compromising detector stability, thus, enabling sufficient gain for operation with higher pressures with nitrogen. A pressure of $1.8\;\si{bar}$ was achieved while maintaining the operating voltage below $6\;\si{\kilo\volt}$ thanks to the use of a DLC-coated ACHINOS with eleven $1\;\si{\milli\meter}$ diameter anodes~\cite{Giomataris:2022bvz}. A $30\;\si{\centi\meter}$ in diameter spherical proportional counter, instrumented with this ACHINOS, was irradiated with neutrons from an $^{241}$Am-$^9$Be source.  In addition to the fast neutrons directly from the source, thermal neutrons were studied by placing the source inside a graphite stack. The results are presented in Fig.~\ref{fig:neutronSpectra}, with Fig.~\ref{fig:neutron_thermal1p8} showing the recorded amplitude distribution with and without the neutron source present inside the graphite stack, and Fig.~\ref{fig:neutron_fast1p5} demonstrating the detection of fast neutrons. In both cases, peaks from the calibration sources -- $^{210}$Po and $^{222}$Rn -- are visible.

\begin{figure}[h]
\centering
\vspace{-0.5cm}
\subfigure[\label{fig:fig12a}]{\includegraphics[width=0.30\linewidth]{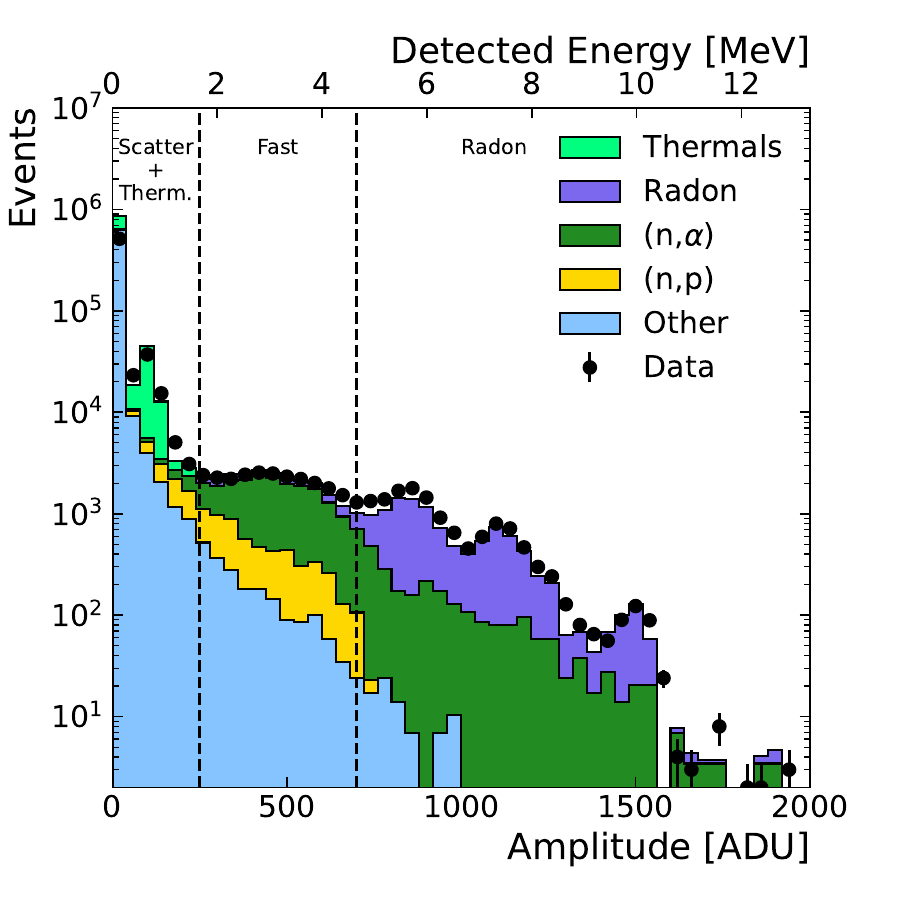}}
\subfigure[\label{fig:fig12b}]{\includegraphics[width=0.30\linewidth]{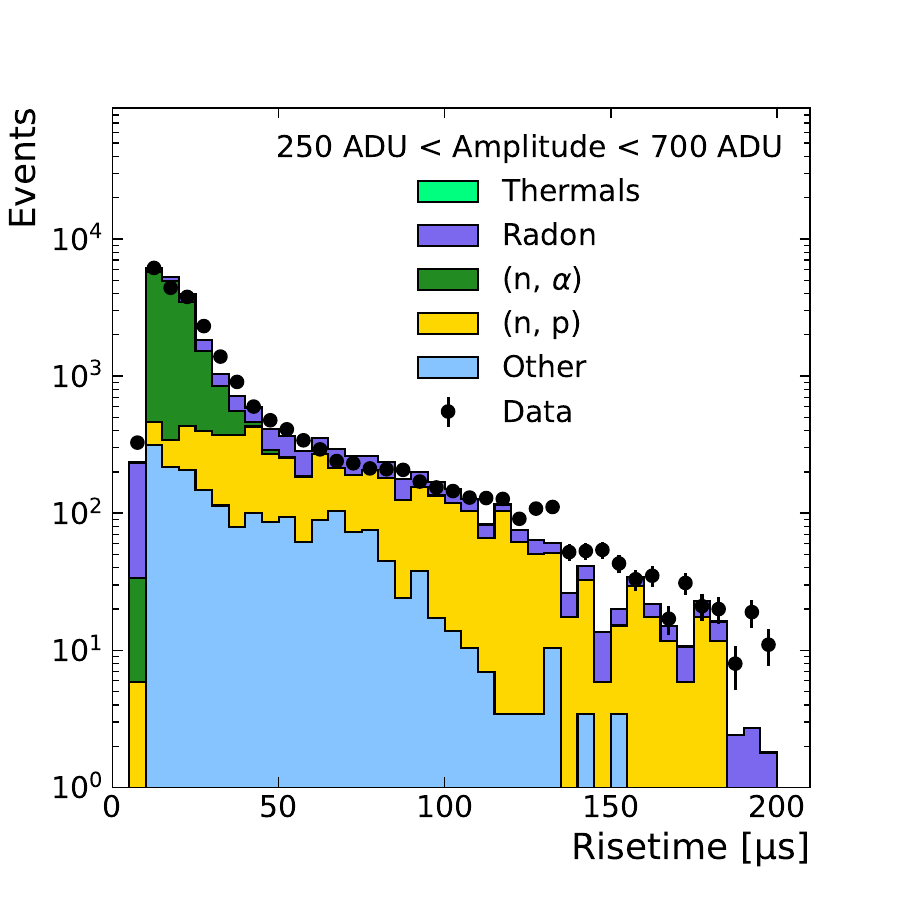}}
\subfigure[\label{fig:fig13}]{\includegraphics[width=0.30\linewidth]{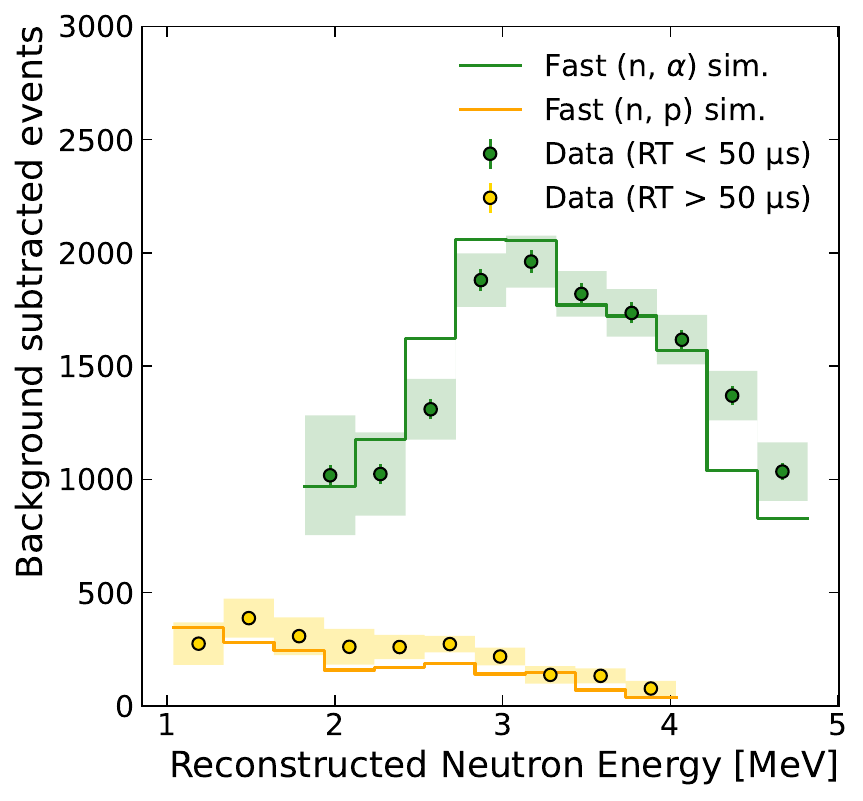}}
\vspace{-0.3cm}
\caption{\subref{fig:fig12a} Fast neutrons amplitude distribution from $^{241}$Am-$^{9}$Be compared to simulations.  \subref{fig:fig12b} Pulse risetime distribution for events in \subref{fig:fig12a} categorised as "fast".
\subref{fig:fig13} Background subtracted data for the neutron reconstructed energy compared simulation. Events are attributed to the $\left(n,p\right)$ and $\left(n,\alpha\right)$ reactions based on their risetime.From Ref.~\cite{Giomataris:2022bvz}.
\label{fig:fig12}}
\vspace{-0.3cm}
\end{figure}

The measured detector response has also been compared to predictions from the simulation framework. Figure~\ref{fig:fig12a} shows the amplitude distribution measured for fast neutrons when the detector was filled with $1.5\;\si{bar}$ nitrogen compared to the simulated signal and background contributions. In addition to the signals measured from the (n,p) and (n,$\alpha$) reactions and the $^{222}$Rn present in the gas, the dominant background (denoted as `other') comes from a combination of photons produced from neutron capture on the detector vessel and neutron elastic scattering in the gas. 
A scaling factor of 1.7 was applied to the (n,p) reaction cross section in order to match the observed data. The simulation was used to select the region dominated by fast neutron interactions, between $250\;\si{ADU}$ (analogue-to-digital units, which is proportional to recorded energy) and $700\;\si{ADU}$ for subsequent investigation. Figure~\ref{fig:fig12b} shows the pulse risetime for events in that region, again compared to that of the simulations signal and background contributions. Thanks to the predictive power of the simulation, the risetime could be used to differentiate the $\left(n,p\right)$- and $\left(n,\alpha\right)$-induced events, allowing for the difference in the reaction Q-value to be applied to the energy reconstruction. 
Figure~\ref{fig:fig13} compares the measured and simulated energy distribution following this correction, demonstrating the power of the simulation framework. 

\begin{wrapfigure}{R}{0.32\linewidth}
  \centering
  \vspace{-0.40cm}
\includegraphics[width=0.99\linewidth]{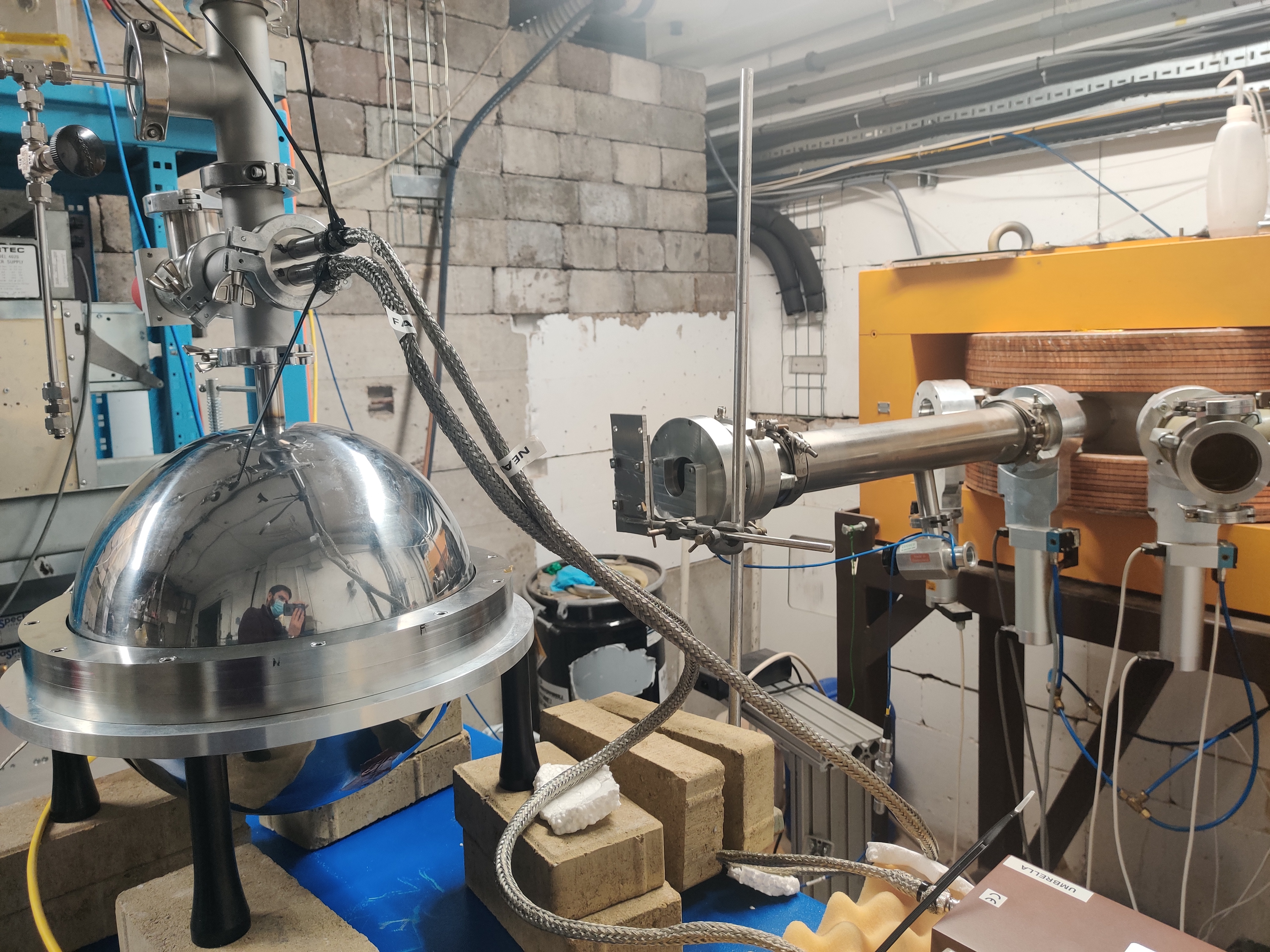}
  \vspace{-0.65cm}
\caption{Spherical proportional counter at the MC40 cyclotron.\label{fig:cyclotronSetup}}
\vspace{-1.25cm}
\end{wrapfigure}

The same detector was employed at the MC40 cyclotron at the University of Birmingham~\cite{Giomataris:2022kxw}. 
Neutrons with energies up to $8\;\si{\mega\eV}$ were produced through the $^9$Be(d,n)$^{10}$B reaction. The experimental set-up is shown in Fig.~\ref{fig:cyclotronSetup}.

Thanks to the ongoing development of the spherical proportional counter for direct dark matter searches, the potential to utilise the detector for neutron spectroscopy has been realised. This will not only provide an invaluable tool for in situ neutron background measurements in NEWS-G, but also opens the possibility of industrial applications of the detector. Continued development for the detector to be operated at even greater pressures with nitrogen are in hand, with the remaining challenges to the full realisation of the detector to such applications coming from the need to simplify and package the detector read-out and acquisition.

\section{Summary}
\label{sec:summary}

Over a period of 20 years the spherical proportional counter has been transformed from an idea to a tool for the exploration of the particle nature of the Dark Matter in the universe. The progress achieved in the instrumentation of the detector enabled its use for the search of light Dark Matter candidates, in a previously unconstrained region of the parameter space. Great strides have been made towards the efficient and stable operation of large-size, high-pressure, spherical proportional counters, and  the methodology for constructing extremely radiopure detectors is now demonstrated and can be readily scaled-up in terms of detector volume. The signal read-out has been revolutionised and three-dimensional detector fiducialisation is now achieved. Reliable and predictive simulation tools are established, while the understanding of the microphysics in the gas mixtures of relevance is significantly improved. The potential of spherical proportional counters in applications beyond fundamental research is also established.

Looking to the future, instrumentation and data analysis techniques will continue to be improved for spherical proportional counter exploitation in the most challenging industrial applications, such as neutron spectroscopy. Nevertheless, the technology and expertise for the construction of large-scale, high-radiopurity, spherical proportional counters to explore the sub-GeV light Dark Matter landscape down to the neutrino fog in the coming years is already available.

\section*{Acknowledgements}
The majority of developments on the spherical proportional counter have been achieved in the context of international scientific collaborations employing this novel detector for fundamental science, including NEWS\=/G and R2D2. It was through this collective effort that this significant progress was made. The creativity, inventiveness, and leadership of Dr.~I.~Giomataris and Prof.~G.~Gerbier have been key throughout the years. 
The collaboration with world-leading deep underground science facilities --  Laboratoire Souterraine de Modane, SNOLAB, and Boulby Underground Laboratory -- provided the ideal environment for our work. 

This research has been supported by a broad mixture of funding sources world-wide, including  the Canada Research Chairs program, the Canada Foundation for Innovation, the Arthur B. McDonald Canadian Astroparticle Physics Research Institute, the Natural Sciences and Engineering Research Council of Canada, the French National Research Agency (ANR-15-CE31-0008), the European Union’s Horizon 2020 research and innovation programme (Marie Skłodowska-Curie Grant Agreements 
841261-DarkSPHERE, 895168-neutronSPHERE, and 101026519-GaGARin). Support from the UK
Research and Innovation—Science and Technology Facilities Council (UKRI-STFC)  (Grants ST/S000860/1, ST/V006339/1, ST/W005611/1, ST/W000652/1, ST/X005976/1), the UKRI Horizon Europe Underwriting scheme (GA101066657/Je-S EP/X022773/1) and the the Royal Society International Exchanges Scheme (IES\textbackslash R3\textbackslash 170121) is acknowledged.

Every effort has been made to include all relevant developments and results, framed in the broader context of the research field. We recognise that in an effort to keep this review at a reasonable length, we may have not been able to do justice to all the exciting outcomes obtained over more than 20 years.

PK and KN acknowledge support by UKRI-STFC (ST/X005976/1).
KN acknowledges support by the Deutsche Forschungsgemeinschaft (DFG, German Research Foundation) under Germany's Excellence Strategy – EXC 2121 "Quantum Universe“ – 390833306.

\bibliographystyle{hieeetr}
\bibliography{bibliography.bib}

\begin{thebibliography}{100}

\bibitem{Bertone:2010zza}
J.~Silk {\em et~al.}, {\em Particle Dark Matter: Observations, Models and
  Searches}.
\newblock Cambridge: Cambridge Univ. Press, 2010.

\bibitem{Clowe:2006eq}
D.~Clowe {\em et~al.}, ``{A direct empirical proof of the existence of dark
  matter},'' {\em Astrophys. J. Lett.}, vol.~648, p.~L109, 2006.

\bibitem{Planck2015}
{Planck Collaboration}, ``Planck 2015 results - xiii. cosmological
  parameters,'' {\em Astron. Astrophys.}, vol.~594, p.~A13, 2016.

\bibitem{Battaglieri:2017aum}
M.~Battaglieri {\em et~al.}, ``{US Cosmic Visions: New Ideas in Dark Matter
  2017: Community Report},'' 2017, 1707.04591.

\bibitem{Billard:2021uyg}
J.~Billard {\em et~al.}, ``{Direct detection of dark matter\textemdash{}APPEC
  committee report*},'' {\em Rept. Prog. Phys.}, vol.~85, p.~056201, 2022.

\bibitem{Liu:2017drf}
J.~Liu, X.~Chen, and X.~Ji, ``{Current status of direct dark matter detection
  experiments},'' {\em Nature Phys.}, vol.~13, p.~212, 2017.

\bibitem{Goodman:1984dc}
M.~W. Goodman and E.~Witten, ``{Detectability of Certain Dark Matter
  Candidates},'' {\em Phys. Rev. D}, vol.~31, p.~3059, 1985.

\bibitem{Evans:2018bqy}
N.~W. Evans, C.~A.~J. O'Hare, and C.~McCabe, ``{Refinement of the standard halo
  model for dark matter searches in light of the Gaia Sausage},'' {\em Phys.
  Rev. D}, vol.~99, p.~023012, 2019.

\bibitem{MarrodanUndagoitia:2015veg}
T.~Marrod\'an~Undagoitia and L.~Rauch, ``{Dark matter direct-detection
  experiments},'' {\em J. Phys. G}, vol.~43, p.~013001, 2016.

\bibitem{Schumann:2019eaa}
M.~Schumann, ``{Direct Detection of WIMP Dark Matter: Concepts and Status},''
  {\em J. Phys. G}, vol.~46, p.~103003, 2019.

\bibitem{Ianni:2017vqi}
A.~Ianni, ``{Review of technical features in underground laboratories},'' {\em
  Int. J. Mod. Phys. A}, vol.~32, p.~1743001, 2017.

\bibitem{LZ:2022lsv}
J.~Aalbers {\em et~al.}, ``{First Dark Matter Search Results from the
  LUX-ZEPLIN (LZ) Experiment},'' {\em Phys. Rev. Lett.}, vol.~131, p.~041002,
  2023.

\bibitem{PandaX-4T:2021bab}
Y.~Meng {\em et~al.}, ``{Dark Matter Search Results from the PandaX-4T
  Commissioning Run},'' {\em Phys. Rev. Lett.}, vol.~127, p.~261802, 2021.

\bibitem{XENON:2018voc}
E.~Aprile {\em et~al.}, ``{Dark Matter Search Results from a One Ton-Year
  Exposure of XENON1T},'' {\em Phys. Rev. Lett.}, vol.~121, p.~111302, 2018.

\bibitem{PandaX-II:2017hlx}
X.~Cui {\em et~al.}, ``{Dark Matter Results From 54-Ton-Day Exposure of
  PandaX-II Experiment},'' {\em Phys. Rev. Lett.}, vol.~119, p.~181302, 2017.

\bibitem{LUX:2016ggv}
D.~S. Akerib {\em et~al.}, ``{Results from a search for dark matter in the
  complete LUX exposure},'' {\em Phys. Rev. Lett.}, vol.~118, p.~021303, 2017.

\bibitem{DEAP:2019yzn}
R.~Ajaj {\em et~al.}, ``{Search for dark matter with a 231-day exposure of
  liquid argon using DEAP-3600 at SNOLAB},'' {\em Phys. Rev. D}, vol.~100,
  p.~022004, 2019.

\bibitem{Agnes:2018ves}
P.~Agnes {\em et~al.}, ``{Low-Mass Dark Matter Search with the DarkSide-50
  Experiment},'' {\em Phys. Rev. Lett.}, vol.~121, p.~081307, 2018.

\bibitem{Migdal1941}
A.~B. Migdal, ``Ionization of atoms accompanying $\alpha$-and $\beta$-decay,''
  {\em J. Phys. USSR}, vol.~4, p.~449, 1941.

\bibitem{Bernabei:2007jz}
R.~Bernabei {\em et~al.}, ``{On electromagnetic contributions in WIMP
  quests},'' {\em Int. J. Mod. Phys. A}, vol.~22, p.~3155, 2007.

\bibitem{Ibe:2017yqa}
M.~Ibe, W.~Nakano, Y.~Shoji, and K.~Suzuki, ``{Migdal Effect in Dark Matter
  Direct Detection Experiments},'' {\em J. High Energy Phys.}, vol.~03, p.~194,
  2018.

\bibitem{Dolan:2017xbu}
M.~J. Dolan, F.~Kahlhoefer, and C.~McCabe, ``{Directly detecting sub-GeV dark
  matter with electrons from nuclear scattering},'' {\em Phys. Rev. Lett.},
  vol.~121, p.~101801, 2018.

\bibitem{Kouvaris:2016afs}
C.~Kouvaris and J.~Pradler, ``{Probing sub-GeV Dark Matter with conventional
  detectors},'' {\em Phys. Rev. Lett.}, vol.~118, p.~031803, 2017.

\bibitem{Aprile:2019jmx}
E.~Aprile {\em et~al.}, ``{Search for Light Dark Matter Interactions Enhanced
  by the Migdal Effect or Bremsstrahlung in XENON1T},'' {\em Phys. Rev. Lett.},
  vol.~123, p.~241803, 2019.

\bibitem{LUX:2018akb}
D.~S. Akerib {\em et~al.}, ``{Results of a Search for Sub-GeV Dark Matter Using
  2013 LUX Data},'' {\em Phys. Rev. Lett.}, vol.~122, p.~131301, 2019.

\bibitem{LZ:2023poo}
J.~Aalbers {\em et~al.}, ``{Search for new physics in low-energy electron
  recoils from the first LZ exposure},'' {\em Phys. Rev. D}, vol.~108,
  p.~072006, 2023.

\bibitem{Xu:2023wev}
J.~Xu {\em et~al.}, ``{Search for the Migdal effect in liquid xenon with
  keV-level nuclear recoils},'' {\em Phys. Rev. D}, vol.~109, p.~L051101, 2024.

\bibitem{Araujo:2022wjh}
H.~M. Ara\'ujo {\em et~al.}, ``{The MIGDAL experiment: Measuring a rare atomic
  process to aid the search for dark matter},'' {\em Astropart. Phys.},
  vol.~151, p.~102853, 2023.

\bibitem{Adams:2022zvg}
D.~Adams {\em et~al.}, ``{Measuring the Migdal effect in semiconductors for
  dark matter detection},'' {\em Phys. Rev. D}, vol.~107, p.~L041303, 2023.

\bibitem{Nakamura:2020kex}
K.~D. Nakamura {\em et~al.}, ``{Detection capability of the Migdal effect for
  argon and xenon nuclei with position-sensitive gaseous detectors},'' {\em
  Prog. Theor. Exp. Phys.}, vol.~2021, p.~013C01, 2021.

\bibitem{Bell:2021ihi}
N.~F. Bell {\em et~al.}, ``{Observing the Migdal effect from nuclear recoils of
  neutral particles with liquid xenon and argon detectors},'' {\em Phys. Rev.
  D}, vol.~105, p.~096015, 2022.

\bibitem{EDELWEISS:2019vjv}
{EDELWEISS Collaboration}, ``{Searching for low-mass dark matter particles with
  a massive Ge bolometer operated above-ground},'' {\em Phys. Rev. D}, vol.~99,
  p.~082003, 2019.

\bibitem{CDEX:2019hzn}
{CDEX Collaboration}, ``{Constraints on Spin-Independent Nucleus Scattering
  with sub-GeV Weakly Interacting Massive Particle Dark Matter from the CDEX-1B
  Experiment at the China Jinping Underground Laboratory},'' {\em Phys. Rev.
  Lett.}, vol.~123, p.~161301, 2019.

\bibitem{XENON:2019zpr}
E.~Aprile {\em et~al.}, ``{Search for Light Dark Matter Interactions Enhanced
  by the Migdal Effect or Bremsstrahlung in XENON1T},'' {\em Phys. Rev. Lett.},
  vol.~123, p.~241803, 2019.

\bibitem{SENSEI:2020dpa}
{SENSEI Collaboration}, ``{SENSEI: Direct-Detection Results on sub-GeV Dark
  Matter from a New Skipper-CCD},'' {\em Phys. Rev. Lett.}, vol.~125,
  p.~171802, 2020.

\bibitem{CDEX:2021cll}
Z.~Z. Liu {\em et~al.}, ``{Studies of the Earth shielding effect to direct dark
  matter searches at the China Jinping Underground Laboratory},'' {\em Phys.
  Rev. D}, vol.~105, p.~052005, 2022.

\bibitem{EDELWEISS:2022ktt}
E.~Armengaud {\em et~al.}, ``{Search for sub-GeV dark matter via the Migdal
  effect with an EDELWEISS germanium detector with NbSi transition-edge
  sensors},'' {\em Phys. Rev. D}, vol.~106, p.~062004, 2022.

\bibitem{Bell:2023uvf}
N.~F. Bell {\em et~al.}, ``{Exploring light dark matter with the Migdal effect
  in hydrogen-doped liquid xenon},'' {\em Phys. Rev. D}, vol.~109, p.~L091902,
  2024.

\bibitem{Angloher:2015ewa}
G.~Angloher {\em et~al.}, ``{Results on light dark matter particles with a
  low-threshold CRESST-II detector},'' {\em Eur. Phys. J.}, vol.~C76, p.~25,
  2016.

\bibitem{Abdelhameed:2019hmk}
A.~Abdelhameed {\em et~al.}, ``{First results from the CRESST-III low-mass dark
  matter program},'' {\em Phys. Rev. D}, vol.~100, p.~102002, 2019.

\bibitem{CRESST:2022jig}
G.~Angloher {\em et~al.}, ``{Probing spin-dependent dark matter interactions
  with $^6$Li: CRESST Collaboration},'' {\em Eur. Phys. J. C}, vol.~82, p.~207,
  2022.

\bibitem{Agnese:2018gze}
{SuperCDMS Collaboration}, ``{Search for Low-Mass Dark Matter with CDMSlite
  Using a Profile Likelihood Fit},'' {\em Phys. Rev. D}, vol.~99, p.~062001,
  2019.

\bibitem{Agnese:2016cpb}
{SuperCDMS Collaboration}, ``{Projected Sensitivity of the SuperCDMS SNOLAB
  experiment},'' {\em Phys. Rev. D}, vol.~95, p.~082002, 2017.

\bibitem{Lindhard1963}
J.~Lindhard, V.~Nielsen, M.~Scharff, and P.~Thomsen, ``Integral equations
  governing radiation effects,'' {\em Kong.~Dan.~Vid.~Sel.~Mat.~Fys.~Med.},
  vol.~33, 1963.

\bibitem{Essig:2011nj}
R.~Essig, J.~Mardon, and T.~Volansky, ``{Direct Detection of Sub-GeV Dark
  Matter},'' {\em Phys. Rev. D}, vol.~85, p.~076007, 2012.

\bibitem{Cowan:1956rrn}
C.~L. Cowan {\em et~al.}, ``{Detection of the free neutrino: A Confirmation},''
  {\em Science}, vol.~124, p.~103, 1956.

\bibitem{Davis:1964hf}
R.~Davis, ``{Solar neutrinos. II: Experimental},'' {\em Phys. Rev. Lett.},
  vol.~12, p.~303, 1964.

\bibitem{Davis:1968cp}
R.~Davis, Jr., D.~S. Harmer, and K.~C. Hoffman, ``{Search for neutrinos from
  the sun},'' {\em Phys. Rev. Lett.}, vol.~20, p.~1205, 1968.

\bibitem{Trimble:1973ca}
V.~Trimble and F.~Reines, ``{The solar neutrino problem - a progress(?)
  report},'' {\em Rev. Mod. Phys.}, vol.~45, p.~1, 1973.

\bibitem{Haxton:1995hv}
W.~C. Haxton, ``{The solar neutrino problem},'' {\em Ann. Rev. Astron.
  Astrophys.}, vol.~33, p.~459, 1995.

\bibitem{Nygren:1974nfi}
D.~R. Nygren, ``{The Time Projection Chamber: A New 4 pi Detector for Charged
  Particles},'' {\em eConf}, vol.~C740805, p.~58, 1974.

\bibitem{BONVICINI1994438}
G.~Bonvicini, ``A solar neutrino experiment with neutrino energy resolution,''
  {\em Nuclear Physics B - Proceedings Supplements}, vol.~35, p.~438, 1994.

\bibitem{Gorodetzky:1999ty}
P.~Gorodetzky {\em et~al.}, ``{Identification of solar neutrinos by individual
  electron counting in HELLAZ},'' {\em Nucl. Instrum. Meth. A}, vol.~433,
  p.~554, 1999.

\bibitem{Dolbeau:2005kt}
J.~Dolbeau {\em et~al.}, ``{The solar neutrino HELLAZ project},'' {\em Nucl.
  Phys. B Proc. Suppl.}, vol.~138, p.~94, 2005.

\bibitem{Super-Kamiokande:1998kpq}
Y.~Fukuda {\em et~al.}, ``{Evidence for oscillation of atmospheric
  neutrinos},'' {\em Phys. Rev. Lett.}, vol.~81, p.~1562, 1998.

\bibitem{SNO:2001kpb}
Q.~R. Ahmad {\em et~al.}, ``{Measurement of the rate of $\nu_e+d \to p+p+e^-$
  interactions produced by $^8$B solar neutrinos at the Sudbury Neutrino
  Observatory},'' {\em Phys. Rev. Lett.}, vol.~87, p.~071301, 2001.

\bibitem{Giomataris:2003bp}
Y.~Giomataris and J.~D. Vergados, ``{Neutrino properties studied with a triton
  source using large TPC detectors},'' {\em Nucl. Instrum. Meth. A}, vol.~530,
  p.~330, 2004.

\bibitem{Vergados:2009ei}
J.~D. Vergados, F.~T. Avignone, III, and I.~Giomataris, ``{Coherent Neutral
  Current Neutrino-Nucleus Scattering at a Spallation Source: A Valuable
  Experimental Probe},'' {\em Phys. Rev. D}, vol.~79, p.~113001, 2009.

\bibitem{Dedes:2009bk}
A.~Dedes, I.~Giomataris, K.~Suxho, and J.~D. Vergados, ``{Searching for
  Secluded Dark Matter via Direct Detection of Recoiling Nuclei as well as Low
  Energy Electrons},'' {\em Nucl. Phys. B}, vol.~826, p.~148, 2010.

\bibitem{Vergados:2011gia}
J.~D. Vergados, Y.~Giomataris, and Y.~N. Novikov, ``{Probing the fourth
  neutrino existence by neutral current oscillometry in the spherical gaseous
  TPC},'' {\em Nucl. Phys. B}, vol.~854, p.~54, 2012.

\bibitem{Vergados:2011na}
J.~D. Vergados, Y.~Giomataris, and Y.~N. Novikov, ``{Novel way to search for
  sterile neutrinos},'' {\em Phys. Rev. D}, vol.~85, p.~033003, 2012.

\bibitem{Giomataris:2005fx}
Y.~Giomataris and J.~D. Vergados, ``{A Network of neutral current spherical
  TPC's for dedicated supernova detection},'' {\em Phys. Lett. B}, vol.~634,
  p.~23, 2006.

\bibitem{Giomataris:1995fq}
Y.~Giomataris, P.~Rebourgeard, J.~P. Robert, and G.~Charpak, ``{MICROMEGAS: A
  High granularity position sensitive gaseous detector for high particle flux
  environments},'' {\em Nucl. Instrum. Meth. A}, vol.~376, p.~29, 1996.

\bibitem{Collar:2000jq}
J.~I. Collar and Y.~Giomataris, ``{Possible low-background applications of
  MICROMEGAS detector technology},'' {\em Nucl. Instrum. Meth. A}, vol.~471,
  p.~254, 2000.

\bibitem{Myers:1990sk}
S.~Myers and E.~Picasso, ``{The Design, construction and commissioning of the
  CERN Large Electron Positron collider},'' {\em Contemp. Phys.}, vol.~31,
  p.~387, 1990.

\bibitem{Giomataris:2005bb}
I.~Giomataris {\em et~al.}, ``{NOSTOS experiment and new trends in rare event
  detection},'' {\em Nucl. Phys. B Proc. Suppl.}, vol.~150, p.~208, 2006.

\bibitem{Aune:2004bc}
S.~Aune {\em et~al.}, ``{NOSTOS: A new low-energy neutrino experiment},'' in
  {\em {5th International Workshop on the Identification of Dark Matter}},
  p.~607, 2004.

\bibitem{Aune:2005is}
S.~Aune {\em et~al.}, ``{NOSTOS: A Spherical TPC to detect low energy
  neutrinos},'' {\em AIP Conf. Proc.}, vol.~785, p.~110, 2005.

\bibitem{Aune:2005hv}
S.~Aune {\em et~al.}, ``{Progress on a spherical TPC for low energy neutrino
  detection},'' {\em J. Phys. Conf. Ser.}, vol.~39, p.~281, 2006.

\bibitem{Giomataris:2008ap}
I.~Giomataris {\em et~al.}, ``{A Novel large-volume Spherical Detector with
  Proportional Amplification read-out},'' {\em J. Instrum.}, vol.~3, p.~P09007,
  2008.

\bibitem{Essig:2013lka}
R.~Essig {\em et~al.}, ``{Working Group Report: New Light Weakly Coupled
  Particles},'' 1311.0029.

\bibitem{Petraki:2013wwa}
K.~Petraki and R.~R. Volkas, ``{Review of asymmetric dark matter},'' {\em Int.
  J. Mod. Phys.}, vol.~A28, p.~1330028, 2013.

\bibitem{Zurek:2013wia}
K.~M. Zurek, ``{Asymmetric Dark Matter: Theories, Signatures, and
  Constraints},'' {\em Phys. Rept.}, vol.~537, p.~91, 2014.

\bibitem{Tulin:2017ara}
S.~Tulin and H.-B. Yu, ``{Dark Matter Self-interactions and Small Scale
  Structure},'' {\em Phys. Rept.}, vol.~730, p.~1, 2018.

\bibitem{Fitzpatrick:2012ix}
A.~L. Fitzpatrick {\em et~al.}, ``{The Effective Field Theory of Dark Matter
  Direct Detection},'' {\em J. Cosmol. Astropart. Phys.}, vol.~02, p.~004,
  2013.

\bibitem{Cerdeno:2018bty}
D.~G. Cerde\~{n}o, A.~Cheek, E.~Reid, and H.~Schulz, ``{Surrogate Models for
  Direct Dark Matter Detection},'' {\em J. Cosmol. Astropart. Phys.},
  vol.~1808, p.~011, 2018.

\bibitem{Gerbier:2014jwa}
G.~Gerbier {\em et~al.}, ``{NEWS : a new spherical gas detector for very low
  mass WIMP detection},'' 2014, 1401.7902.

\bibitem{Nikolopoulos:2020vma}
K.~Nikolopoulos, ``{Search for light dark matter with NEWS-G},'' {\em J.
  Instrum.}, vol.~15, p.~C06034, 2020.

\bibitem{Bougamont:2010mj}
E.~Bougamont {\em et~al.}, ``{Ultra low energy results and their impact to dark
  matter and low energy neutrino physics},'' {\em Journal of Modern Physics},
  vol.~3, p.~57, 2012.

\bibitem{Savvidis:2016wei}
I.~Savvidis {\em et~al.}, ``{Low energy recoil detection with a spherical
  proportional counter},'' {\em Nucl. Instrum. Meth. A}, vol.~877, p.~220,
  2018.

\bibitem{Katsioulas:2022cqe}
I.~Katsioulas {\em et~al.}, ``{ACHINOS: a multi-anode read-out for position
  reconstruction and tracking with spherical proportional counters},'' {\em J.
  Instrum.}, vol.~17, p.~C08025, 2022.

\bibitem{Balogh:2020nmo}
{NEWS-G Collaboration}, ``{Copper electroplating for background suppression in
  the NEWS-G experiment},'' {\em Nucl. Instrum. Meth. A}, vol.~988, p.~164844,
  2021.

\bibitem{Arnaud:2017bjh}
{NEWS-G Collaboration}, ``{First results from the NEWS-G direct dark matter
  search experiment at the LSM},'' {\em Astropart. Phys.}, vol.~97, p.~54,
  2018.

\bibitem{Knights:2019tmx}
P.~Knights, ``{Gas and copper purity investigations for NEWS-G},'' {\em J.
  Phys. Conf. Ser.}, vol.~1312, p.~012009, 2019.

\bibitem{NEWS-G:2023qwh}
{NEWS-G Collaboration}, ``{Exploring light dark matter with the DarkSPHERE
  spherical proportional counter electroformed underground at the Boulby
  Underground Laboratory},'' {\em Phys. Rev. D}, vol.~108, p.~112006, 2023.

\bibitem{GEANT4:2002zbu}
S.~Agostinelli {\em et~al.}, ``{GEANT4--a simulation toolkit},'' {\em Nucl.
  Instrum. Meth. A}, vol.~506, p.~250, 2003.

\bibitem{Allison:2016lfl}
J.~Allison {\em et~al.}, ``{Recent developments in Geant4},'' {\em Nucl.
  Instrum. Meth. A}, vol.~835, p.~186, 2016.

\bibitem{Bouclier:1996im}
R.~Bouclier {\em et~al.}, ``{The Gas electron multiplier (GEM)},'' {\em IEEE
  Transactions on Nuclear Science}, vol.~44, pp.~646--650, 1997.

\bibitem{Katsioulas:2019sui}
I.~Katsioulas {\em et~al.}, ``{Development of a simulation framework for
  spherical proportional counters},'' {\em J. Instrum.}, vol.~15, p.~C06013,
  2020.

\bibitem{Veenhof:1993hz}
R.~Veenhof, ``{Garfield, a drift chamber simulation program},'' {\em Conf.
  Proc. C}, vol.~9306149, p.~66, 1993.

\bibitem{Veenhof:1998tt}
R.~Veenhof, ``{GARFIELD, recent developments},'' {\em Nucl. Instrum. Methods
  Phys. Res. A}, vol.~419, p.~726, 1998.

\bibitem{garfield}
H.~Schindler, ``Garfield++ user guide.''
  \url{https://garfieldpp.web.cern.ch/garfieldpp/documentation/UserGuide.pdf},
  2020.

\bibitem{Smirnov:2005yi}
I.~B. Smirnov, ``{Modeling of ionization produced by fast charged particles in
  gases},'' {\em Nucl. Instrum. Meth. A}, vol.~554, p.~474, 2005.

\bibitem{Biagi:1999nwa}
S.~F. Biagi, ``{Monte Carlo simulation of electron drift and diffusion in
  counting gases under the influence of electric and magnetic fields},'' {\em
  Nucl. Instrum. Meth. A}, vol.~421, p.~234, 1999.

\bibitem{gmsh}
C.~Geuzaine and J.-F. Remacle, ``Gmsh: A 3-d finite element mesh generator with
  built-in pre- and post-processing facilities,'' {\em Int. J. Numer. Methods
  Eng.}, vol.~79, p.~1309, 2009.

\bibitem{Nikolopoulos:2011zza}
K.~Nikolopoulos, P.~Bhattacharya, V.~Chernyatin, and R.~Veenhof, ``{Electron
  transparency of a micromegas mesh},'' {\em J. Instrum.}, vol.~6, p.~P06011,
  2011.

\bibitem{Schindler:2012wta}
H.~Schindler, {\em {Microscopic Simulation of Particle Detectors}}.
\newblock PhD thesis, Vienna, Tech. U., Atominst., 2012.

\bibitem{Pfeiffer:2018yam}
D.~Pfeiffer {\em et~al.}, ``{Interfacing Geant4, Garfield++ and Degrad for the
  Simulation of Gaseous Detectors},'' {\em Nucl. Instrum. Meth. A}, vol.~935,
  p.~121, 2019.

\bibitem{Herd:2023hmu}
D.~Herd {\em et~al.}, ``{First operation of an ACHINOS-equipped spherical
  proportional counter with individual anode read-out},'' {\em J. Instrum.},
  vol.~19, p.~P01018, 2024.

\bibitem{Bouet:2020lbp}
R.~Bouet {\em et~al.}, ``{R2D2 spherical TPC: first energy resolution
  results},'' {\em J. Instrum.}, vol.~16, p.~P03012, 2021.

\bibitem{Giomataris:2022bvz}
I.~Giomataris {\em et~al.}, ``{Neutron spectroscopy with a high-pressure
  nitrogen-filled spherical proportional counter},'' {\em Nucl. Instrum. Meth.
  A}, vol.~1049, p.~168124, 2023.

\bibitem{Amoroso:2020llb}
A.~Amoroso {\em et~al.}, ``{PARSIFAL: A toolkit for triple-GEM parametrized
  simulation},'' {\em Comput. Phys. Commun.}, vol.~295, p.~109000, 2024.

\bibitem{Quemener:2021bzz}
G.~Qu\'em\'ener and S.~Salvador, ``{OuroborosBEM: a fast multi-GPU microscopic
  Monte Carlo simulation for gaseous detectors and charged particle
  dynamics},'' {\em J. Instrum.}, vol.~17, p.~P01020, 2022.

\bibitem{neep-gpu}
T.~Neep, K.~Nikolopoulos, and M.~Slater, ``Gpu acceleration of garfield++,''
  {\em in preparation}.

\bibitem{Katsioulas:2018pyh}
I.~Katsioulas {\em et~al.}, ``{A sparkless resistive glass correction electrode
  for the spherical proportional counter},'' {\em J. Instrum.}, vol.~13,
  p.~P11006, 2018.

\bibitem{Savvidis:2010zz}
I.~Savvidis {\em et~al.}, ``{Underground low flux neutron background
  measurements in LSM using a large volume (1m**3) spherical proportional
  counter},'' {\em J. Phys. Conf. Ser.}, vol.~203, p.~012030, 2010.

\bibitem{giomatarisrd512020}
I.~Giomataris, ``The spherical proportional gaseous detector and comparisons,''
  in {\em RD51 Collaboration Meeting and the topical workshop on “New
  Horizons in TPCs”}, 2020.

\bibitem{Giganon:2017isb}
A.~Giganon {\em et~al.}, ``{A multiball read-out for the spherical proportional
  counter},'' {\em J. Instrum.}, vol.~12, p.~P12031, 2017.

\bibitem{Giomataris:2020rna}
I.~Giomataris {\em et~al.}, ``{A resistive ACHINOS multi-anode structure with
  DLC coating for spherical proportional counters},'' {\em J. Instrum.},
  vol.~15, p.~11, 2020.

\bibitem{shockley}
W.~Shockley, ``Currents to conductors induced by a moving point charge,'' {\em
  J. Appl. Phys.}, vol.~9, p.~635, 1938.

\bibitem{ramo}
S.~{Ramo}, ``Currents induced by electron motion,'' {\em Proceedings of the
  IRE}, vol.~27, pp.~584--585, 1939.

\bibitem{Meregaglia:2017nhx}
A.~Meregaglia {\em et~al.}, ``{Study of a spherical Xenon gas TPC for
  neutrinoless double beta detection},'' {\em J. Instrum.}, vol.~13, p.~P01009,
  2018.

\bibitem{Katsioulas:2021usd}
{R2D2 Collaboration}, ``{Status of the R2D2 project: A future neutrinoless
  double beta decay experiment},'' {\em J. Phys. Conf. Ser.}, vol.~2105,
  p.~012016, 2021.

\bibitem{Bouet:2022kav}
R.~Bouet {\em et~al.}, ``{Simultaneous scintillation light and charge readout
  of a pure argon filled Spherical Proportional Counter},'' {\em Nucl. Instrum.
  Meth. A}, vol.~1028, p.~166382, 2022.

\bibitem{Katsioulas:2021pux}
I.~Katsioulas, P.~Knights, and K.~Nikolopoulos, ``{Ionisation quenching factors
  from W-values in pure gases for rare event searches},'' {\em Astropart.
  Phys.}, vol.~141, p.~102707, 2022.

\bibitem{Ahlen:1987mn}
S.~P. Ahlen {\em et~al.}, ``{Limits on Cold Dark Matter Candidates from an
  Ultralow Background Germanium Spectrometer},'' {\em Phys. Lett. B}, vol.~195,
  p.~603, 1987.

\bibitem{Aprile:2018dbl}
E.~Aprile {\em et~al.}, ``{Dark Matter Search Results from a One Ton-Year
  Exposure of XENON1T},'' {\em Phys. Rev. Lett.}, vol.~121, p.~111302, 2018.

\bibitem{Akerib:2016vxi}
D.~Akerib {\em et~al.}, ``{Results from a search for dark matter in the
  complete LUX exposure},'' {\em Phys. Rev. Lett.}, vol.~118, p.~021303, 2017.

\bibitem{lux2016}
M.~Carmona-Benitez {\em et~al.}, ``First results of the lux dark matter
  experiment,'' {\em Nuclear and Particle Physics Proceedings}, vol.~273-275,
  pp.~309--313, 2016.
\newblock 37th International Conference on High Energy Physics (ICHEP).

\bibitem{zeplin2012}
D.~Akimov, ``Wimp-nucleon cross-section results from the second science run of
  zeplin-iii,'' {\em Phys. Lett. B}, vol.~709, p.~14, 2012.

\bibitem{Joo:2018hom}
H.~Joo {\em et~al.}, ``{Quenching factor measurement for NaI(Tl) scintillation
  crystal},'' {\em Astropart. Phys.}, vol.~108, pp.~50--56, 2019.

\bibitem{xenon_qf}
E.~Aprile {\em et~al.}, ``Scintillation response of liquid xenon to low energy
  nuclear recoils,'' {\em Phys. Rev. D}, vol.~72, p.~072006, 2005.

\bibitem{silicon_qf}
A.~R. Sattler, ``Ionization produced by energetic silicon atoms within a
  silicon lattice,'' {\em Phys. Rev.}, vol.~138, p.~A1815, 1965.

\bibitem{germanium_qf}
A.~R. Sattler, F.~L. Vook, and J.~M. Palms, ``Ionization produced by energetic
  germanium atoms within a germanium lattice,'' {\em Phys. Rev.}, vol.~143,
  p.~588, 1966.

\bibitem{Muraz:2016upt}
J.~Muraz {\em et~al.}, ``{A table-top ion and electron beam facility for
  ionization quenching measurement and gas detector calibration},'' {\em Nucl.
  Instrum. Meth. A}, vol.~832, p.~214, 2016.

\bibitem{santos2008ionization}
D.~Santos {\em et~al.}, ``Ionization quenching factor measurement of helium
  4,'' 2008.

\bibitem{Tampon:2017mpm}
B.~Tampon {\em et~al.}, ``{Ionization Quenching Factor measurement of 1 keV to
  25 keV protons in Isobutane gas mixture},'' {\em EPJ Web Conf.}, vol.~153,
  p.~01014, 2017.

\bibitem{NEWS-G:2022fym}
{NEWS-G Collaboration}, ``{Measurements of the ionization efficiency of protons
  in methane},'' {\em Eur. Phys. J. C}, vol.~82, p.~1114, 2022.

\bibitem{NEWS-G:2021mhf}
{NEWS-G Collaboration}, ``{Quenching factor measurements of neon nuclei in neon
  gas},'' {\em Phys. Rev. D}, vol.~105, p.~052004, 2022.

\bibitem{ZIEGLER20101818}
J.~F. Ziegler, M.~Ziegler, and J.~Biersack, ``Srim - the stopping and range of
  ions in matter (2010),'' {\em Nucl. Instrum. Meth. B}, vol.~268, p.~1818,
  2010.

\bibitem{VDNguyen_1980}
V.~D. Nguyen {\em et~al.}, ``Recent experimental results on w-values for heavy
  particles,'' {\em Phys. Med. Biol.}, vol.~25, p.~509, 1980.

\bibitem{EWaibel_1992}
E.~Waibel and B.~Grosswendt, ``W values and other transport data on low energy
  electrons in tissue equivalent gas,'' {\em Phys. Med. Biol.}, vol.~37,
  p.~1127, 1992.

\bibitem{Abgrall:2013rze}
N.~Abgrall {\em et~al.}, ``{The Majorana Demonstrator Neutrinoless Double-Beta
  Decay Experiment},'' {\em Adv. High Energy Phys.}, vol.~2014, p.~365432,
  2014.

\bibitem{Hoppe2009JRADIOANALNUCLCHEM}
E.~W. Hoppe {\em et~al.}, ``Microscopic evaluation of contaminants in
  ultra-high purity copper,'' {\em J. Radioanal. Nucl. Chem.}, vol.~282,
  p.~315, 2009.

\bibitem{Overman2012TECHREPmajoranaCopper}
N.~Overman {\em et~al.}, ``Majorana electroformed copper mechanical analysis,''
  Tech. Rep. PNNL-21315, Pacific Northwest National Laboratory, 2012.

\bibitem{Overman2015}
N.~R. Overman {\em et~al.}, ``Mechanical property anisotropy in ultra-thick
  copper electrodeposits,'' {\em Appl. Phys. A}, vol.~120, p.~1181, 2015.

\bibitem{bard1985standard}
A.~Bard, R.~Parsons, and J.~Jordan, {\em Standard Potentials in Aqueous
  Solution}.
\newblock Monographs in Electroanalytical Chemistry and Electrochemistr, Taylor
  \& Francis, 1985.

\bibitem{atkins1997crc}
P.~Atkins, {\em Physical Chemistry, 6th edition}.
\newblock W.H. Freeman and Company, New York, 1997.

\bibitem{lide2006crc}
D.~Lide, {\em CRC Handbook of Chemistry and Physics, 87th Edition}.
\newblock Taylor \& Francis, 2006.

\bibitem{haynes2011crc}
W.~Haynes, {\em CRC Handbook of Chemistry and Physics, 92nd Edition}.
\newblock CRC Press, 2011.

\bibitem{Vitale:2021xrm}
A.~Vitale {\em et~al.}, ``{A preliminary investigation into the
  electrodeposition and synthesis of radiopure Copper-Chromium
  Copper\textendash{}Chromium alloys for rare-event physics detector
  systems},'' {\em Nucl. Instrum. Meth. A}, vol.~1003, p.~165291, 2021.

\bibitem{NEWS-G:2022kon}
{NEWS-G Collaboration}, ``{The NEWS-G detector at SNOLAB}.'' 2022.

\bibitem{Mavrokoridis:2011wv}
K.~Mavrokoridis {\em et~al.}, ``{Argon Purification Studies and a Novel Liquid
  Argon Re-circulation System},'' {\em J. Instrum.}, vol.~6, p.~P08003, 2011.

\bibitem{Knights:2021}
K.~Altenm\"uller {\em et~al.}, ``{Purification Efficiency and Radon Emanation
  of Gas Purifiers used with Pure and Binary Gas Mixtures for Gaseous Dark
  Matter Detectors},'' in {\em {2021 IEEE Nuclear Science Symposium (NSS) and
  Medical Imaging Conference (MIC) and 28th International Symposium on
  Room-Temperature Semiconductor Detectors}}, 2021.

\bibitem{pobrienMastersThesis}
P.~O'Brien, ``Optimization of processing parameters and development of a radon
  trapping system for the news-g dark matter detector,'' 2021.

\bibitem{Cowan:2010js}
G.~Cowan, K.~Cranmer, E.~Gross, and O.~Vitells, ``{Asymptotic formulae for
  likelihood-based tests of new physics},'' {\em Eur. Phys. J. C}, vol.~71,
  p.~1554, 2011.
\newblock [Erratum: Eur.Phys.J.C 73, 2501 (2013)].

\bibitem{SuperCDMS:2017nns}
R.~Agnese {\em et~al.}, ``{Low-mass dark matter search with CDMSlite},'' {\em
  Phys. Rev. D}, vol.~97, p.~022002, 2018.

\bibitem{Collar:2018ydf}
J.~I. Collar, ``{Search for a nonrelativistic component in the spectrum of
  cosmic rays at Earth},'' {\em Phys. Rev. D}, vol.~98, p.~023005, 2018.

\bibitem{CRESST:2022dtl}
G.~Angloher {\em et~al.}, ``{Testing spin-dependent dark matter interactions
  with lithium aluminate targets in CRESST-III},'' {\em Phys. Rev. D},
  vol.~106, p.~092008, 2022.

\bibitem{Behnke:2016lsk}
E.~Behnke {\em et~al.}, ``{Final Results of the PICASSO Dark Matter Search
  Experiment},'' {\em Astropart. Phys.}, vol.~90, pp.~85--92, 2017.

\bibitem{PICO:2019vsc}
C.~Amole {\em et~al.}, ``{Dark Matter Search Results from the Complete Exposure
  of the PICO-60 C$_3$F$_8$ Bubble Chamber},'' {\em Phys. Rev. D}, vol.~100,
  p.~022001, 2019.

\bibitem{Yellin:2002xd}
S.~Yellin, ``{Finding an upper limit in the presence of unknown background},''
  {\em Phys. Rev. D}, vol.~66, p.~032005, 2002.

\bibitem{Baxter:2021pqo}
D.~Baxter {\em et~al.}, ``{Recommended conventions for reporting results from
  direct dark matter searches},'' {\em Eur. Phys. J. C}, vol.~81, p.~907, 2021.

\bibitem{Billard:2013qya}
J.~Billard, L.~Strigari, and E.~Figueroa-Feliciano, ``{Implication of neutrino
  backgrounds on the reach of next generation dark matter direct detection
  experiments},'' {\em Phys. Rev. D}, vol.~89, p.~023524, 2014.

\bibitem{NEWS-G:2024jms}
{NEWS-G Collaboration}, ``{Search for light dark matter with NEWS-G at the LSM
  using a methane target},'' 2024.

\bibitem{Cox:2022ekg}
P.~Cox, M.~J. Dolan, C.~McCabe, and H.~M. Quiney, ``{Precise predictions and
  new insights for atomic ionization from the Migdal effect},'' {\em Phys. Rev.
  D}, vol.~107, p.~035032, 2023.

\bibitem{Lovesey:1982}
S.~W. {Lovesey}, C.~D. {Bowman}, and R.~G. {Johnson}, ``{Electron excitation in
  atoms and molecules by neutron-nucleus scattering},'' {\em Zeitschrift fur
  Physik B Condensed Matter}, vol.~47, p.~137, 1982.

\bibitem{Blanco:2022pkt}
C.~Blanco {\em et~al.}, ``{Molecular Migdal effect},'' {\em Phys. Rev. D},
  vol.~106, p.~115015, 2022.

\bibitem{Essig:2017kqs}
R.~Essig, T.~Volansky, and T.-T. Yu, ``{New Constraints and Prospects for
  sub-GeV Dark Matter Scattering off Electrons in Xenon},'' {\em Phys. Rev. D},
  vol.~96, p.~043017, 2017.

\bibitem{XENON:2019gfn}
E.~Aprile {\em et~al.}, ``{Light Dark Matter Search with Ionization Signals in
  XENON1T},'' {\em Phys. Rev. Lett.}, vol.~123, p.~251801, 2019.

\bibitem{PandaX-II:2021nsg}
C.~Cheng {\em et~al.}, ``{Search for Light Dark Matter-Electron Scatterings in
  the PandaX-II Experiment},'' {\em Phys. Rev. Lett.}, vol.~126, p.~211803,
  2021.

\bibitem{Essig:2015cda}
R.~Essig, M.~Fernandez-Serra, J.~Mardon, A.~Soto, T.~Volansky, and T.-T. Yu,
  ``{Direct Detection of sub-GeV Dark Matter with Semiconductor Targets},''
  {\em J. High Energy Phys.}, vol.~05, p.~046, 2016.

\bibitem{Hamaide:2021hlp}
L.~Hamaide and C.~McCabe, ``{Fueling the search for light dark matter-electron
  scattering with spherical proportional counters},'' {\em Phys. Rev. D},
  vol.~107, p.~063002, 2023.

\bibitem{Burdin:2014xma}
S.~Burdin {\em et~al.}, ``{Non-collider searches for stable massive
  particles},'' {\em Phys. Rept.}, vol.~582, p.~1, 2015.

\bibitem{Raby:1997pb}
S.~Raby, ``{Gauge mediated SUSY breaking at an intermediate scale},'' {\em
  Phys. Rev. D}, vol.~56, p.~2852, 1997.

\bibitem{Hardy:2014mqa}
E.~Hardy, R.~Lasenby, J.~March-Russell, and S.~M. West, ``{Big Bang Synthesis
  of Nuclear Dark Matter},'' {\em J. High Energy Phys.}, vol.~06, p.~011, 2015.

\bibitem{Ponton:2019hux}
E.~Pont\'on, Y.~Bai, and B.~Jain, ``{Electroweak Symmetric Dark Matter
  Balls},'' {\em J. High Energy Phys.}, vol.~09, p.~011, 2019.

\bibitem{Lehmann:2019zgt}
B.~V. Lehmann, C.~Johnson, S.~Profumo, and T.~Schwemberger, ``{Direct detection
  of primordial black hole relics as dark matter},'' {\em J. Cosmol. Astropart.
  Phys.}, vol.~10, p.~046, 2019.

\bibitem{Adhikari:2021fum}
D.~Collaboration, ``{First Direct Detection Constraints on Planck-Scale Mass
  Dark Matter with Multiple-Scatter Signatures Using the DEAP-3600 Detector},''
  {\em Phys. Rev. Lett.}, vol.~128, p.~011801, 2022.

\bibitem{Clark:2020mna}
M.~Clark {\em et~al.}, ``{Direct Detection Limits on Heavy Dark Matter},'' {\em
  Phys. Rev. D}, vol.~102, p.~123026, 2020.

\bibitem{Bhoonah:2020fys}
A.~Bhoonah, J.~Bramante, B.~Courtman, and N.~Song, ``{Etched plastic searches
  for dark matter},'' {\em Phys. Rev. D}, vol.~103, p.~103001, 2021.

\bibitem{Acevedo:2021tbl}
J.~F. Acevedo, J.~Bramante, and A.~Goodman, ``{Old rocks, new limits: excavated
  ancient mica searches for dark matter},'' {\em J. Cosmol. Astropart. Phys.},
  vol.~11, p.~085, 2023.

\bibitem{Peccei:1977hh}
R.~D. Peccei and H.~R. Quinn, ``{CP Conservation in the Presence of
  Instantons},'' {\em Phys. Rev. Lett.}, vol.~38, p.~1440, 1977.

\bibitem{Dienes:1999gw}
K.~R. Dienes, E.~Dudas, and T.~Gherghetta, ``{Invisible axions and large radius
  compactifications},'' {\em Phys. Rev. D}, vol.~62, p.~105023, 2000.

\bibitem{Chang:1999si}
S.~Chang, S.~Tazawa, and M.~Yamaguchi, ``{Axion model in extra dimensions with
  TeV scale gravity},'' {\em Phys. Rev. D}, vol.~61, p.~084005, 2000.

\bibitem{DiLella:2002ea}
L.~DiLella and K.~Zioutas, ``{Observational evidence for gravitationally
  trapped massive axion(-like) particles},'' {\em Astropart. Phys.}, vol.~19,
  pp.~145--170, 2003.

\bibitem{VazquezdeSolaFernandez:2020xcp}
F.~Vazquez~de Sola~Fernandez, {\em {Solar KK axion search with NEWS-G}}.
\newblock PhD thesis, Queen's U., Kingston, 2020.

\bibitem{NEWS-G:2021vfh}
{NEWS-G Collaboration}, ``{Solar Kaluza-Klein axion search with NEWS-G},'' {\em
  Phys. Rev. D}, vol.~105, p.~012002, 2022.

\bibitem{Akimov:2017ade}
{COHERENT Collaboration}, ``{Observation of Coherent Elastic Neutrino-Nucleus
  Scattering},'' {\em Science}, vol.~357, p.~1123, 2017.

\bibitem{Katsioulas:2016fet}
I.~Katsioulas, {\em {Study and detection of low energy neutrinos}}.
\newblock PhD thesis, Aristotle U., Thessaloniki, 2016.

\bibitem{Vidal:2021tmt}
M.~Vidal, {\em {Quenching factor measurement of neon nuclei in neon gas and
  study of the feasibility of detecting coherent elastic neutrino-nucleus
  scattering at a nuclear reactor using a spherical proportional counter}}.
\newblock PhD thesis, Queen's U., Kingston, 2021.

\bibitem{Heusser:2015ifa}
G.~Heusser {\em et~al.}, ``{GIOVE - A new detector setup for high sensitivity
  germanium spectroscopy at shallow depth},'' {\em Eur. Phys. J. C}, vol.~75,
  p.~531, 2015.

\bibitem{Buck:2020opf}
C.~Buck {\em et~al.}, ``{A novel experiment for coherent elastic neutrino
  nucleus scattering: CONUS},'' {\em J. Phys. Conf. Ser.}, vol.~1342,
  p.~012094, 2020.

\bibitem{Vergados:2005ny}
J.~D. Vergados and Y.~Giomataris, ``{Dedicated supernova detection by a network
  of neutral current spherical tpc's},'' {\em Phys. Atom. Nucl.}, vol.~70,
  p.~140, 2007.

\bibitem{Lang:2016zhv}
R.~F. Lang, C.~McCabe, S.~Reichard, M.~Selvi, and I.~Tamborra, ``{Supernova
  neutrino physics with xenon dark matter detectors: A timely perspective},''
  {\em Phys. Rev. D}, vol.~94, p.~103009, 2016.

\bibitem{Brooks:2002oyj}
F.~D. Brooks and H.~Klein, ``{Neutron spectrometry\textemdash{}historical
  review and present status},'' {\em Nucl. Instrum. Meth. A}, vol.~476, p.~1,
  2002.

\bibitem{Pietropaolo:2020frm}
A.~Pietropaolo {\em et~al.}, ``{Neutron detection techniques from
  \ensuremath{\mu}eV to GeV},'' {\em Phys. Rept.}, vol.~875, p.~1, 2020.

\bibitem{KOUZES20101035}
R.~T. Kouzes {\em et~al.}, ``Neutron detection alternatives to 3he for national
  security applications,'' {\em Nucl. Instrum. Meth. A}, vol.~623, p.~1035,
  2010.

\bibitem{Piquemal:2012fs}
F.~Piquemal, ``{Modane underground laboratory: Status and project},'' {\em Eur.
  Phys. J. Plus}, vol.~127, p.~110, 2012.

\bibitem{Bougamont:2015jzx}
E.~Bougamont {\em et~al.}, ``{Neutron spectroscopy with the Spherical
  Proportional Counter based on nitrogen gas},'' {\em Nucl. Instrum. Meth. A},
  vol.~847, p.~10, 2017.

\bibitem{Giomataris:2022kxw}
I.~Giomataris {\em et~al.}, ``{Neutron spectroscopy: The case of the spherical
  proportional counter},'' {\em Nucl. Instrum. Meth. A}, vol.~1045, p.~167590,
  2023.

\end{thebibliography}

\end{document}